\documentclass[3p,twoside,fleqn,sort&compress]{elsarticle}
\usepackage{amssymb,amsmath,latexsym}
\usepackage[varg]{pxfonts}
\usepackage{mathrsfs}
\usepackage{pst-all}
\font\tenscr=rsfs10  scaled \magstep0 \font\sevenscr=rsfs7 scaled \magstep0
\font\fivescr=rsfs5 scaled \magstep0 \skewchar\tenscr='177 \skewchar\sevenscr='177
\skewchar\fivescr='177
\newfam\scrfam \textfont\scrfam=\tenscr \scriptfont\scrfam=\sevenscr
\scriptscriptfont\scrfam=\fivescr
\def\mathscr{\fam\scrfam}
\renewcommand{\cal}{\mathscr}
\font\stenscr=rsfs5  scaled \magstep3
\newfam\sscrfam \textfont\sscrfam=\stenscr
\def\smathscr{\fam\sscrfam}
\newcommand{\scal}{\smathscr}
\newtheorem{theorem}{Theorem}[section]
\newtheorem{lemma}[theorem]{Lemma}
\newtheorem{corollary}[theorem]{Corollary}
\newtheorem{remark}[theorem]{Remark}

\newtheorem{example}[theorem]{Example}
\newtheorem{proposition}[theorem]{Proposition}

\def\endproof{\qed\endtrivlist}
\expandafter\let\csname endproof*\endcsname=\endproof

\def\qedsymbol{\ifmmode\bgroup\else$\bgroup\aftergroup$\fi
  \vcenter{\hrule\hbox{\vrule height.6em\kern.6em\vrule}\hrule}\egroup}
\def\qed{\ifmmode\else\unskip\nobreak\fi\quad\qedsymbol}

\newcommand{\ind}{\mbox{\rm ind}}

\newcommand{\im}{\qopname\relax{no}{Im}}

\usepackage{color}

\begin{document}

\journal{\qquad}

\title{\LARGE\bf Bisimulations for fuzzy automata\tnoteref{t1}}
\tnotetext[t1]{Research supported by Ministry  of Science and Technological Development, Republic
of Serbia, Grant No. 174013}
\author[fsmun]{Miroslav \'Ciri\'c\corref{cor}}
\ead{miroslav.ciric@pmf.edu.rs}

\author[fsmun]{Jelena Ignjatovi\'c}
\ead{jelena.ignjatovic@pmf.edu.rs}

\author[tfc]{Nada Damljanovi\'c}
\ead{nada@tfc.kg.ac.rs}

\author[fsmun]{Milan Ba\v si\'c}
\ead{basic\symbol{95}milan@yahoo.com}

\cortext[cor]{Corresponding author. Tel.: +38118224492; fax: +38118533014.}

\address[fsmun]{University of Ni\v s, Faculty of Sciences and Mathematics, Vi\v segradska 33, P. O. Box 224, 18000 Ni\v s, Serbia}
\address[tfc]{University of Kragujevac, Technical faculty in \v Ca\v cak, Svetog Save 65, P. O. Box 131, 32000 \v Ca\v cak, Serbia}

\begin{abstract}\small
Bisimulations have been widely used in many areas of computer science to model equivalence between various~systems, and to reduce
the number of states of these systems, whereas uniform fuzzy relations have recently been intro\-duced as a means to model the fuzzy equivalence between elements of two possible different sets.~Here we use~the~con\-junction of these two concepts as a powerful tool in the study of equivalence between fuzzy automata.~We prove~that a uniform fuzzy relation between fuzzy automata $\scal A$ and $\scal B$ is a forward bisimulation if and only if its kernel and co-kernel are forward bisimulation fuzzy equivalence relations on $\scal A$ and $\scal B$ and there is a special isomorphism between factor fuzzy automata with respect to these fuzzy equivalence relations.~As a consequence we get that fuzzy auto\-mata $\scal A$~and $\scal B$ are UFB-equivalent,~i.e., there is a uniform forward bisimulation between them, if and only if there is a special isomorphism between the~factor fuzzy automata of $\scal A$ and $\scal B$ with respect~to~their greatest forward bisimulation fuzzy equivalence
relations.~This result~re\-duces the problem of testing UFB-equivalence to the problem of testing isomorphism of fuzzy automata, which is closely related to the well-known graph isomorphism problem.~We prove some similar results for backward-forward bisimulations, and we point to fundamental differences.~Because of the duality with~the studied concepts, backward and forward-backward bisimulations are not considered separately.~Finally, we give a comprehensive overview of various concepts on deterministic, nondeterministic, fuzzy, and weighted automata, which are related to bisimulations.


\end{abstract}

\maketitle

\section{Introduction}\label{sectionIntroduction}

Study of fuzzy automata and languages was initiated in 1960s by Santos \cite{Santos.68,Santos.72,Santos.76}, Wee \cite{Wee.67}, Wee and~Fu \cite{Wee.Fu.69}, and Lee and Zadeh \cite{Lee.Zadeh.69}. From late 1960s until early 2000s mainly fuzzy automata and languages~with membership values in the G\"odel structure have been considered (see for example \cite{DP.80,GSG.77,MM.02}).~The idea of studying~fuzzy automata with membership values in some structured abstract set comes back to W. Wechler \cite{Wechler.78}, and in recent~years~researcher's  attention has been aimed mostly to fuzzy automata with membership values in complete residuated lattices, lattice-ordered moinoids, and other kinds of lattices.~Fuzzy automata taking membership values in a complete residuated
lattice were first studied by D. W. Qiu in \cite{Qiu.01,Qiu.02}, where some basic concepts
were discussed, and later, D. W. Qiu and his coworkers have carried out~extensive
research of these fuzzy automata (cf.~\cite{Qiu.04,Qiu.06,WQ.10,XQ.09a,XQ.09b,XQL.09,XQLF.07}).~From a different point
of view, fuzzy automata~with membership values in a complete residuated lattice were
studied by J. Ignjatovi\'c, M. \'Ciri\'c and their co\-work\-ers in \cite{CSIP.07,CSIP.10,IC.10,ICB.08,ICB.09,ICBP.10,SCI.10}.~Fuzzy automata with membership values in a lattice-ordered monoid~were inves\-tigated by Y. M. Li and others \cite{LL.07,LL.06,LP.05,3Li.06}, fuzzy automata over other types of lattices were the subject of
\cite{DV.10,Li.11,LP.07,KSY.07,KL.07,P.04a,P.04b,PK.04,PZ.08}, and automata which generalize fuzzy automata over any type of lattices, as well as weighted automata over semirings, have been studied recently in \cite{CDIV.10,DSV.10,JIC.11}.~Notably, fuzzy automata have been used to simulate discrete event systems (see \cite{LY.02,Qiu.05,QL.09}).~In the present paper, we
will study bisimulations and related properties concerning fuzzy automata over a complete residuated lattice.

Bisimulations are generally considered as one of the most important contributions of concurrency~theory to computer science, although they have been discovered not only in computer science, but also (and roughly at the same time) in modal logic and set theory.~They are employed today in a number of areas of computer science, such as functional languages, object-oriented languages, types, data types, domains, databases, compiler optimizations, program analysis, verification tools, etc.~For more information about bisimulations we refer to \cite{DPP.04,GPP.03,LV.95,Milner.89,Milner.99,RM-C.00,Sang.09}.

In concurrency theory, bisimulations were introduced by Milner \cite{Milner.80} and Park \cite{Park.81} as a means for testing behavioural equivalence among processes, but they have been also very successfully exploited to reduce~the state-space of processes.~The most common structures on which bisimulations have been studied are labelled transition systems.~They are
essentially labelled directed graphs, or nondeterministic automata, when initial and terminal states are added.~Bisimulations have been mostly used to model equivalence between states~of the same system, and to reduce
the number of states of this system.~In particular, many algorithms have been proposed to compute the greatest bisimulation equivalence relation on a given labelled graph or a labeled transition system, and the faster ones are based on the crucial equivalence between the greatest bisimulation equivalence
relation and the relational coarsest partition problem (cf.~\cite{DPP.04,GPP.03,KS.90,RT.08,PT.87}).~Bisimulations between states of two distinct systems have been  much less studied, probably due to lack of an appropriate~concept of a relation between two distinct sets
which would behave like an equivalence.~The most often, bisimulations have been considered either as arbitrary relations (which have shown oneself to be too general), or as functions (which are too special).~However, a
more suitable concept has appeared recently in \cite{CIB.09}, the~concept of a uniform relation, and in the fuzzy framework the concept of a uniform fuzzy rela\-tion.

The original intention of the authors in \cite{CIB.09} was to introduce uniform fuzzy relations as a basis for defining such concept of a fuzzy function which would provide a correspondence between fuzzy functions and fuzzy equivalence
relations, analogous to the correspondence between crisp functions and crisp~equiva\-lence relations.~This was done, but also, it turned out that uniform fuzzy relations
establish natural relationships between fuzzy partitions of two sets, some kind of ``uniformity'' between these fuzzy partitions.~Roughly speaking,~uniform fuzzy relations can be conceived as fuzzy equivalence relations which relate elements of two possibly
different~sets.~In~[19], uniform fuzzy relations were employed to solve some systems of fuzzy relation equations, systems that have important applications in approximate reasoning, especially in fuzzy control.~Then in \cite{ICB.09}, they were used to define and study fuzzy homomorphisms and fuzzy relational morphisms of algebras, and to establish relationships between fuzzy homomorphisms, fuzzy relational morphisms, and fuzzy congru\-ences, analogous to relationships between homomorphisms, relational morphisms, and congruences in classical algebra.~In the same paper, fuzzy relational morphisms were also applied to deterministic fuzzy automata.~We~will see later that fuzzy~relational morphisms are the same as forward bisimulations (in the terminology used in this paper) when these two concepts are considered in the context of deterministic fuzzy automata.

The main aim of this paper is to show that the conjunction of two concepts, uniform fuzzy~relations~and bisimulations, provides a very powerful tool in the study of equivalence between fuzzy automata,~as well as between some related structures.~In this symbiosis, uniform fuzzy relations serve as fuzzy~equivalence
relations which relate elements of two possibly
different sets, while bisimulations ensure compati\-bility with the~transitions, initial and terminal states of fuzzy automata.~Our second goal is to try~to~propose some names that could better regulate very confused terminology in dealing with simulations and bisimulations.~And third, we want to demonstrate the algebraic way of looking at this issue, and to connect the basic concepts of our study with the fundamental algebraic concepts of a homomorphism, congruence and relational morphism.

Our main results are the following.~We define two kinds of simulations, forward and backward simulations, and considering four possible cases when a fuzzy relation and its inverse are forward or backward simu\-lations, we define four types of bisimulations.~Two~types of bisimulations are homotypic, where both the considered fuzzy relation and its inverse are forward or backward simulations (forward and backward bisimulations), and two types are heterotypic, where one of them is a forward and the other is a backward simulation (backward-forward and forward-backward bisimulations).~As forward and backward bisimula\-tions, and backward-forward and forward-backward bisimulations, are dual concepts, our article deals~only with forward and backward-forward bisimulations.~First we examine forward bisimulations.~We~prove~that if there is a forward bisimulation between two fuzzy automata, then there is the greatest one and it is a~partial fuzzy function (cf.~Theorem~\ref{th:GFB}).~This result emphasizes the importance of bisimulations that are uniform fuzzy relations, and we turn to uniform forward bisimulations.~Given any uniform fuzzy relation $\varphi $ between fuzzy automata $\cal A$ and $\cal B$, we prove that $\varphi $ is a forward simulation if and only if its kernel $E_A^\varphi $ and co-kernel $E_B^\varphi $ are forward bisimulation fuzzy equivalence relations on $\cal A$ and $\cal B$, and there is a special isomorphism between related factor fuzzy automata (cf.~Theorem~\ref{th:ufbr}).~In addition, we prove other results that relate bisimulations, bisimulation fuzzy equivalence relations and factor fuzzy automata, and resemble to homomorphism and isomorphism theorems, and other algebraic results connecting homomorphisms, congruences, and relational morphisms (cf.~Theorems~\ref{th:nat.uff} and \ref{th:G:E}).~We also provide another way to define uniform forward bisimulations, where the inequalities that appear in the original definition are replaced by equalities (cf.~Theorem~\ref{th:ufbreqcond}), and determine necessary and sufficient conditions for the existence of a uniform forward bisimulation with a given kernel and co-kernel (cf.~Theorem~\ref{th:ufb.ex}).

Further, two automata $\cal A$ and $\cal B$ are defined to be UFB-equivalent if there is a uniform forward bisimulation between $\cal A$ and $\cal B$.~Our main result asserts that if two fuzzy automata $\cal A$ and $\cal B$ are UFB-equivalent, then there is a uniform forward bisimulation between them whose kernel and co-kernel are the greatest forward bisimulations on $\cal A$ and $\cal B$ (cf.~Theorem~\ref{th:UFBeq.great}).~This result and its consequences provide a way to test whether two given fuzzy automata $\cal A$ and $\cal B$ are UFB-equivalent.~First, we have to compute the greatest forward bisimulation fuzzy equivalence relations $E$ on $\cal A$ and $F$ on $\cal B$.~In numerous cases this can be done effectively using the algorithm~developed in \cite{CSIP.10}.~After that, we construct factor fuzzy automata ${\cal A}/E$ and ${\cal B}/F$, and check if there is an isomorphism between them that satisfies certain condition.~But, even when we are able to~effec\-tively compute the greatest forward bisimulations $E$ and $F$ and construct the factor fuzzy automata ${\cal A}/E$ and ${\cal B}/F$, it may be a very hard problem to determine whether there is an isomorphism between ${\cal A}/E$ and ${\cal B}/F$.~This problem is closely related to the well-known graph isomorphism problem.

In our discussion of backward-forward bisimulations we emphasize the similarities and fundamental differences between them and forward bisimulations.~Finally, we give a comprehensive overview of various concepts on deter\-ministic, nondeterministic, fuzzy, and weighted automata, which are related to the algebraic concepts of a homomorphism, congruence and relational morphism.~In particular, we discuss relationships between these concepts and the concepts of bisimulations.

The paper is organized as follows.~In Section \ref{sectionPreliminaries} we give definitions of basic notions and notation~concern\-ing fuzzy sets and relations, and in Section \ref{sectionAutomata} we define basic notions and notation concerning fuzzy~auto\-mata.~Section \ref{sectionUniform} contains a different presentation of some results from \cite{CIB.09} on uniform fuzzy relations,~as well as some new results on these relations.~In Section \ref{sectionBisimulations} we define four types of bisimulations and we discuss the main properties of forward bisimulations.~Section \ref{sectionUFB} provides characterization results for uniform~forward bisimulations.~Then, in Section \ref{sectionUFBeq} we define UFB-equi\-valence between fuzzy automata, prove the main characterization result for UFB-equivalent fuzzy automata, and consider the problem of testing UFB-equi\-valence.~Section \ref{sectionBFB} discuss basic properties of backward-forward bisimulations and points to similarities and fundamental differences between them and forward bisimulations.~Finally, Section \ref{sec:related} gives an overview of various concepts related to bisimulations.

It is worth mentioning that from a different point of view problems of equivalence, state reduction and minimization  of fuzzy automata have been discussed in a number of papers.~More information on these issues will be given in Section \ref{sec:related}.

\section{Preliminaries}\label{sectionPreliminaries}

The terminology and basic notions in this section are according to \cite{Bel.02,BV.05,KY.95}.

We will use complete residuated lattices as the structures of membership (truth) values.~Residuated lattices are a
very general algebraic structure and generalize many  algebras with very important applications (see for example \cite{Hajek.98,Hohle.95,Qiu.07}).~A {\it residuated lattice\/} is an algebra ${\cal L}=(L,\land
,\lor , \otimes ,\to , 0, 1)$ such that
\begin{itemize}
\parskip=-2pt
\item[{\rm (L1)}] $(L,\land ,\lor , 0, 1)$ is a lattice with the least element $0$ and the
greatest element~$1$,
\item[{\rm (L2)}] $(L,\otimes ,1)$ is a commutative monoid with the unit $1$,
\item[{\rm (L3)}] $\otimes $ and $\to $ form an {\it adjoint pair\/}, i.e., they satisfy the
{\it adjunction property\/}: for all $x,y,z\in L$,
\begin{equation}\label{eq:adj}
x\otimes y \leqslant z \ \Leftrightarrow \ x\leqslant y\to z .
\end{equation}
\end{itemize}
If, in addition, $(L,\land ,\lor , 0, 1)$ is a complete lattice, then ${\cal L}$ is called a {\it
complete residuated lattice\/}.~Emphasizing their monoidal structure, in some sources residuated lattices are called
integral, commutative, residuated $\ell $-monoids \cite{Hohle.95}.

The operations $\otimes $ (called {\it multiplication\/}) and $\to
$ (called {\it residuum\/}) are intended for modeling the conjunction and implication of the
corresponding logical calculus, and supremum ($\bigvee $) and infimum ($\bigwedge $) are intended
for modeling of the existential and general quantifier, respectively. An operation $\leftrightarrow $ defined
by
\begin{equation}\label{eq:bires}
x\leftrightarrow y = (x\to y) \land (y\to x),
\end{equation}
called {\it biresiduum\/} (or {\it biimplication\/}), is used for modeling the equivalence of truth
values.

It can be easily verified that with respect to $\leqslant $, $\otimes $ is isotonic
in both arguments, $\to $ is isotonic in the second and antitonic in the first
argument, and for any $x,y,z\in L$ and any $\{x_i\}_{i\in I},\{y_i\}_{i\in
I}\subseteq L$, the following hold:
\begin{eqnarray}
&& x\leftrightarrow y \leqslant x\otimes z\leftrightarrow y\otimes z ,\label{eq:bir.mult} \\
&& \Bigl(\bigvee_{i\in I}x_i\Bigr)\otimes x=\bigvee_{i\in I}(x_i\otimes x) ,\label{eq:inf.dist} \\
&& \bigwedge_{i\in I}(x_1\leftrightarrow y_i)\leqslant \Bigl(\bigwedge_{i\in I}x_i\Bigr)\leftrightarrow
\Bigl(\bigwedge_{i\in I}y_i\Bigr) \label{eq:bir.inf.inf}\\
&& \bigwedge_{i\in I}(x_i\leftrightarrow y_i)\leqslant \Bigl(\bigvee_{i\in I}x_i\Bigr)\leftrightarrow
\Bigl(\bigvee_{i\in I}y_i\Bigr). \label{eq:bir.inf.sup}
\end{eqnarray}
For other properties of complete residuated lattices we refer to \cite{Bel.02,BV.05}.

The most studied and applied structures of truth values, defined
on the real unit interval $[0,1]$ with $x\land y =\min
(x,y)$ and $x\lor y =\max (x,y)$, are the {\it {\L}ukasiewicz
structure\/} (where $x\otimes y = \max(x+y-1,0)$, $x\to y=
\min(1-x+y,1)$), the {\it Goguen} ({\it product\/}) {\it
structure\/} ($x\otimes y = x\cdot y$, $x\to y= 1$ if $x\leqslant y$,
and~$=y/x$ otherwise), and the {\it G\"odel structure\/} ($x\otimes
y = \min(x,y)$, $x\to y= 1$ if $x\leqslant y$, and $=y$
otherwise).~More~generally, an algebra $([0,1],\land ,\lor ,
\otimes,\to , 0, 1)$ is a complete~resi\-duated lattice if and
only if $\otimes $ is a left-continuous t-norm and the residuum is
defined by $x\to y = \bigvee \{u\in [0,1]\,|\, u\otimes x\leqslant y\}$ (cf.~\cite{BV.05}).~Another
important set of truth values is the set
$\{a_0,a_1,\ldots,a_n\}$, $0=a_0<\dots <a_n=1$, with $a_k\otimes
a_l=a_{\max(k+l-n,0)}$ and $a_k\to a_l=a_{\min(n-k+l,n)}$. A
special case of the latter algebras is the two-element Boolean
algebra of classical logic with the support $\{0,1\}$.~The only
adjoint pair on the two-element Boolean algebra consists of the
classical conjunction and implication operations.~This structure
of truth values we call the {\it Boolean structure\/}.

In the further text $\cal L$ will be a complete residuated lattice.~A {\it
fuzzy subset\/} of a set $A$ {\it over\/} ${\cal L}$, or~simply a {\it fuzzy
subset\/} of $A$, is any function from $A$ to $L$.~Ordinary crisp subsets~of~$A$ are considered as fuzzy subsets of $A$ taking membership values in the set
$\{0,1\}\subseteq L$.~Let $f$ and $g$ be two
fuzzy subsets of $A$.~The {\it equality\/} of $f$ and $g$ is defined as the
usual equality of functions, i.e., $f=g$ if and only if $f(x)=g(x)$, for every
$x\in A$. The {\it inclusion\/} $f\leqslant g$ is also defined pointwise:~$f\leqslant g$ if
and only if $f(x)\leqslant g(x)$, for every $x\in A$.~Endowed with this partial order the set ${\cal F}(A)$ of all fuzzy subsets of $A$ forms a complete residuated lattice, in which the
meet (intersection) $\bigwedge_{i\in I}f_i$ and the join (union) $\bigvee_{i\in I}f_i$ of an
arbitrary family $\{f_i\}_{i\in I}$ of fuzzy subsets of $A$ are functions from $A$ to $L$
defined by
\[
\left(\bigwedge_{i\in I}f_i\right)(x)=\bigwedge_{i\in I}f_i(x), \qquad \left(\bigvee_{i\in
I}f_i\right)(x)=\bigvee_{i\in I}f_i(x),
\]
and the {\it product\/} $f\otimes g$ is a fuzzy subset defined by $f\otimes g
(x)=f(x)\otimes g(x)$, for every $x\in A$.~

The {\it crisp~part\/} of a fuzzy subset $f$ of $A$ is a crisp
subset $\widehat f=\{a\in A\,|\, f(a)=1\}$ of $A$.~We~will~also consider $\widehat f$ as a
function $\widehat f:A\to L$ defined by $\widehat f(a)=1$, if $f(a)=1$, and $\widehat f(a)=0$,
if $f(a)<1$.

Let $A$ and $B$ be non-empty sets.~A {\it fuzzy relation between sets\/} $A$ and $B$ (in this order) is any function from $A\times B$ to~$L$, that is to say, any fuzzy subset of $A\times B$, and the equality, inclusion (ordering), joins and meets of fuzzy relations
are defined as for fuzzy sets.~In particular, a {\it fuzzy relation on a set\/} $A$ is any function from $A\times A$ to $L$, i.e., any fuzzy subset of $A\times A$.~To highlight the difference between fuzzy relations between two sets and those on a set, fuzzy relations between two sets will be usually denoted by small Greek letters, and fuzzy relations on a set by capital Latin letters.~The {\it inverse\/} of a fuzzy relation $\varphi \in {\cal F}(A\times B)$ is a fuzzy relation
$\varphi^{-1}\in {\cal F}(B\times A)$ defined by $\varphi^{-1}(b,a)=\varphi (a,b)$, for all $a\in A$ and $b\in B$.~A {\it crisp relation\/} is a fuzzy relation which takes values only in the set $\{0,1\}$, and if $\varphi $ is a crisp relation of $A$ to $B$, then expressions ``$\varphi (a,b)=1$'' and ``$(a,b)\in \varphi $'' will have the same meaning.

For non-empty sets $A$, $B$ and $C$, and fuzzy relations $\varphi\in {\cal F}(A\times B)$ and $\psi \in {\cal F}(B\times C)$, their {\it composition\/}~$\varphi\circ \psi$ is an fuzzy relation from ${\cal F}(A\times C)$ defined by
\begin{equation}\label{eq:comp.rr}
(\varphi \circ \psi )(a,c)=\bigvee_{b\in B}\,\varphi(a,b)\otimes \psi(b,c),
\end{equation}
for all $a\in A$ and $c\in C$.~If $\varphi $ and $\psi $ are crisp relations, then $\varphi\circ \psi $ is an ordinary composition of relations,~i.e.,
\[
\varphi\circ \psi=\left\{(a,c)\in A\times C\mid (\exists b\in B)\, (a,b)\in \varphi\ \&\ (b,c)\in \psi\right\},
\]
and if $\varphi $ and $\psi $ are functions, then $\varphi\circ \psi $ is an ordinary composition of functions, i.e.,
$(\varphi\circ \psi )(a)=\psi (\varphi(a))$, for every $a\in A$.~Next, if $f\in {\cal F}(A)$, $\varphi\in {\cal F}(A\times B)$ and $g\in {\cal F}(B)$, the {\it compositions\/} $f\circ \varphi$ and $\varphi\circ g$ are fuzzy subsets of $B$ and $A$, respectively, which are defined by
\begin{equation}\label{eq:comp.sr}
(f \circ \varphi)(b)=\bigvee_{a\in A}\,f(a)\otimes \varphi(a,b),\quad
(\varphi \circ g)(a)=\bigvee_{b\in B}\,\varphi(a,b)\otimes g(b),
\end{equation}
for every $a\in A$ and $b\in B$.

In particular, for $f,g\in {\cal F}(A)$ we write
\begin{equation}\label{eq:comp.ss}
f \circ g =\bigvee_{a\in A}\,f(a)\otimes g(a) .
\end{equation}
The value $f\circ g$ can be interpreted as the "degree of overlapping" of $f$ and $g$.~In particular, if $f$ and $g$ are crisp sets and $\varphi $ is a crisp relation, then
\[
f \circ \varphi = \left\{b\in B\mid (\exists a\in f)\, (a,b)\in\varphi \right\}, \ \ \
\varphi\circ g = \left\{a\in A\mid (\exists b\in g)\, (a,b)\in\varphi \right\}.
\]

Let $A$, $B$, $C$ and $D$ be non-empty sets. Then for any $\varphi_1\in {\cal F}(A\times B)$, $\varphi_2\in {\cal F}(B\times C)$ and $\varphi_3\in {\cal F}(C\times D)$~we~have
\begin{align}
&(\varphi_{1}\circ \varphi_{2})\circ \varphi_{3} = \varphi_{1}\circ (\varphi_{2}\circ \varphi_{3}), \label{eq:comp.as}
\end{align}
and for $\varphi_{0}\in {\cal F}(A\times B)$,\, $\varphi_{1},\varphi_{2}\in {\cal F}(B\times C)$ and $\varphi_{3}\in {\cal F}(C\times D)$~we~have that
\begin{align}
&\varphi_{1}\leqslant \varphi_{2}\ \ \text{implies}\ \ \varphi_1^{-1}\leqslant \varphi_2^{-1},\ \ \varphi_{0}\circ \varphi_{1} \leqslant  \varphi_{0}\circ \varphi_{2}, \ \ \text{and}\ \ \varphi_{1}\circ \varphi_{3} \leqslant  \varphi_{2}\circ \varphi_{3}. \label{eq:comp.mon}
\end{align}
Further, for any $\varphi\in {\cal F}(A\times B)$, $\psi\in {\cal F}(B\times C)$, $f\in {\cal F}(A)$, $g\in {\cal F}(B)$ and $h\in {\cal F}(C)$
we can easily verify~that
\begin{equation}\label{eq:rs.comp.as}
(f\circ \varphi)\circ \psi=f\circ (\varphi\circ \psi), \quad (f\circ \varphi)\circ g=f\circ (\varphi\circ g),\quad (\varphi\circ \psi)\circ h=\varphi\circ (\psi\circ h)
\end{equation}
and consequently, the parentheses~in (\ref{eq:rs.comp.as}) can be omitted, as well as the parentheses~in (\ref{eq:comp.as}).

Finally, for all $\varphi,\varphi_i\in {\cal F}(A\times B)$ ($i\in I$) and $\psi,\psi_i\in {\cal F}(B\times C)$ ($i\in I$) we have that
\begin{align}
&(\varphi\circ \psi)^{-1} = \psi^{-1}\circ \varphi^{-1}, \label{eq:comp.inv} \\
& \varphi\circ \bigl(\bigvee_{i\in I}\psi_i\bigr) = \bigvee_{i\in I}(\varphi\circ \psi_i) , \ \
\bigl(\bigvee_{i\in I}\varphi_i\bigr)\circ \psi = \bigvee_{i\in I}(\varphi_i\circ \psi), \label{eq:comp.sup} \\
&\bigl(\bigvee_{i\in I}\varphi_i\bigr)^{-1} = \bigvee_{i\in I}\varphi_i^{-1}. \label{eq:sup.inv}
\end{align}

We note that if $A$, $B$ and $C$ are finite sets of cardinality $|A|=k$, $|B|=m$ and $|C|=n$, then $\varphi \in {\cal F}(A\times B)$ and $\psi \in {\cal F}(B\times C)$ can be treated as $k\times m$ and $m\times n$ fuzzy matrices over $\cal L$, and $\varphi\circ \psi$ is the matrix~pro\-duct. Analogously, for $f\in {\cal F}(A)$ and $g\in {\cal F}(B)$ we can treat $f\circ \varphi$ as the product of a $1\times k$ matrix $f$ and a $k\times m$ matrix $\varphi $,
and $\varphi\circ g$ as the product of an $k\times m$ matrix $R$ and an $m\times 1$ matrix $g^t$
(the transpose~of~$g$).

A fuzzy relation $E  $ on a set $A$ is%
\begin{itemize}\parskip=0pt
\itemindent=5pt
\item[(R)] {\it reflexive\/} if $E(a,a)=1$, for every $a\in A$;
\item[(S)] {\it symmetric\/} if $E(a,b)=E(b,a)$, for all $a,b\in A$;
\item[(T)] {\it transitive\/} if $E(a,b)\otimes E(b,c)\leqslant E(a,c)$, for all
$a,b,c\in A$.
\end{itemize}
If $E$ is reflexive and transitive, then $E\circ E=E$.
A fuzzy relation on $A$ which is reflexive, symmetric and transitive is called a
{\it fuzzy equivalence relation\/}. With respect to the ordering of fuzzy
relations, the set ${\cal E}(A)$ of all fuzzy equivalence relations on a set $A$ is a complete
lattice, in which the meet coincide with the ordinary intersection of fuzzy relations, but in the
general case, the join in ${\cal E}(A)$ does not coincide with the ordinary union of fuzzy
relations.

For a fuzzy~equi\-valence relation
$E  $ on $A$ and $a\in A$ we define a fuzzy subset $E _a$ of $A$ by:
\[
E _a(x)=E  (a,x),\ \ \text{for every}\ x\in A .
\]
We call $E _a$ an {\it equivalence class\/} of $E$ deter\-mined by $a$.~The set ${A/E}  =\{E _a\,|\, a\in A\}$ is called the {\it
factor set\/} of $A$ with respect to $E$ (cf. \cite{Bel.02,CIB.07}).
Cardinality of the factor set ${A/E}  $, in notation $\ind (E)$, is called the {\it index\/}~of~$E$.
The same notation we use for crisp
equivalence relations, i.e., for an equivalence relation $\pi $ on $A$, the related factor set is denoted
by $A/\pi $, the equivalence class of an element $a\in A$ is denoted by $\pi_a$, and the index~of~$\pi $~is denoted by $\ind (\pi)$.

The following properties of fuzzy equivalence relations will be useful in the later work.

\begin{lemma}\label{le:feq}
Let $E$ be a fuzzy equivalence relation on a set $A$ and let $\widehat E$ be its crisp part.~Then
$\widehat E$ is a crisp~equivalence relation on $A$, and for every $a,b\in A$ the following
conditions are equivalent:
\begin{itemize}\parskip-2pt
\item[{\rm (i)}] $E(a,b)=1$;
\item[{\rm (ii)}] $E_a=E_b$;
\item[{\rm (iii)}] $\widehat E_a=\widehat E_b$;
\item[{\rm (iv)}] $E(a,c)=E(b,c)$, for every $c\in A$.
\end{itemize}
Consequently, $\ind(E)=\ind(\widehat E)$.
\end{lemma}
Note that $\widehat E_a$ denotes the crisp equivalence class of $\widehat E$ determined by $a$.

A fuzzy equivalence relation $E$ on a set $A$ is called a {\it fuzzy
equality\/} if for all $x,y\in A$, $E(x,y)=1$ implies $x=y$. In other words, $E$
is a fuzzy equality if and only if its crisp part $\widehat E$ is a crisp
equality.~For~a~fuzzy~equivalence relation $E$ on a set $A$, a fuzzy relation $\widetilde E$ defined
on the factor set ${A/E}$ by
\[
\widetilde E(E_x,E_y)= E(x,y) ,
\]
for all $x,y\in A$, is well-defined, and it is a fuzzy equality on ${A/E}$.

\section{Fuzzy automata}\label{sectionAutomata}

In this paper we study fuzzy automata with membership values in complete residuated lattices.

In the rest of the paper, if not noted otherwise, let ${\cal L}$ be a complete residuated lattice and let $X$ be an (finite) alphabet.
A {\it fuzzy automaton over\/} $\cal L$ and $X$ , or simply a {\it fuzzy automaton\/}, is a quadruple
${\cal A}=(A,\delta^A,\sigma^A,\tau^A )$, where~$A$ is a non-empty set, called  the
{\it set of states\/}, $\delta^A:A\times X\times A\to L$~is~a
fuzzy subset~of $A\times X\times A$, called the {\it fuzzy transition function\/}, and $\sigma^A: A\to L$ and $\tau^A : A\to L$ are the fuzzy subsets of~$A$, called the {\it fuzzy set of initial states} and the {\it fuzzy set terminal states}, respectively.~We can
interpret $\delta^A (a,x,b)$ as the degree~to~which an~input letter $x\in X$~causes~a~transition from a state $a\in A$ into a
state $b\in A$, whereas we can interpret $\sigma^A(a)$ and $\tau^A(a)$ as the degrees to which $a$ is respectively an input state and a terminal state. For methodological reasons we sometimes allow the set of states $A$ to be infinite.~A~fuzzy auto\-maton whose set of states is finite is called a
{\it fuzzy finite automaton\/}.~

Let $X^*$ denote the free monoid over the alphabet $X$, and  let $e\in X^*$ be the
empty word. The function~$\delta^A $ can be extended~up to a function
$\delta^A_*:A\times X^*\times A\to L$ as follows: If $a,b\in A$, then
\begin{equation}\label{eq:delta.e}
\delta^A_*(a,e ,b)=\begin{cases}\ 1, & \text{if}\ a=b, \\ \ 0, & \mbox{otherwise,}
\end{cases}
\end{equation}
and if $a,b\in A$, $u\in X^*$ and $x\in X$, then
\begin{equation}\label{eq:delta.x}
\delta^A_*(a,ux ,b)= \bigvee _{c\in A} \delta^A_*(a,u,c)\otimes \delta^A (c,x,b).
\end{equation}

By (\ref{eq:inf.dist}) and Theorem 3.1 \cite{LP.05} (see also \cite{Qiu.01,Qiu.02,Qiu.06}), we have that
\begin{equation}\label{eq:delta.uv}
\delta^A_*(a,uv,b)= \bigvee _{c\in A} \delta^A_*(a,u,c)\otimes \delta^A_*(c,v,b),
\end{equation}
for all $a,b\in A$ and $u,v\in X^*$, i.e., if $w=x_1\cdots x_n$, for $x_1,\ldots ,x_n\in X$,
then
\begin{equation}\label{eq:delta.x1xn}
\delta^A_*(a,w,b)= \bigvee _{(c_1,\ldots ,c_{n-1})\in A^{n-1}}
\delta^A(a,x_1,c_1)\otimes \delta^A(c_1,x_2,c_2) \otimes\cdots \otimes \delta^A(c_{n-1},x_n,b).
\end{equation}
Intuitively, the product $\delta^A(a,x_1,c_1)\otimes \delta^A(c_1,x_2,c_2) \otimes\cdots \otimes
\delta^A(c_{n-1},x_n,b)$ represents the degree to which the input word $w$ causes a transition from a
state $a$ into a state $b$ through the sequence of intermediate states $c_1,\ldots ,c_{n-1}\in A$,
and $\delta^A_*(a,w,b)$ represents the supremum of degrees of all possible transitions from $a$ into
$b$ caused by $w$. Also, we can visualize~a fuzzy finite automaton $\cal A$ representing it as a
labelled directed graph whose nodes are~states of $\cal A$, and an edge from a node $a$ into a node
$b$ is labelled by pairs of the form $x/\delta^{A}(a,x,b)$, for any $x\in X$.

If $\delta^A $ is a crisp subset of $A\times X\times A$, that is, $\delta^A:A\times X\times A\to
\{0,1\}$, and $\sigma^A$ and $\tau^A$ are crisp subsets~of~$A$, then $\cal A$ is an ordinary
{\it nondeterministic automaton\/}.~In other words, nondeterministic automata are fuzzy automata
over the Boolean structure.~If $\delta^A $ is a function of $A\times X$ into~$A$, $\sigma^A$ is
a one-element crisp subset of $A$, that is, $\sigma^A=\{a_0\}$, for some $a_0\in A$, and $\tau^A$ is a fuzzy subset of $A$, then
$\cal A$ is called a {\it deterministic fuzzy automaton\/}, and it is denoted by ${\cal A}=(A,\delta^A,a_0,\tau^A )$.~In \cite{CDIV.10,DSV.10} the name {\it crisp-deterministic\/} was used.~For more information on deterministic fuzzy automata we refer to \cite{Bel.02a,IC.10,ICB.08,ICBP.10,LP.05}.~Evidently, if $\delta^A$ is a crisp subset of $A\times X\times A$, or a function of $A\times X$ into $A$, then $\delta^A_*$ is also a crisp subset of $A\times X^*\times A$, or a function of $A\times X^*$ into $A$, respectively.~A~deterministic fuzzy automaton ${\cal A}=(A,\delta^A,a_0,\tau^A )$, where $\tau^A$ is a crisp subset of $A$, is an ordinary deterministic automaton.

If for any $u\in X^*$ we define a fuzzy relation $\delta^A_u$ on $A$ by
\begin{equation}\label{eq:trans.rel}
\delta^A_u (a,b) = \delta^A_* (a,u,b) ,
\end{equation}
for all $a,b\in A$, called the {\it fuzzy transition relation\/} determined by $u$, then (\ref{eq:delta.uv}) can be written as
\begin{equation}\label{eq:delta.uv.2}
\delta^A_{uv}= \delta^A_u\circ \delta^A_v,
\end{equation}
for all $u,v\in X^*$.

The {\it reverse fuzzy automaton} of a fuzzy automaton ${\cal A}=(A,\delta^A, \sigma^A ,\tau^A )$ is defined as the fuzzy automaton $\bar{{\cal A}}=(A,\bar{\delta}^A,\bar{\sigma}^A
,\bar{\tau}^A )$ whose fuzzy transition function and fuzzy sets of initial and terminal states are defined by
$\bar{\delta}^A(a_1,x,a_2)=\delta^A(a_2,x,a_1)$ for all $a_1,a_2\in A$ and $x\in X$, $\bar{\sigma}^A=\tau^A$ and $\bar{\tau}^A=\sigma^A$.~In other words, $\bar{\delta}_x^A= (\delta_x^A)^{-1}$, for each $x\in X$.

A {\it fuzzy language\/} in $X^*$ over ${\cal L}$, or
briefly a {\it fuzzy language\/}, is any fuzzy subset of~$X^*$, i.e., any function from
$X^*$ to $L$.~A {\it fuzzy language recognized by a fuzzy automaton\/} ${\cal A}=(A,\delta^A , \sigma^A,\tau^A )$,
denoted as $L({\cal A})$, is a fuzzy language in ${\cal F}(X^*)$ defined by
\begin{equation}\label{eq:recog}
L({\cal A})(u) = \bigvee_{a,b\in A} \sigma^{A} (a)\otimes \delta_*^A(a,u,b)\otimes \tau^A(b) ,
\end{equation}
or equivalently,
\begin{equation}\label{eq:recog.comp}
\begin{aligned}
L({\cal A})(e) &= \sigma^A\circ \tau^A ,\\
L({\cal A})(u) &= \sigma^A\circ \delta_{x_1}^A\circ \delta_{x_2}^A\circ \cdots
\circ \delta_{x_n}^A\circ \tau^A ,
\end{aligned}
\end{equation}
for any $u=x_1x_2\dots x_n\in X^+$, where $x_1,x_2,\ldots ,x_n\in X$.~In other words, the equality (\ref{eq:recog}) means that the~membership degree of the word
$u$~to~the fuzzy language $L({\cal A})$ is equal to the degree to which $\cal A$ recognizes or accepts
the word $u$.~Using notation from (\ref{eq:comp.sr}), and
the second equality in (\ref{eq:comp.as}), we can state (\ref{eq:recog}) as
\begin{equation}\label{eq:recog.comp}
L({\cal A})(u) = \sigma^A \circ \delta^A_u\circ \tau^A .
\end{equation}
Fuzzy automata $\cal A$~and~$\cal B$ are called {\it language-equivalent\/}, or sometimes just {\it equivalent\/}, if $L({\cal A})=L({\cal B})$.

Let ${\cal A}=(A,\delta^A,\sigma^A, \tau^A)$ be a fuzzy automaton and let $E$  a fuzzy equivalence~relation on~$A$.~Without any restriction on $E$, we~can~define a fuzzy transition function
$\delta^{A/E}:(A/E)\times X\times (A/E)\to L$~by
\begin{equation}\label{eq:dE1}
\delta^{A/E}(E_a,x,E_b) = \bigvee_{a',b'\in A} E(a,a')\otimes \delta^A
(a',x,b')\otimes E(b',b)
\end{equation}
or equivalently,
\begin{equation}\label{eq:dE2}
\delta^{A/E}(E_a,x,E_b) = (E\circ \delta^A_x\circ E)(a,b) = E_a\circ \delta^A_x\circ E_b,
\end{equation}
for any $a,b\in A$ and $x\in X$.

We can also define a fuzzy set
$\sigma^{A/E}\in {\cal F}(A/E)$ of initial states and a fuzzy set $\tau^{A/E}\in {\cal
F}(A/E)$ of terminal states by
\begin{gather}
\sigma^{A/E}(E_a) = \bigvee_{a'\in A}\sigma^A (a')\otimes E(a',a) = (\sigma^A\circ
E)(a) = \sigma^A\circ E_a , \label{eq:sE} \\
\tau^{A/E}(E_a) = \bigvee_{a'\in A}E(a,a')\otimes \tau^A(a') = (E\circ
\tau^A)(a) = E_a\circ\tau^A, \label{eq:tE}
\end{gather}
for any $a\in A$.

Evidently, $\delta^{A/E}$, $\sigma^{A/E}$ and $\tau^{A/E}$ are well-defined and
${\cal A}/E=(A/E,\delta^{A/E},\sigma^{A/E} ,\tau^{A/E})$ is a~fuzzy automaton, called the
{\it factor fuzzy automaton\/} of $\cal A$ with respect to $E$.

Let us note that the fuzzy language $L({{\cal A}/E})$ recognized by the factor fuzzy automaton ${\cal A}/E$ is given by
\begin{align}
&L({{\cal A}/E})(e)= \sigma^A\circ E\circ \tau^A ,\\
&L({{\cal A}/E})(u)= \sigma^A\circ E\circ \delta^A_{x_1}\circ E\circ \delta^A_{x_2}\circ E \circ \cdots
\circ E\circ \delta^A_{x_n}\circ E\circ \tau^A ,
\end{align}
whereas the fuzzy language $L({\cal A})$ recognized by the fuzzy automaton ${\cal A}$ is given by
\begin{align}
&L({\cal A})(e) = \sigma^A\circ \tau ^A,\\
&L({\cal A})(u) = \sigma^A\circ \delta^A_{x_1}\circ \delta^A_{x_2}\circ \cdots
\circ \delta^A_{x_n}\circ \tau ^A,
\end{align}
for any $u=x_1x_2\dots x_n\in X^+$, where $x_1,x_2,\ldots ,x_n\in X$.~Therefore, fuzzy automata $\cal A$ and ${\cal A}/E$ recognize the same fuzzy language if and only if the fuzzy equivalence relation $E$ is a solution to a system of fuzzy relation equations
\begin{equation}\label{eq:sFRE}
\begin{aligned}
&\sigma^A\circ \tau^A = \sigma^A\circ E\circ \tau^A, \\
&\sigma^A\circ \delta_{x_1}^A\circ \delta_{x_2}^A\circ \cdots \circ
\delta_{x_n}^A\circ \tau^A = \sigma^A\circ E\circ \delta_{x_1}^A\circ E\circ \delta_{x_2}^A\circ E \circ \cdots
\circ E\circ \delta_{x_n}^A\circ E\circ \tau^A,
\end{aligned}
\end{equation}
for all $n\in \Bbb N$ and $x_1,x_2,\ldots ,x_n\in X$.~We call (\ref{eq:sFRE}) the {\it general~system\/}.

Let ${\cal A}=(A,\delta^A ,\sigma^A,\tau^A)$ and ${\cal B}=(B,\delta^B ,\sigma^B,\tau^B)$ be fuzzy automata.~A function $\varphi:A\longrightarrow B$ is called an {\it isomorphism} between $\cal A$ and $\cal B$ if it is bijective and for all $a,a_1,a_2\in A$ and $x\in X$, the following is true:
\begin{align}
&\delta_x^A(a_1,a_2)=\delta_x^B(\varphi(a_1),\varphi(a_2)),\\
&\sigma^A(a)=\sigma^B(\varphi(a)),\\
&\tau^A(a)=\tau^B(\varphi(a)).
\end{align}
If there exists an isomorphism between $\cal A$ and $\cal B$, then we say that $\cal A$ and $\cal B$ are {\it isomorphic\/} fuzzy automata. Clearly, the inverse of an isomorphism of fuzzy automata is also an isomorphism, as well as the composition of two isomorphisms.

Cardinality of a fuzzy automaton ${\cal A}=(A,\delta^A,\sigma^A,\tau ^A)$, in notation $|{\cal A}|$, is defined as the cardinality of its set of states $A$.~A fuzzy automaton ${\cal A}$ is called {\it minimal fuzzy automaton} of a language $f\in {\cal F}$ if it recognizes~$f$ and $|{\cal A}|\leqslant |{\cal A}'|$, for any fuzzy automaton ${\cal A}'$ recognizing $f$.~A minimal fuzzy automaton recognizing a~given fuzzy language $f$ is not necessarily unique up to an isomorphism (cf.~Example \ref{ex:lang.UFB}).~This is also true for nondeterministic automata.~Further, minimization of fuzzy automata, as well as of nondeterministic ones is computationally hard.~For that reason, we will be interested in constructing minimal automata in some special subclasses of the class of all fuzzy automata recognizing a given fuzzy language $f$.

\section{Uniform fuzzy relations}\label{sectionUniform}

In this section we recall some notions, notation and results from \cite{CIB.09}, concerning
uniform fuzzy relations and related~con\-cepts.

Let $A$ and $B$ be non-empty sets and let $E $ and $F $ be fuzzy equivalence
relations on
$A$ and $B$,
respectively. If a fuzzy relation $\varphi \in {\cal F}(A\times
B)$ satisfies
\begin{itemize}\itemindent8pt
\item[(EX1)] $\varphi (a_1,b)\otimes E(a_1,a_2)\leqslant \varphi(a_2,b)$, for all $a_1,a_2\in A$ and $b\in B$,
\end{itemize}
then it is called  {\it extensional with respect to\/}~$E$, and if it satisfies
\begin{itemize}\itemindent8pt
\item[(EX2)] $\varphi (a,b_1)\otimes F(b_1,b_2)\leqslant \varphi(a,b_2)$, for all $a\in A$ and $b_1,b_2\in B$,
\end{itemize}
then it is called  {\it extensional with respect to\/}~$F$.~If $\varphi $ is extensional with respect to~$E$ and $F$, and it satisfies
\begin{itemize}\itemindent8pt
\item[(PFF)] $\varphi(a,b_1)\otimes \varphi (a,b_2)\leqslant F(b_1,b_2)$, for all $a\in A$ and $b_1,b_2\in B$,
\end{itemize}
then it is called a {\it partial fuzzy function\/} with respect to~$E$ and $F$.

Partial fuzzy functions were introduced by Klawonn \cite{Klawonn.00}, and studied also by Demirci
\cite{Dem.00,Demirci.03b}. By the adjunction property and symmetry, conditions (EX1) and (EX2) can be also written as
\begin{itemize}\itemindent8pt
\item[(EX1')] $E(a_1,a_2)\leqslant \varphi (a_1,b)\leftrightarrow \varphi(a_2,b)$, for all $a_1,a_2\in A$ and $b\in B$;
\item[(EX2')] $F(b_1,b_2)\leqslant \varphi (a,b_1)\leftrightarrow \varphi(a,b_2)$, for all $a\in A$ and $b_1,b_2\in B$.
\end{itemize}

For any fuzzy relation $\varphi \in {\cal F}(A\times B)$ we can define a
fuzzy equivalence relation $E_A^\varphi $ on~$A$~by
\begin{equation}\label{eq:a.phi.1}
E_A^\varphi(a_1,a_2)=\bigwedge_{b\in B} \varphi(a_1,b)\leftrightarrow \varphi (a_2,b),
\end{equation}
for all $a_1,a_2\in A$, and a fuzzy equivalence relation $E_B^\varphi $ on $B$ by
\begin{equation}\label{eq:b.phi.1}
E_B^\varphi(b_1,b_2)=\bigwedge_{a\in A} \varphi(a,b_1)\leftrightarrow \varphi (a,b_2),
\end{equation}
for all $b_1,b_2\in B$.~They will be called {\it fuzzy equivalence relations\/} on $A$ and $B$
{\it induced by\/} $\varphi $, and in particular, $E_A^\varphi $ will be called the {\it kernel\/} of $\varphi $, and
$E_B^\varphi $ the {\it co-kernel\/} of $\varphi $.~According to (EX1')~and (EX2'), $E_A^\varphi $ and $E_B^\varphi $ are the
greatest $\cal $fuzzy equivalence relations on $A$ and $B$, respectively, such that $\varphi $ is
extensional with respect to~them.

A fuzzy relation $\varphi \in {\cal F}(A\times B)$ is called just a {\it partial fuzzy function\/} if it is a
partial fuzzy function with respect to $E_A^\varphi $~and~$E_B^\varphi $ \cite{CIB.09}. Partial fuzzy functions
were characterized in \cite{CIB.09} as follows:

\begin{theorem}\label{th:PFF}
Let $A$ and $B$ be non-empty sets and let $\varphi \in {\cal F}(A\times B)$ be a fuzzy relation. Then the~fol\-low\-ing conditions are equivalent:
\begin{itemize}\parskip=0pt
\item[{\rm (i)}] $\varphi $ is a partial fuzzy function;
\item[{\rm (ii)}] $\varphi^{-1}$ is a partial fuzzy function;
\item[{\rm (iii)}] $\varphi^{-1}\circ \varphi\leqslant E_B^\varphi $;
\item[{\rm (iv)}] $\varphi \circ \varphi^{-1}\leqslant E_A^\varphi $;
\item[{\rm (v)}] $\varphi \circ \varphi^{-1}\circ \varphi \leqslant \varphi$.
\end{itemize}
\end{theorem}

\begin{proof}
All conditions of this theorem are just another formulation of the conditions of
Theorem 3.1 \cite{CIB.09}.
\end{proof}

A fuzzy relation $\varphi \in {\cal F}(A\times B)$ is called an $\cal L$-{\it function\/} if
for each $a\in A$ there exists $b\in B$ such that $\varphi (a,b)=1$ \cite{Demirci.05a}, and it is called {\it
surjective\/} if for each $b\in B$ there exists $a\in A$ such that $\varphi (a,b)=1$, i.e., if
$\varphi^{-1}$ is an $\cal L$-function. For a surjective fuzzy relation $\varphi \in {\cal F}(A\times B)$ we also say that it is a fuzzy relation of $A$ {\it onto\/} $B$. If
$\varphi $ is an $\cal L$-function and it is surjective, i.e., if both $\varphi $ and
$\varphi^{-1}$ are $\cal L$-functions, then $\varphi $ is called a {\it surjective $\cal
L$-function\/}.

Let us note that a fuzzy relation $\varphi\in {\cal F}(A\times B)$ is an $\cal L$-function if and only if there exists a function $\psi :A\to B$ such that $\varphi (a,\psi(a))=1$, for all $a\in
A$ (cf.~\cite{Demirci.03b,Demirci.05a}). A function $\psi $ with this property we will call a {\it crisp
description\/} of $\varphi $, and we will denote by $CR(\varphi)$ the set of all such functions.

An $\cal L$-function which is a partial fuzzy function with respect to~$E$ and $F$ is called a {\it
perfect~fuzzy function\/} with respect to~$E$ and $F$.~Perfect fuzzy functions were introduced and
studied by Demirci \cite{Dem.00,Demirci.03b}.~A fuzzy relation $\varphi \in
{\cal F}(A\times B)$ which is a perfect fuzzy function with respect to $E_A^\varphi $~and~$E_B^\varphi $ will be called just a {\it perfect fuzzy function\/}.

Let $A$ and $B$ be non-empty sets and let $E$ be a fuzzy equivalence relation on $B$. An ordinary function $\psi :A\to B$ is called {\it $E$-surjective\/} if for any $b\in B$ there exists $a\in A$ such that $E(\psi(a),b)=1$.~In other words,~$\psi $ is $E$-surjective if and only if $\psi \circ E^\sharp $ is an ordinary surjective function
of $A$ onto $B/E$, where $E^\sharp :B\to B/E$~is a function given by $E^\sharp (b)=E_b$,
for each $b\in B$. It is clear that $\psi $ is an $E$-surjective function if and only if~its image $\im \psi $ has a non-empty intersection with every equivalence class of the
crisp equivalence relation~$\ker (E)$.

Let $A$ and $B$ be non-empty sets and let $\varphi\in {\cal F}(A\times B)$ be a partial fuzzy function.
If, in addition, $\varphi $ is a surjective $\cal L$-function, then it will be called a {\it
uniform fuzzy relation\/} \cite{CIB.09}. In other words, a uniform fuzzy relation is a perfect fuzzy function
having the additional property that it is surjective.~A uniform fuzzy
relation that~is also a crisp relation is called a {\it uniform relation\/}.~The following characterizations of uniform fuzzy relations provided in \cite{CIB.09} will be used in the further text.

\begin{theorem}\label{th:ufr} Let $A$ and $B$ be non-empty sets and let $\varphi \in {\cal F}(A\times B)$
be a fuzzy relation. Then the~follow\-ing conditions are equivalent:
\begin{itemize}\parskip=0pt
\item[{\rm (i)}] $\varphi $ is a uniform fuzzy relation;
\item[{\rm (ii)}] $\varphi^{-1}$ is a uniform fuzzy relation;
\item[{\rm (iii)}] $\varphi $ is a surjective $\cal L$-function and
\begin{equation}\label{eq:3phi}
\varphi\circ \varphi^{-1}\circ \varphi =\varphi ;
\end{equation}
\item[{\rm (iv)}] $\varphi $ is a surjective $\cal L$-function and
\begin{equation}\label{eq:EA.phi}
E_A^\varphi
= \varphi\circ \varphi^{-1};
\end{equation}
\item[{\rm (v)}] $\varphi $ is a surjective $\cal L$-function and
\begin{equation}\label{eq:EB.phi}
E_B^\varphi = \varphi^{-1}\circ \varphi;
\end{equation}
\item[{\rm (vi)}] $\varphi $ is an $\cal L$-function, and for all $\psi\in
CR(\varphi )$, $a\in A$ and $b\in B$ we have that
$\psi $ is $E_B^\varphi $-surjective and
\begin{equation}\label{eq:phi.EB}
\varphi (a,b)=E_B^\varphi (\psi(a),b);
\end{equation}
\item[{\rm (vii)}] $\varphi $ is an $\cal L$-function, and for all $\psi\in
CR(\varphi )$ and $a_1,a_2\in A$ we have that
$\psi $ is $E_B^\varphi $-surjective and
\begin{equation}\label{eq:phi.EA}
\varphi (a_1,\psi(a_2))=E_A^\varphi (a_1,a_2).
\end{equation}
\end{itemize}
\end{theorem}

\begin{proof}
Conditions (i), (ii), (vi) and (vii) are the same as in Theorem 3.3 \cite{CIB.09},
whereas (iv) and (v) are just another formulation of conditions (v) and (vi)
of Theorem 3.3 \cite{CIB.09}.~The implication (iii)$\Rightarrow $(i) follows directly
by Theorem \ref{th:PFF}. Thus, we have to prove only (i)$\Rightarrow $(iii).

If $\varphi $ is a uniform fuzzy relation, then it is a surjective $\cal L$-function,
and by Theorem \ref{th:PFF} we obtain that $\varphi\circ\varphi^{-1}\circ\varphi
\leqslant \varphi $, so it remains to prove the opposite inequality.~Consider arbitrary
$\psi\in CR(\varphi )$, $a\in A$~and $b\in B$. Then $\varphi (a,\psi(a))=\varphi^{-1}(\psi(a),a)=1$,
and\[
\begin{aligned}
\varphi (a,b)&=\varphi (a,\psi(a))\otimes \varphi^{-1}(\psi(a),a)\otimes \varphi (a,b) \\
&\leqslant \bigvee_{a'\in A,b'\in B}\varphi(a,b')\otimes \varphi^{-1}(b',a')\otimes \varphi(a',b)= (\varphi
\circ \varphi^{-1}\circ \varphi) (a,b) .
\end{aligned}
\]
Therefore, $\varphi \leqslant \varphi \circ \varphi^{-1}\circ \varphi$, and we have proved
that $\varphi\circ\varphi^{-1}\circ\varphi = \varphi $.
\end{proof}

\begin{corollary}\label{cor:ufr} {\rm \cite{CIB.09}}
Let $A$ and $B$ be non-empty sets, and let $\varphi \in {\cal F}(A\times B)$
be a uniform~fuzzy relation.~Then for all $\psi\in
CR(\varphi )$ and $a_1,a_2\in A$ we have that
\begin{equation}\label{eq:cor.ufr}
E_A^\varphi (a_1,a_2) = E_B^\varphi (\psi(a_1),\psi(a_2)).
\end{equation}
\end{corollary}

\begin{remark}\label{rem:ufr}\rm
Let $A$ and $B$ be non-empty sets. According to Theorem \ref{th:ufr}, a fuzzy relation $\varphi \in {\cal F}(A\times B)$ is a uniform fuzzy relation if and only if its inverse relation $\varphi^{-1}$ is a uniform fuzzy relation.

Moreover, by (iv) and (v) of Theorem \ref{th:ufr}, we have that the kernel of $\varphi^{-1}$ is the co-kernel of $\varphi $, and conversely, the co-kernel of $\varphi^{-1}$ is the kernel of $\varphi $, that is
\[
E_B^{\varphi^{-1}}=E_B^\varphi \quad\text{and}\quad E_A^{\varphi^{-1}}=E_A^\varphi .
\]
\end{remark}

The next lemmas will be very useful in our further work.

\begin{lemma}\label{le:tpsi}
Let $A$ and $B$ be  non-empty sets, let $\varphi \in {\cal F}(A\times B)$ be a
uniform fuzzy relation,~let $E=E_A^\varphi $ and $F=E_B^\varphi $, and let
a function $\widetilde\varphi :A/E\to B/F$ be defined by
\begin{equation}\label{eq:tphi}
\widetilde\varphi (E_a)= F_{\psi(a)}, \ \ \text{for any $a\in A$ and $\psi \in CR(\varphi )$.}
\end{equation}
Then $\widetilde \varphi $ is a well-defined function {\rm ({\it it does not depend on the choice of $\psi \in CR(\varphi )$ and $a\in A$\/})}, it is a bijective function  of $A/E$ onto $B/F$, and $(\widetilde\varphi)^{-1}=\widetilde{\varphi^{-1}}$.
\end{lemma}

\begin{proof}
Let $a_1,a_2\in A$ such that $E_{a_1}=E_{a_2}$ and let $\psi \in CR(\varphi)$.~By Corollary \ref{cor:ufr} we have  $F(\psi(a_1),\psi(a_2))=E(a_1,a_2)=1$,
so $F_{\psi(a_1)}=F_{\psi(a_2)}$.~On the other hand, let $a\in A$ and
$\psi_1,\psi_2\in CR(\varphi )$.~Then by Theorem~\ref{th:ufr} it follows
that $F(\psi_1(a),\psi_2(a))\geqslant  \varphi(a,\psi_1(a))\otimes \varphi(a,\psi_2(a))=1$,
and again, we~obtain $F_{\psi(a_1)}=F_{\psi(a_2)}$. Hence, $\widetilde\varphi
$ is well-defined, i.e., it does not depend on the choice of $\psi\in CR(\varphi)$
and a~repre\-sentative of a fuzzy equivalence class of $E$.

Next, for any $\psi \in CR(\varphi)$, $\xi\in CR(\varphi^{-1})$, $a\in
A$ and $b\in B$, by (vi) and (vii) of Theorem \ref{th:ufr} we obtain
\[
E(a,\xi(\psi(a)))=\varphi^{-1}(\psi(a),a)=\varphi(a,\psi(a))=1, \ \
F(b,\psi(\xi(b)))=\varphi(\xi(b),b)=\varphi^{-1}(b,\xi(b))=1,
\]
and hence, $E_a=E_{\xi(\psi(a))}$ and $F_b=E_{\psi(\xi(a))}$. By this it
follows that $(\widetilde\varphi)^{-1}=\widetilde{\varphi^{-1}}$.
\end{proof}

\begin{lemma}\label{le:UFR.order}
Let $A$ and $B$ be  non-empty sets, and let $\varphi_1,\varphi_2\in {\cal F}(A\times B)$ be uniform fuzzy relations.~Then the~following conditions are equivalent:
\begin{itemize}\parskip0pt
\item[{\rm (i)}] $\varphi_1\leqslant \varphi_2$;
\item[{\rm (ii)}] $\varphi_1^{-1}\leqslant \varphi_2^{-1}$;
\item[{\rm (iii)}] $CR(\varphi_1)\subseteq CR(\varphi_2)$ \ and \ $E_A^{\varphi_1}\leqslant E_A^{\varphi_2}$;
\item[{\rm (iv)}] $CR(\varphi_1)\subseteq CR(\varphi_2)$ \ and \ $E_B^{\varphi_1}\leqslant E_B^{\varphi_2}$.
\end{itemize}
\end{lemma}

\begin{proof} The equivalence (i)$\Leftrightarrow $(ii) is obvious.

(i)$\Rightarrow $(iii).~Let $\varphi_1\leqslant \varphi_2$.~If $\psi \in CR(\varphi_1)$, then for every $a\in A$ we have that $1=\varphi_1(a,\psi(a))\leqslant \varphi_2(a,\psi(a))$, what~implies
$\varphi_2(a,\psi(a))=1$, and hence, $\psi\in CR(\varphi_2)$.

Further, consider arbitrary $a_1,a_2\in A$.~By (vii) of Theorem \ref{th:ufr}, for any $\psi \in CR(\varphi_1)\subseteq CR(\varphi_2)$~we have that
\[
E_A^{\varphi_1}(a_1,a_2)=\varphi_1(a_1,\psi(a_2))\leqslant \varphi_2(a_1,\psi(a_2))=E_A^{\varphi_2}(a_1,a_2),
\]
and therefore, $E_A^{\varphi_1}\leqslant E_A^{\varphi_2}$.

(iv)$\Rightarrow $(i).~Consider arbitrary $a\in A$, $b\in B$, and $\psi \in CR(\varphi_1)\subseteq~CR(\varphi_2)$.~According to (vi) of Theorem \ref{th:ufr}, we have that
\[
\varphi_1(a,b)=E_B^{\varphi_1}(\psi(a),b)\leqslant E_B^{\varphi_2}(\psi(a),b)=\varphi_2(a,b),
\]
and we conclude that $\varphi_1\leqslant \varphi_2$.

In view of Remark \ref{rem:ufr}, by the same arguments used in the proof of (i)$\Rightarrow $(iii) we can prove (ii)$\Rightarrow $(iv), and by the same arguments used in the proof of (iv)$\Rightarrow $(i) we can prove (iii)$\Rightarrow $(ii).
\end{proof}

As a direct consequence of the previous lemma we obtain the following corollary which shows that a uniform fuzzy relation is uniquely determined by its crisp representation and kernel, as well as by its crisp representation and co-kernel.

\begin{corollary}\label{le:UFR.equality}
Let $A$ and $B$ be  non-empty sets, and let $\varphi_1,\varphi_2\in {\cal F}(A\times B)$ be uniform fuzzy relations.~Then the following conditions are equivalent:
\begin{itemize}\parskip0pt
\item[{\rm (i)}] $\varphi_1= \varphi_2$;
\item[{\rm (ii)}] $\varphi_1^{-1}= \varphi_2^{-1}$;
\item[{\rm (iii)}] $CR(\varphi_1)= CR(\varphi_2)$ \ and \ $E_A^{\varphi_1}= E_A^{\varphi_2}$;
\item[{\rm (iv)}] $CR(\varphi_1)= CR(\varphi_2)$ \ and \ $E_B^{\varphi_1}= E_B^{\varphi_2}$.
\end{itemize}
\end{corollary}

We know that the composition of two uniform fuzzy relations need not be a uniform fuzzy relation (cf. Example 6.1 \cite{CIB.09}).~However, if the co-kernel of the first factor of the composition is contained in the~kernel~of the second factor, then the composition is uniform, as the following lemma shows.

\begin{lemma}\label{le:UFR.comp}
Let $A$, $B$ and $C$ be non-empty sets, and let $\varphi_1\in {\cal F}(A\times B)$ and $\varphi_2\in {\cal F}(B\times C)$.
\begin{itemize}\parskip-2pt
\item[{\rm (a)}] If $\varphi_1$ and $\varphi_2$ are surjective $\cal L$-functions, then $\varphi_1\circ \varphi_2$ is also a surjective $\cal L$-function.
\item[{\rm (b)}] If $\varphi_1$ and $\varphi_2$ are uniform fuzzy relations such that $E_B^{\varphi_1}\leqslant E_B^{\varphi_2}$, then $\varphi_1\circ \varphi_2$ is also a uniform fuzzy relation.
\end{itemize}
\end{lemma}

\begin{proof}
(a) This is a simple consequence of the fact that for arbitrary $\psi_1\in CR(\varphi_1)$ and $\psi_2\in CR(\varphi_2)$ we have $\psi_1\circ\psi_2\in CR(\varphi_1\circ\varphi_2)$.

(b) The assumption $E_B^{\varphi_1}\leqslant E_B^{\varphi_2}$ and Theorem \ref{th:ufr} yield
\[
\varphi_2\circ \varphi_2^{-1}\circ \varphi_1^{-1}\circ \varphi_1= E_B^{\varphi_2}\circ E_B^{\varphi_1}=E_B^{\varphi_2}=\varphi_2\circ \varphi_2^{-1},
\]
and for $\varphi=\varphi_1\circ\varphi_2$ we have
\[
\varphi\circ \varphi^{-1}\circ\varphi = \varphi_1\circ\varphi_2\circ \varphi_2^{-1}\circ \varphi_1^{-1}\circ \varphi_1\circ \varphi_2 = \varphi_1\circ \varphi_2\circ \varphi_2^{-1}\circ \varphi_2=\varphi_1\circ\varphi_2=\varphi .
\]
Together with (a) this implies that $\varphi=\varphi_1\circ\varphi_2$ is a uniform fuzzy relation.
\end{proof}

\section{Bisimulations for fuzzy automata}\label{sectionBisimulations}

Now we are ready to move to consideration of bisimulations for fuzzy automata.~In this section we give definitions and discuss the basic properties of bisimulations.

Let ${\cal A}=(A,\delta^A,\sigma^A,\tau^A)$ and ${\cal B}=(B,\delta^B,\sigma^B,\tau^B)$ be fuzzy automata, and let $\varphi\in {\cal F}(A\times B)$ be a non-empty fuzzy relation. We call $\varphi $ a {\it forward simulation\/} if
\begin{align}\label{eq:fsi0}
&\sigma^A\leqslant \sigma^B\circ\varphi^{-1},\\
\label{eq:fs0}
&\varphi^{-1}\circ \delta_x^A\leqslant \delta_x^B\circ \varphi^{-1},\quad \text{for every $x\in X$},\\
\label{eq:fst0}
&\varphi^{-1}\circ\tau^A\leqslant \tau^B,
\end{align}
and a {\it backward simulation\/} if
\begin{align}\label{eq:bsi0}
&\sigma^A\circ\varphi\leqslant \sigma^B,\\
\label{eq:bs0}
&\delta_x^A\circ \varphi \leqslant \varphi\circ \delta_x^B ,\quad \text{for every $x\in X$},\\
\label{eq:bst0}
&\tau^A\leqslant\varphi\circ\tau^B.
\end{align}
Furthermore, we call $\varphi $ a {\it forward bisimulation\/} if both $\varphi $ and $\varphi^{-1}$ are forward simulations, that is, if $\varphi $ satisfies (\ref{eq:fsi0})--(\ref{eq:fst0}) and
\begin{align}\label{eq:fsi0i}
&\sigma^B\leqslant \sigma^A\circ\varphi,\\
\label{eq:fs0i}
&\varphi\circ \delta_x^B\leqslant \delta_x^A\circ \varphi,\quad \text{for every $x\in X$},\\
\label{eq:fst0i}
&\varphi\circ\tau^B\leqslant \tau^A,
\end{align}
and a {\it backward bisimulation\/}, if both $\varphi $ and $\varphi^{-1}$ are backward simulations, i.e., if
$\varphi $ satisfies (\ref{eq:bsi0})--(\ref{eq:bst0}) and
\begin{align}\label{eq:bsi0i}
&\sigma^B\circ\varphi^{-1}\leqslant \sigma^A,\\
\label{eq:bs0i}
&\delta_x^B\circ \varphi^{-1} \leqslant \varphi^{-1}\circ \delta_x^A ,\quad \text{for every $x\in X$},\\
\label{eq:bst0i}
&\tau^B\leqslant\varphi^{-1}\circ\tau^A.
\end{align}
Also, if $\varphi $ is a forward simulation and $\varphi^{-1}$ is a backward simulation, i.e., if
\begin{align}\label{eq:fbsi0}
&\sigma^A= \sigma^B\circ\varphi^{-1},\\
\label{eq:fbs0}
&\varphi^{-1}\circ \delta_x^A= \delta_x^B\circ \varphi^{-1},\quad \text{for every $x\in X$},\\
\label{eq:fbst0}
&\varphi^{-1}\circ\tau^A= \tau^B,
\end{align}
then $\varphi $ is called a {\it forward-backward bisimulation\/}, or briefly a {\it FB-bisimulation\/}, and if $\varphi $ is a backward simulation and $\varphi^{-1}$ is a forward simulation, i.e., if
\begin{align}\label{eq:bfsi0}
&\sigma^A\circ\varphi= \sigma^B,\\
\label{eq:bfs0}
&\delta_x^A\circ \varphi = \varphi\circ \delta_x^B ,\quad \text{for every $x\in X$},\\
\label{eq:bfst0}
&\tau^A=\varphi\circ\tau^B.
\end{align}
then $\varphi $ is called a {\it backward-forward bisimulation\/}, or briefly a {\it BF-bisimulation\/}.~For the sake of simplicity, we~will call $\varphi $ just a {\it simulation\/} if $\varphi $ is either a forward or a backward simulation, and just a {\it bisimulation\/} if $\varphi $ is any of the four types of bisimulations defined above.~Moreover, forward and backward bisimulations will be called {\it homotypic\/}, whereas backward-forward and forward-backward bisimulations will be called {\it heterotypic\/}.

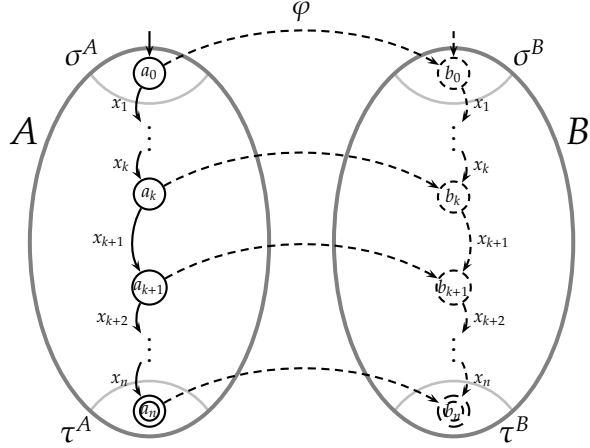
\begin{figure}
\begin{center}
\psset{unit=0.4cm}
\begin{pspicture}(-10,-7.2)(10,5)
\psarc[linecolor=lightgray,linewidth=1pt](-5,6){2.4}{215}{325}
\psarc[linecolor=lightgray,linewidth=1pt](-5,-8){2.4}{35}{145}
\psarc[linecolor=lightgray,linewidth=1pt](5,6){2.4}{215}{325}
\psarc[linecolor=lightgray,linewidth=1pt](5,-8){2.4}{35}{145}
\psellipse[linecolor=gray,linewidth=1.5pt](-5,-1)(4,6.5)
\psellipse[linecolor=gray,linewidth=1.5pt](5,-1)(4,6.5)
\rput(-9.1,2.7){\Large $A$}
\rput(9.1,2.7){\Large $B$}
\rput(-7.2,5.3){\large $\sigma^A$}
\rput(7.5,5.3){\large $\sigma^B$}
\rput(-7.4,-7.2){\large $\tau^A$}
\rput(7.0,-7.2){\large $\tau^B$}
\pnode(-5,6){AI}
\cnode(-5,4.6){.55}{A0}
\rput(A0){\scriptsize $a_0$}
\cnode[linewidth=0,linecolor=white](-5,2.5){.55}{A1}
\rput(-5,2.7){$\vdots$}
\cnode(-5,0.6){.55}{Ak}
\rput(Ak){\scriptsize $a_k$}
\cnode(-5,-2.5){.6}{Ak1}
\rput(Ak1){\scriptsize $a_{k+1}$}
\cnode[linewidth=0,linecolor=white](-5,-4.5){.55}{Ak2}
\rput(-5,-4.3){$\vdots$}
\cnode[doubleline=true,doublesep=1.5pt](-5,-6.6){.55}{An}
\rput(An){\scriptsize $a_n$}
\pnode(5,6){BI}
\cnode[linestyle=dashed,dash=3pt 2pt](5,4.6){.55}{B0}
\rput(B0){\scriptsize $b_0$}
\cnode[linewidth=0,linecolor=white](5,2.5){.55}{B1}
\rput(5,2.7){$\vdots$}
\cnode[linestyle=dashed,dash=3pt 2pt](5,0.5){.55}{Bk}
\rput(Bk){\scriptsize $b_k$}
\cnode[linestyle=dashed,dash=3pt 2pt](5,-2.5){.6}{Bk1}
\rput(Bk1){\scriptsize $b_{k+1}$}
\cnode[linewidth=0,linecolor=white](5,-4.5){.55}{Bk2}
\rput(5,-4.3){$\vdots$}
\cnode[doubleline=true,doublesep=1.5pt,linestyle=dashed,dash=3pt 2pt](5,-6.6){.55}{Bn}
\rput(Bn){\scriptsize $b_n$}
\ncline{->}{AI}{A0}
\ncarc[arcangle=-30]{->}{A0}{A1}\Bput[2pt]{\scriptsize $x_1$}
\ncarc[arcangle=-30]{->}{A1}{Ak}\Bput[2pt]{\scriptsize $x_k$}
\ncarc[arcangle=-30]{->}{Ak}{Ak1}\Bput[2pt]{\scriptsize $x_{k+1}$}
\ncarc[arcangle=-30]{->}{Ak1}{Ak2}\Bput[2pt]{\scriptsize $x_{k+2}$}
\ncarc[arcangle=-30]{->}{Ak2}{An}\Bput[2pt]{\scriptsize $x_n$}
\ncline[linestyle=dashed,dash=3pt 2pt]{->}{BI}{B0}
\ncarc[arcangle=30,linestyle=dashed,dash=3pt 2pt]{->}{B0}{B1}\Aput[2pt]{\scriptsize $x_1$}
\ncarc[arcangle=30,linestyle=dashed,dash=3pt 2pt]{->}{B1}{Bk}\Aput[2pt]{\scriptsize $x_k$}
\ncarc[arcangle=30,linestyle=dashed,dash=3pt 2pt]{->}{Bk}{Bk1}\Aput[2pt]{\scriptsize $x_{k+1}$}
\ncarc[arcangle=30,linestyle=dashed,dash=3pt 2pt]{->}{Bk1}{Bk2}\Aput[2pt]{\scriptsize $x_{k+2}$}
\ncarc[arcangle=30,linestyle=dashed,dash=3pt 2pt]{->}{Bk2}{Bn}\Aput[2pt]{\scriptsize $x_n$}
\ncarc[arcangle=30,linestyle=dashed,dash=3pt 2pt]{->}{A0}{B0}\Aput[3pt]{$\varphi$}
\ncarc[arcangle=30,linestyle=dashed,dash=3pt 2pt]{->}{Ak}{Bk}
\ncarc[arcangle=30,linestyle=dashed,dash=3pt 2pt]{->}{Ak1}{Bk1}
\ncarc[arcangle=30,linestyle=dashed,dash=3pt 2pt]{->}{An}{Bn}
\end{pspicture}\\
\caption{Forward and backward simulation}\label{fig:FBS}
\end{center}
\end{figure}

The meaning of forward and backward simulations can be best explained in the case when $\cal A$ and~$\cal B$~are  nondeterministic (Boolean) automata.~For this purpose we will use the diagram shown in Figure 1.~Let~$\varphi $~be~a forward simulation between $\cal A$ and $\cal B$ and let $a_0,a_1,\ldots ,a_n$ be an arbitrary successful run of the automaton $\cal A$ on a word $u=x_1x_2\cdots x_n$ ($x_1,x_2,\ldots ,x_n\in X$), i.e., a sequence of states of $\cal A$ such that $a_0\in \sigma^A$,
$(a_k,a_{k+1})\in \delta^A_{x_{k+1}}$, for $0\leqslant k\leqslant n-1$, and $a_n\in \tau^A$.~According to (\ref{eq:fsi0}), there is an initial state $b_0\in \sigma^B$ such that $(a_0,b_0)\in\varphi $.~Suppose that for some $k$, $0\le k\le n-1$, we have built a sequence of states $b_0,b_1,\ldots ,b_k$ such that
$(b_{i-1},b_i)\in \delta^B_{x_i}$ and $(a_i,b_i)\in \varphi $, for each $i$, $1\le i\le k$.~Then $(b_k,a_{k+1})\in \varphi^{-1}\circ \delta^A_{x_{k+1}}$, and by~(\ref{eq:fs0}) we obtain that
$(b_k,a_{k+1})\in \delta^B_{x_{k+1}}\circ \varphi^{-1}$, so there exists $b_{k+1}\in B$ such that $(b_k,b_{k+1})\in \delta^B_{x_{k+1}}$ and $(a_{k+1},b_{k+1})\in \varphi $.~Therefore, we have successively built a sequence $b_0,b_1,\ldots ,b_n$ of states of $\cal B$ such that $b_0\in \sigma^B$, $(b_k,b_{k+1})\in \delta^B_{x_{k+1}}$, for every $k$, $0\le k\le n-1$, and $(a_k,b_k)\in \varphi$, for each $k$, $0\le k\le n$.~Moreover, by~(\ref{eq:fst0}) we obtain that $b_n\in\tau^B$.~Thus, the sequence $b_0,b_1,\ldots ,b_n$ is a successful run of the automaton $\cal B$ on the word $u$ which simulates the original run $a_0,a_1,\ldots ,a_n$ of $\cal A$ on~$u$.~In contrast to forward simulations, where we build the sequence $b_0,b_1,\ldots ,b_n$ moving forward, starting with $b_0$ and ending with $b_n$, in the case of backward simulations we build this sequence moving backward, starting with $b_n$ and ending with $b_0$.~In a similar way we can understand forward and backward simulations between arbitrary fuzzy automata, taking into account degrees of possibility of transitions and degrees of relationship.

Evidently, bisimulation is a fuzzy relation which realizes simulation of a fuzzy automaton by another, and its inverse realizes the reverse simulation.~It is worth noting that bisimulation is a more restrictive~concept than the \emph{two-way simulation\/}, by which we mean a pair of simulations of a fuzzy automaton to~another, and vice versa, which are not necessarily mutually inverse.~Namely, the following example demonstrates a pair of fuzzy automata $\cal A$ and $\cal B$ such that there are forward simulations between $\cal A$ and $\cal B$ and between $\cal B$ and $\cal A$, but there is not any forward bisimulation between $\cal A$ and $\cal B$.

\begin{example}\label{ex:two-way}\rm
Let $\cal L$ be the G\"odel structure, and let ${\cal A}=(A,\delta^A,\sigma^A,\tau^A)$ and ${\cal B}=(B,\delta^B,\sigma^B,\tau^B)$ be fuzzy automata over $\cal L$ and $X=\{x,y\}$ with $|A|=3$, $|B|=2$ and
\[
\begin{aligned}
&\sigma^A=\begin{bmatrix} 0 & 0 & 1 \end{bmatrix}, \qquad &&\delta_x^A=\begin{bmatrix}
1 & 0.3 & 0.4 \\ 0.5 & 1 & 0.3 \\ 0.4 & 0.6 & 0.7 \end{bmatrix}, \qquad &&\delta_y^A=\begin{bmatrix}
0.5 & 0.6 & 0.2 \\ 0.6 & 0.3 & 0.4 \\ 0.7 & 0.7 & 1 \end{bmatrix},\qquad &&\tau^A=\begin{bmatrix}
1 \\ 1 \\ 1 \end{bmatrix}, \\
&\sigma^B=\begin{bmatrix} 0.7 & 1 \end{bmatrix}, \qquad &&\delta_x^B=\begin{bmatrix}
1 & 0.6 \\ 0.6 & 0.7 \end{bmatrix}, \qquad &&\delta_y^B=\begin{bmatrix}
0.6 & 0.6 \\ 0.7 & 1 \end{bmatrix},\qquad &&\tau^B=\begin{bmatrix}
1 \\ 1 \end{bmatrix}.
\end{aligned}
\]
We have that fuzzy relations
\[
\varphi = \begin{bmatrix} 1 & 0.7 \\ 1 & 0.7 \\ 0.6 & 1 \end{bmatrix}, \qquad
\psi = \begin{bmatrix} 1 & 1 & 0.7 \\ 0.6 & 0.6 & 1 \end{bmatrix},
\]
are respectively the greatest forward simulation between $\cal A$ and $\cal B$, and vice versa, but there is not any forward bisimulation between $\cal A$ and $\cal B$.

Indeed, let $\alpha $ be an arbitrary fuzzy relation between $A$ and $B$ such that $\varphi\leqslant \alpha $, i.e.,
\[
\alpha = \begin{bmatrix} 1 & a \\ 1 & b \\ c & 1 \end{bmatrix},\quad \text{with}\ a\geqslant 0.7,\ b\geqslant 0.7,\ \text{and}\ c\geqslant 0.6 .
\]
If $\alpha^{-1}\circ \delta_x^A\leqslant \delta_x^B\circ \alpha^{-1}$, then we easily obtain that $a\leqslant 0.7$, $b\leqslant 0.7$, and $c\leqslant 0.7$, which means that $a=b=0.7$ and $0.6\leqslant c\leqslant 0.7$.~Moreover, if $\alpha^{-1}\circ \delta_y^A\leqslant \delta_y^B\circ \alpha^{-1}$, then we obtain that $c\leqslant 0.6$, and hence, $c=0.6$.~Therefore, we have proved that $\alpha=\varphi $, which means that $\varphi $ is the greatest forward simulation between $\cal A$ and $B$.~In a similar way we show that $\psi $ is the greatest forward simulation between $\cal B$ and $A$.

Suppose now that $\beta $ is an arbitrary forward bisimulation between $\cal A$ and $\cal B$.~Then $\beta \leqslant \varphi $ and $\beta^{-1}\leqslant \psi $, i.e., $\beta \leqslant \varphi \land \psi^{-1}$, which implies
\[
\sigma^A\circ \beta \leqslant \sigma^A\circ(\varphi \land \psi^{-1})= \begin{bmatrix}
0 & 0 & 1 \end{bmatrix}\circ \begin{bmatrix} 1 & 0.6 \\ 1 & 0.6 \\ 0.6 & 1 \end{bmatrix} =
\begin{bmatrix} 0.6 & 1 \end{bmatrix} < \begin{bmatrix} 0.7 & 1 \end{bmatrix} = \sigma^B .
\]
This contradicts our assumption that $\beta $ is a forward bisimulation.~Hence, there is not any
forward bisimulation between $\cal A$ and $\cal B$.
\end{example}

In numerous papers dealing with simulations and bisimulations mostly forward simulations and forward~bisimulations have been studied.~They have been usually called just simulations and bisimulations, or {\it strong simulations\/} and {\it strong bisimulations\/} (cf.~\cite{Milner.89,Milner.99,RM-C.00}), and the greatest bisimulation equivalence relation has been usually called the {\it bisimilarity\/}.~Automata between which there is a bisimulation have been often called {\it bisimilar\/}.~Distinction between forward and backward simulations, and forward and backward bisimulations, has been made, for instance, in \cite{Buch.08,HKPV.98,LV.95} (for various kinds of automata), but~less or more these concepts differ from the concepts having the same name which are considered here.~More~similar to our concepts of forward and backward simulations and bisimulations are those studied in \cite{Brihaye.07}, and in \cite{HMM.07,HMM.09} (for tree automata).

The following lemma can be easily proved by induction.

\begin{lemma}\label{le:delta.u}
If condition {\rm (\ref{eq:fs0})} or condition {\rm (\ref{eq:bs0})} holds for every
letter $x\in X$, then it also holds if we replace~the letter $x$ by an arbitrary
word $u\in X^*$.

\end{lemma}

The following theorem presents some of the most important properties of simulations and bisimulations.

\begin{theorem}\label{th:Lang}
Let ${\cal A}=(A,\delta^A,\sigma^A,\tau^A)$ and ${\cal B}=(B,\delta^B,\sigma^B,\tau^B)$ be fuzzy automata and
let $\varphi\in {\cal F}(A\times B)$ be a fuzzy relation.~Then
\begin{itemize}\parskip-2pt
\item[{\rm (A)}] If $\varphi$ is a simulation, then $L({\cal A})\leqslant L({\cal B})$.
\item[{\rm (B)}] If $\varphi$ is a bisimulation, then $L({\cal A})=L({\cal B})$.
\end{itemize}
\end{theorem}

\begin{proof}
(A) Let $\varphi$ be a forward simulation. Then by (\ref{eq:fsi0})--(\ref{eq:fst0}) and Lemma \ref{le:delta.u}, for arbitrary $u\in X^*$, we obtain
\[
L({\cal A})(u)=\sigma^A\circ\delta_u^A\circ\tau^A\leqslant \sigma^B\circ\varphi^{-1}\circ\delta_u^A\circ\tau^A\leqslant \sigma^B\circ\delta_u^B\circ \varphi^{-1}\circ\tau^A\leqslant\sigma^B\circ\delta_u^B\circ\tau^B=L({\cal B})(u).
\]
Therefore, $L({\cal A})(u)\leqslant L({\cal B})(u)$. Similarly we prove the case when $\varphi$ is a backward simulation.

(B) This is a direct consequence of the definition of four types of bisimulations and the~statement (A).
\end{proof}

Let us note that the converse of the previous theorem does not necessarily hold.~Namely, the following example shows that two fuzzy automata may be language-equivalent, but there is not any kind of simulation and bisimulation between them.

\begin{example}\label{ex:no.sim}\rm
Let $\cal L$ be the G\"odel structure, and let ${\cal A}=(A,\delta^A,\sigma^A,\tau^A)$ and ${\cal B}=(B,\delta^B,\sigma^B,\tau^B)$ be fuzzy automata over $\cal L$ and $X=\{x\}$ with $|A|=3$, $|B|=2$ and
\[
\sigma^A=\begin{bmatrix} 0 & 0.5 & 1 \end{bmatrix}, \quad \delta_x^A=\begin{bmatrix}
1 & 0 & 0 \\ 0 & 0 & 1 \\ 0 & 0 & 0 \end{bmatrix},\quad \tau^A=\begin{bmatrix}
0 \\ 0 \\ 0.6 \end{bmatrix},\qquad \sigma^B=\begin{bmatrix} 0.6 & 0 \end{bmatrix}, \quad \delta_x^B=\begin{bmatrix} 0 & 1 \\ 0 & 0 \end{bmatrix}, \quad \tau^B=\begin{bmatrix}
1 \\ 0.5 \end{bmatrix}.
\]\pagebreak

\noindent
Then $L({\cal A})(e)=L({\cal B})(e)=0.6$, $L({\cal A})(x)=L({\cal B})(x)=0.5$, and $L({\cal A})(x^k)=L({\cal B})(x^k)=0$, for each $k\in \mathbb{N}$, $k\geqslant 2$. Thus, $\cal A$ and $\cal B$ are language-equivalent fuzzy automata.

We will show that there is not any kind of simulation and bisimulation between $\cal A$ and $\cal B$.~Consider an arbitrary fuzzy relation
\[
\varphi = \begin{bmatrix} a_{11} & a_{12} \\ a_{21} & a_{22} \\ a_{31} & a_{32}\end{bmatrix}
\]
between $A$ and $B$.~If $\sigma^A\leqslant \sigma^B\circ\varphi^{-1}$ holds, we have that $1\leqslant 0.6\land a_{31}\leqslant 0.6$, so we have obtained a contradiction. Thus, we conclude that there is not any forward simulation between $\cal A$ and $\cal B$.~Also, if $\delta_x^A\circ \varphi \leqslant \varphi\circ \delta_x^B$ holds, then we easily obtain that $a_{31}=0$, and if $\tau^A\leqslant \varphi \circ \tau^B$, then $0.6 \leqslant a_{31}\lor (a_{32}\land 0.5)=a_{32}\land 0.5\leqslant 0.5$, and again, we have a contradiction.~Thus, there is not any backward simulation between $\cal A$ and $\cal B$.~Since there is not any kind of simulation between $\cal A$ and $\cal B$, we conclude that there is not any kind of bisimulations between $\cal A$ and $\cal B$.

In a similar way we can show that there is not any kind of simulation between $\cal B$ and $\cal A$.
\end{example}

The proof of the next lemma follows directly by definitions of forward and backward simulations and the dual fuzzy automaton.

\begin{lemma}\label{reverse}
Let ${\cal A}=(A,\delta^A,\sigma^A, \tau^A)$ and ${\cal B}=(B,\delta^B, \sigma^B,\tau^B)$ be fuzzy automata.~A fuzzy relation $\varphi \in {\cal F}(A\times B)$ is a backward simulation between fuzzy automata ${\cal A}$ and ${\cal B}$ if and only if it is a forward simulation between the reverse fuzzy automata $\bar{{\cal A}}$ and $\bar{{\cal B}}$.
\end{lemma}

The preceeding lemma remain true if we replace all occurrences of the term simulation with bisimulation. In view of this lemma, for any statement on forward simulations or bisimulations which~is~univer\-sally~valid (valid for all fuzzy automata) there is the corresponding universally valid statement on backward~simula\-tions or bisimulations.~For that reason, in the sequel we deal only with forward bisimulations.

The next two lemmas can be easily proved using the definition of a forward bisimulation and conditions (\ref{eq:comp.mon}),   (\ref{eq:comp.inv}), (\ref{eq:comp.sup}), and (\ref{eq:sup.inv}).~Note that these lemmas also hold for all types of simulations and bisimulations.

\begin{lemma}\label{Composition}
Let ${\cal A}=(A,\delta^A,\sigma^A,\tau^A)$, ${\cal B}=(B,\delta^B,\sigma^B,\tau^B)$ and ${\cal C}=(C,\delta^C,\sigma^C,\tau^C)$ be fuzzy automata and
let $\varphi_1\in {\cal F}(A\times B)$ and $\varphi_2\in {\cal F}(B\times C)$ be forward bisimulations.

Then $\varphi_1\circ\varphi_2\in {\cal F}(A\times C)$ is also a forward bisimulation.
\end{lemma}

\begin{lemma}\label{le:fb.union}
Let ${\cal A}=(A,\delta^A,\sigma^A,\tau^A)$ and ${\cal B}=(B,\delta^B,\sigma^B,\tau^B)$ be fuzzy automata and
let $\{\varphi_i\}_{i\in I}\subseteq {\cal F}(A\times B)$ be a non-empty family of forward bisimulations between ${\cal A}$ and ${\cal B}$.

Then $\bigvee_{i\in I}\varphi_i$ is also a forward bisimulation between ${\cal A}$ and ${\cal B}$.
\end{lemma}

Now we are ready to state and prove the following fundamental result.

\begin{theorem}\label{th:GFB}
Let ${\cal A}=(A,\delta^A,\sigma^A,\tau^A)$ and ${\cal B}=(B,\delta^B,\sigma^B,\tau^B)$ be fuzzy automata such that there exists at least one forward bisimulation  between ${\cal A}$ and ${\cal B}$. Then there exist the greatest forward bisimulation  between ${\cal A}$ and ${\cal B}$.

Moreover, the greatest forward bisimulation  between ${\cal A}$ and ${\cal B}$ is a partial fuzzy function.
\end{theorem}

\begin{proof}
By the hypothesis of the theorem, the collection of all forward bisimulations  between ${\cal A}$ and ${\cal B}$ is~non-empty.~Let $\varphi $ be the join (union) of all members of this collection.~According to Lemma \ref{le:fb.union}, $\varphi $ is a forward bisimulation, and hence, it is the greatest forward bisimulation  between ${\cal A}$ and ${\cal B}$.

Further, according to Lemma \ref{Composition}, $\varphi \circ \varphi^{-1}\circ \varphi $ is also a forward bisimulation  between ${\cal A}$ and ${\cal B}$, and since $\varphi $ is the greatest one, we conclude that $\varphi \circ \varphi^{-1}\circ \varphi \leqslant \varphi $.~Now, by Theorem \ref{th:PFF} we obtain that $\varphi $ is a partial fuzzy function.
\end{proof}

The previous theorem points to the importance of studying those bisimulations which are partial fuzzy functions.~However, bisimulations are intended to model the equivalence between automata (or~some~rela\-ted systems), and two automata are considered to be equivalent (as a whole) if every state of the first~automaton is equivalent to some state of the second automaton, and vice versa.~In~the context of fuzzy automata this means that a fuzzy relation which we use to model the equivalence has to be a surjective $\cal L$-function (cf. Section \ref{sectionUniform}).~Seeing that partial fuzzy functions which are surjective $\cal L$-functions are exactly uniform fuzzy relations, we can say that the whole discussion points to the significance of studying those~bisimulations which are uniform fuzzy relations.~Just this kind of bisimulations is the main subject in most of this paper.

In numerous articles bisimulations were studied in the context of automata (or similar systems) without fixed initial states, or the definition of a bisimulation did not include any requirement concerning the~initial states.~In such cases there is a bisimulation between any two automata, the empty relation between them. However, here the empty relation is not a forward bisimulation, because it does not satisfy conditions (\ref{eq:fsi0}) and (\ref{eq:fsi0i}) (except if $\sigma^A$ and $\sigma^B$ are empty).~For that reason, in the formulation of Theorem \ref{th:GFB} we were forced to assume that there is at least one forward bisimulation  between ${\cal A}$ and ${\cal B}$.

At the beginning of this section we have defined the concepts of simulations and bisimulations between two fuzzy automata which are generally different.~In the rest of the section we consider the case when the two fuzzy automata are the same, i.e., we discuss bisimulations between a fuzzy automaton and itself.

Let $(A, \delta^A, \sigma^A, \tau^A)$ be an arbitrary fuzzy automaton. If a fuzzy relation $\varphi\in {\cal F}(A\times A)$ is a forward~bisimulation between ${\cal A}$ and itself, it will be called a {\it forward bisimulation on }~${\cal A}$ (analogously we define {\it backward~bisimulations on} ${\cal A}$).~Forward bisimulations on ${\cal A}$ which are fuzzy equivalence relations will be called {\it forward~bisimulation fuzzy equivalence relations}.

Seeing that any fuzzy equivalence relation satisfies condition (\ref{eq:fsi0}) (because of reflexivity and symmetry), we have that a fuzzy equivalence relation $E$ on ${\cal A}$ is a forward bisimulation on ${\cal A}$ if and only if
\begin{align}\label{Aca1}
&E\circ \delta_x^A\leqslant \delta_x^A\circ E,\qquad\text{for each }x\in X, \\
&E\circ \tau^A\leqslant \tau^A. \label{Aca1t}
\end{align}
(condition (\ref{Aca1t}) can be replaced with $E\circ \tau^A= \tau^A$). According to Theorem 4.1 \cite{CSIP.10} (see also Theorem~1~\cite{CSIP.07}), condition (\ref{Aca1}) is equivalent to
\begin{equation}
E\circ\delta_x^A\circ E=\delta_x^A\circ E,\qquad\text{for each }x\in X.
\end{equation}
Similarly, a fuzzy equivalence relation $E$ on ${\cal A}$ is a backward bisimulation on ${\cal A}$ if and only if
\begin{align}\label{Aca2}
&\delta_x^A\circ E\leqslant E\circ \delta_x^A,\qquad\text{for each }x\in X,\\
&\sigma^A\circ E\leqslant \sigma^A,\label{Aca2s}
\end{align}
(again we can replace (\ref{Aca2s}) with $\sigma^A\circ E= \sigma^A$), and we also have that condition (\ref{Aca2}) is equivalent to
\begin{equation}
E\circ\delta_x^A\circ E=E\circ \delta_x^A,\qquad\text{for each }x\in X.
\end{equation}
Let us note that forward and backward bisimulation fuzzy equivalence relations on fuzzy automata were studied in \cite{CSIP.07,CSIP.10,SCI.10}, where they were called right and left invariant fuzzy equivalence relations.~In~a~more general context, some related fuzzy equivalence relations, called right and left regular fuzzy equivalence relations, were studied in \cite{ICB.10}.

According to the results proved in \cite{CSIP.07,CSIP.10,SCI.10} (see also \cite{ICB.10}), any fuzzy automaton has the greatest forward and backward simulations, which are fuzzy quasi-orders, and the greatest forward and backward bisimulations, which are fuzzy equivalence relations.~Algorithms for their computing were provided in the same~papers.~It is worth noting~that many efficient algorithms were also provided for computing the greatest bisimulation equivalence relation and the greatest simulation quasi-order on a labeled transition system or a nondeterministic automaton.~The faster algorithms for computing the greatest bisimulation equivalence relation are based on the crucial equivalence between this problem and the rela\-tional coarsest partition problem (cf.~\cite{DPP.04,GPP.03,KS.90,PT.87,RT.08}).

\section{Uniform forward bisimulations}\label{sectionUFB}

In the previous section we pointed out the importance of uniform forward bisimulations. However, the true picture of their significance give the results proved in this and the next section.

\begin{theorem}\label{th:fb}
Let ${\cal A}=(A,\delta^A,\sigma^A, \tau^A)$ and ${\cal B}=(B,\delta^B,\sigma^B, \tau^B)$ be fuzzy automata and
let $\varphi \in {\cal F}(A\times B)$~be~an~arbi\-trary fuzzy relation.~Then the following is true: \\ [5pt]
{\rm (A)} $\varphi \circ  \varphi^{-1}$ is a forward bisimulation on $\cal A$ and $\varphi^{-1}\circ \varphi$ is a forward bisimulation on $\cal B$, and
\begin{align}
\label{eq:fb1}&\varphi\circ \varphi^{-1}\circ \delta_x^A \leqslant \varphi \circ \delta_x^B\circ
\varphi^{-1}\leqslant \delta_x^A\circ \varphi\circ \varphi^{-1},\qquad\text{for every }x\in X, \\
\label{eq:fb2}&\varphi^{-1}\circ \varphi\circ \delta_x^B \leqslant \varphi^{-1} \circ \delta_x^A\circ
\varphi\leqslant \delta_x^B\circ \varphi^{-1}\circ \varphi,\qquad\text{for every }x\in X,
\end{align}
{\rm (B)} If $\varphi $ is a uniform forward bisimulation, then $\varphi\circ\varphi^{-1}$ is a forward bisimulation fuzzy equivalence relation on ${\cal A}$ and $\varphi^{-1}\circ\varphi$ is a forward bisimulation fuzzy equivalence relation on ${\cal B}$.
\end{theorem}

\begin{proof}
{\rm (A)} The inequalities (\ref{eq:fb1}) and (\ref{eq:fb2}) follow directly by definition of a forward bisimulation and
 compatibility of the ordering of fuzzy relations with respect to the composition of fuzzy relations (cf.~Equation~(\ref{eq:comp.mon})\,).

By (\ref{eq:fb1}) and (\ref{eq:fb2}) it also follows that $\varphi \circ  \varphi^{-1}$ and $\varphi^{-1}\circ \varphi$ are
forward bisimulations on $\cal A$ and $\cal B$.

{\rm (B)} If $\varphi$ is a uniform forward bisimulation, then by Theorem \ref{th:ufr} it follows $\varphi\circ\varphi^{-1}=E_A^{\varphi}$ and $\varphi^{-1}\circ\varphi=E_B^{\varphi}$, and therefore,  $\varphi\circ\varphi^{-1}=E_A^{\varphi}$ and $\varphi^{-1}\circ\varphi=E_B^{\varphi}$ are fuzzy equivalence relations on ${\cal A}$ and ${\cal B}$, respectively.~The
rest of the claim follows from (A).
\end{proof}

It is worth to repeat that $\varphi \in {\cal F}(A\times B)$ is a uniform fuzzy relation if and only if it is a surjective $\cal L$-func\-tion and $E_A^\varphi =\varphi \circ \varphi^{-1}$, or equivalently, if it is a surjective $\cal L$-function and $E_B^\varphi =\varphi^{-1} \circ \varphi$ (cf.~Theorem~\ref{th:ufr}).

Remind that forward and backward bisimulations were defined in Section \ref{sectionBisimulations} by means of in\-equalities.~But, it is often much easier to work with equalities than with inequalities, and the following theorem shows that uniform forward bisimulations can be also defined by means of equalities.

\begin{theorem}\label{th:ufbreqcond}
Let ${\cal A}=(A,\delta^A,\sigma^A,\tau^A)$ and ${\cal B}=(B,\delta^B,\sigma^B,\tau^B)$ be fuzzy automata and
let $\varphi\in {\cal F}(A\times B)$ be a uniform fuzzy relation. Then $\varphi$ is a forward bisimulation if and only if
the following hold:
\begin{align}
\label{eq:1}
 \sigma^A\circ \varphi\circ\varphi^{-1}&=\sigma^B\circ\varphi^{-1},&\ \     \sigma^A\circ\varphi&=\sigma^B\circ\varphi^{-1}\circ\varphi, &\ &\\
 \label{eq:2}
\delta_x^A\circ
\varphi\circ \varphi^{-1}&=\varphi \circ \delta_x^B\circ \varphi^{-1} ,&\ \  \varphi^{-1}\circ \delta_x^A\circ \varphi &= \delta_x^B\circ
\varphi^{-1}\circ \varphi ,&\ \ &\textit{for every }x\in X, \\
\label{eq:3}
 \tau^A&=\varphi\circ\tau^B,&\ \   \varphi^{-1}\circ\tau^A&=\tau^B. &\ &
\end{align}
\end{theorem}

\begin{proof}
Let $\varphi $ be a forward bisimulation between fuzzy automata ${\cal A}$ and ${\cal B}$.~Then $\sigma^A\leqslant \sigma^B\circ\varphi^{-1}$ implies
\[
\sigma^A\circ \varphi\circ\varphi^{-1}\leqslant \sigma^B\circ\varphi^{-1}\circ \varphi\circ\varphi^{-1}=\sigma^B\circ \varphi^{-1},
\]
and $\sigma^B\leqslant \sigma^A\circ\varphi$ implies
$\sigma^B\circ\varphi^{-1}\leqslant \sigma^A\circ\varphi\circ\varphi^{-1}.
 $
Thus, we have $\sigma^A\circ \varphi\circ\varphi^{-1}=\sigma^B\circ\varphi^{-1}$.~Moreover, we have $\sigma^A\circ\varphi=\sigma^A\circ\varphi\circ\varphi^{-1}\circ\varphi=\sigma^B\circ\varphi^{-1}\circ\varphi$.

By (A) of Theorem \ref{th:fb},
for each $x\in X$ we have that
\[
\varphi \circ\delta_x^B\circ\varphi^{-1} \leqslant \delta_x^A\circ \varphi
\circ\varphi^{-1}= \varphi \circ\varphi^{-1}\circ \delta_x^A\circ \varphi \circ\varphi^{-1}
\leqslant\varphi \circ \delta_x^B\circ\varphi^{-1}\circ \varphi \circ\varphi^{-1} = \varphi \circ\delta_x^B\circ\varphi^{-1},
\]
whence $\varphi \circ\delta_x^B\circ\varphi^{-1} = \delta_x^A\circ \varphi
\circ\varphi^{-1}$, and similarly, $\varphi^{-1}\circ\delta_x^A\circ\varphi =
\delta_x^B\circ \varphi^{-1}\circ\varphi$.~On the other hand, by (B) of Theorem \ref{th:fb} we obtain that $\tau^A=\varphi\circ\varphi^{-1}\circ \tau^A\leqslant \varphi\circ\tau^B\leqslant \tau^A$, so $\tau ^A=\varphi\circ\tau^B$, and similarly $\tau^B=\varphi^{-1}\circ\tau^A$.

Conversely, let   (\ref{eq:1})--(\ref{eq:3}) hold. Then $\sigma^A\leqslant \sigma^A\circ\varphi\circ\varphi^{-1}=\sigma^B\circ\varphi^{-1}$, and for every $x\in X$, we have
\[
\varphi^{-1}\circ\delta_x^A\leqslant \varphi^{-1}\circ\delta_x^A\circ\varphi\circ\varphi^{-1} =\delta_x^B\circ\varphi\circ\varphi^{-1}\circ\varphi^{-1}=\delta_x^B\circ\varphi^{-1}.
\]
Clearly, $\varphi^{-1}\circ\tau^A\leqslant\tau^B$, so $\varphi$ is a forward simulation between ${\cal A}$ and ${\cal B}$. Further, $\sigma^B\leqslant\sigma^B\circ\varphi^{-1}\circ\varphi=\sigma^A\circ\varphi$, and for every $x\in X$, we have
\[
\varphi\circ\delta_x^B\leqslant \varphi\circ\delta_x^B\circ\varphi^{-1}\varphi=\delta_x^A\circ\varphi\circ\varphi^{-1}\circ\varphi=\delta_x^A\circ\varphi.
\]
Also, $\varphi\circ\tau^A\leqslant \tau^B$, whence it follows that $\varphi^{-1}$ is a forward simulation between ${\cal B}$ and ${\cal A}$.~Therefore, $\varphi$ is a forward bisimulation between fuzzy automata ${\cal A}$ and ${\cal B}$.
\end{proof}

The following theorem is one of the main results of this article.

\begin{theorem}\label{th:ufbr}
Let ${\cal A}=(A,\delta^A,\sigma^A,\tau^A)$ and ${\cal B}=(B,\delta^B,\sigma^B,\tau^B)$ be fuzzy automata and
let $\varphi \in {\cal F}(A\times B)$ be a uniform fuzzy relation. Then
 $\varphi $ is a forward bisimulation if and only if the following is true:
\begin{itemize}\parskip=0pt
\item[{\rm (i)}] $E_A^\varphi $ is a forward bisimulation
on $\cal A$;
\item[{\rm (ii)}] $E_B^\varphi $ is a forward bisimulation  on $\cal B$;
\item[{\rm (iii)}]  $\widetilde \varphi $ is an isomorphism of\/ factor fuzzy automata ${\cal A}/E_A^\varphi $ and ${\cal B}/E_B^\varphi $.
\end{itemize}
\end{theorem}

\begin{proof}

 For the sake of simplicity set $E=E_A^\varphi $ and $F=E_B^\varphi$.
\smallskip

Let $\varphi $ be a forward bisimulation.~The assertions (i) and (ii)
follow directly by~(iv) and (v) of Theorem~\ref{th:ufr} and Theorem \ref{th:fb}.~According~to~Lemma \ref{le:tpsi},
 $\widetilde\varphi$ is a bijective~function of $A/E$ onto $B/F$.

Next, consider arbitrary $a,a_1,a_2\in A$, $x\in X$ and $\psi \in CR(\varphi )$.~According to
Theorem \ref{th:ufbreqcond}, we have that
\begin{align*}
\sigma^{A/E}(E_a)&=(\sigma^A\circ E)(a)=(\sigma^B\circ\varphi^{-1})(a)=\bigvee_{b\in B}\sigma^B(b)\otimes\varphi(a,b)
                 =\bigvee_{b\in B}\sigma^B(b)\otimes F(\psi(a),b)\\
                 &=\bigvee_{b\in B}\sigma^B(b)\otimes F(b,\psi(a))
                 =(\sigma^B\circ F)(\psi(a))=\sigma^{B/F}(F_{\psi(a)})=\sigma^{B/F}(\widetilde{\varphi}(E_a)),\\
\tau^{A/E}(E_a)&=(\tau^A\circ E)(a)=\tau^A(a)=(\varphi\circ\tau^B)(a)=\bigvee_{b\in B}\varphi(a,b)\otimes\tau^B(b)
=\bigvee_{b\in B}F(\psi(a),b)\otimes\tau^B(b)\\
&=(F\circ\tau^B)(\psi(a))
=\tau^{B/F}(F_{\psi(a)})=\tau^{B/F}(\widetilde{\varphi}(E_a))
\end{align*}
Further, by (i), (\ref{eq:fb1}) and (\ref{eq:3phi}) we obtain that
\[
\delta_x^A\circ E = E\circ \delta_x^A\circ E = \varphi \circ \varphi^{-1}\circ \delta_x^A\circ \varphi \circ \varphi^{-1} \leqslant \varphi \circ \delta_x^B\circ \varphi^{-1}\circ \varphi \circ \varphi^{-1} = \varphi\circ\delta_x^B\circ \varphi^{-1} \leqslant \delta_x^A\circ
E,
\]
and hence, $E\circ \delta_x^A\circ E = \delta_x^A\circ E= \varphi\circ\delta_x^B\circ \varphi^{-1}$.
By this and by (\ref{eq:phi.EB}) it follows that
\[
\begin{aligned}
\delta^{A/E}(E_{a_1},x,E_{a_2}) &= (E\circ \delta_x^A\circ E) (a_1,a_2) \\
&= (\varphi\circ\delta_x^B\circ \varphi^{-1} )(a_1,a_2) \\
& = \bigvee_{b_1,b_2\in B} \varphi(a_1,b_1)\otimes
\delta_x^B(b_1,b_2)\otimes \varphi(a_2,b_2) \\
& = \bigvee_{b_1,b_2\in B} F(\psi(a_1),b_1)\otimes
\delta_x^B(b_1,b_2)\otimes F(\psi(a_2),b_2) \\
& = (F\circ \delta_x^B\circ F) (\psi(a_1),\psi(a_2)) \\
& = \delta^{B/F}(F_{\psi(a_1)},x,F_{\psi(a_2)}) \\
& = \delta^{B/F}(\widetilde\varphi(E_{a_1}),x,\widetilde\varphi(E_{a_2})) ,
\end{aligned}
\]
Therefore, $\widetilde{\varphi}$ is an isomorphism between fuzzy automata  ${\cal A}/E$ and ${\cal B}/F$.

Conversely, let (i), (ii) and (iii) hold. For arbitrary $\psi \in CR(\varphi )$, $\xi \in CR(\varphi^{-1})$, $a,a_1,a_2\in A$, $b_1,b_2\in B$ and $x\in X$ we have that
\begin{align*}
\sigma^A(a)&\leqslant (\sigma^A\circ E)(a)=\sigma^{A/E}(E_a)=\sigma^{B/F}(\widetilde{\varphi}(E_a))=\sigma^{B/F}(F_{\psi(a)})=(\sigma^B\circ F)(\psi(a))\\
&=\bigvee_{b\in B}\sigma^B(b)\otimes F(b,\psi(a))=\bigvee_{b\in B}\sigma^B(b)\otimes F(\psi(a),b)=
\bigvee_{b\in B}\sigma^B(b)\otimes\varphi(a,b)\\
&=\bigvee_{b\in B}\sigma^B(b)\otimes\varphi^{-1}(b,a)=(\sigma^B\circ\varphi^{-1})(a),
\end{align*}
so, $\sigma^A\leqslant \sigma^B\circ\varphi^{-1}$, and in a similar way we prove that $\sigma^B\leqslant \sigma^A\circ\varphi$.
Further,
\[
\begin{aligned}
(E\circ \delta_x^A\circ E)(a_1,a_2)&=  \delta_{}^{A/E}(E_{a_1},x,E_{a_2}) = \delta^{B/F}(\widetilde\varphi(E_{a_1}),x,\widetilde\varphi(E_{a_2})) \\
&= \delta^{B/F}(F_{\psi(a_1)},x,F_{\psi(a_2)}) =  (F\circ \delta_x^B\circ F) (\psi(a_1),\psi(a_2)) ,
\end{aligned}
\]
and similarly,
\[
(F\circ \delta_x^B\circ F)(b_1,b_2)= (E\circ \delta_x^A\circ E) (\xi(b_1),\xi(b_2)).
\]
According to (i) and (ii), $\varphi^{-1}\circ \delta_x^A = \varphi^{-1}\circ E\circ \delta_x^A \leqslant \varphi^{-1}\circ \delta_x^A\circ E$, and now, by (\ref{eq:phi.EB})
and (\ref{eq:phi.EA}), for all $a\in A$ and $b\in B$ we obtain that
\[
\begin{aligned}
(\varphi^{-1}\circ \delta_x^A) (b,a)  &\leqslant (\varphi^{-1} \circ \delta_x^A\circ
E) (b,a)   = \bigvee_{a_1\in A} \varphi^{-1}(b,a_1)\otimes ( \delta_x^A\circ
E)(a_1,a) \\
&= \bigvee_{a_1\in A} E(\xi(b),a_1)\otimes ( \delta_x^A\circ
E)(a_1,a) = (E\circ \delta_x^A\circ E)(\xi(b),a) \\
&=(F\circ \delta_x^B\circ F)(\psi((\xi(b)),\psi(a)) =
\bigvee_{b_1\in B} F(\psi(\xi(b)),b_1)\otimes ( \delta_x^B\circ F)(b_1,\psi(a))\\
&= \bigvee_{b_1\in B} F(b,b_1)\otimes ( \delta_x^B\circ F)(b_1,\psi(a)) =
( F \circ \delta_x^B\circ F )(b,\psi(a)) = (\delta_x^B\circ F )(b,\psi(a))\\
&= \bigvee_{b_2\in B} \delta_x^B(b,b_2)\otimes F(b_2,\psi(a)) = \bigvee_{b_2\in B} \delta_x^B(b,b_2)\otimes \varphi(a,b_2) =(\delta_x^B\circ \varphi^{-1})(b,a),
\end{aligned}
\]
so $\varphi^{-1}\circ \delta_x^A\leqslant \delta_x^B\circ \varphi^{-1}$.~Here we have used
the fact that $F(\psi (\xi (b)),b)=\varphi (\xi(b),b)=\varphi^{-1}(b,\xi (b)) = 1$,
whence $F_{\psi(\xi(b))}=F_b$ and
\[
F(\psi(\xi(b)),b_1)=F_{\psi(\xi(b))}(b_1)=F_{b}(b_1)=F(b,b_1).
\]
In a similar way we show that $\varphi\circ \delta_x^B\leqslant \delta_x^A\circ \varphi$.
Further,
\begin{align*}
\tau^A(a)&=(E\circ\tau^A)(a)=\tau^{A/E}(E_a)=\tau^{B/F}(\widetilde{\varphi}(E_a))=\tau^{B/F}(F_{\psi(a)})=\bigvee_{b\in B} F(\psi(a),b)\otimes\tau^B(b)\\
&=\bigvee_{b\in B}\varphi(a,b)\otimes\tau^B(b)=(\varphi\circ\tau^B)(a),
\end{align*}
so $\tau^A=\varphi\circ\tau^B$, and similarly we get $\varphi^{-1}\circ\tau^A\leqslant \tau^B$. Therefore, $\varphi$ is a uniform forward bisimulation between fuzzy automata ${\cal A}$ and ${\cal B}$.
\end{proof}

\begin{theorem}\label{th:ufb.ex}
 Let ${\cal A}=(A,\delta^A,\sigma^A,\tau^A)$ and ${\cal B}=(B,\delta^B,\sigma^B,\tau^B)$ be fuzzy automata, and
let $E$ be a forward~bisimu\-lation on $\cal A$ and $F$ a forward bisimulation on $\cal B$.

Then there exists a uniform forward bisimulation $\varphi\in {\cal F}(A\times
B)$ such that
\begin{equation}\label{eq:ufb.1}
E_A^\varphi =E\ \ \mbox{and}\ \ E_B^\varphi =F ,
\end{equation}
if and only if there exists an isomorphism $\phi :{\cal
A}/E \to {\cal B}/F$ such that for all $a_1,a_2\in A$ we have
\begin{equation}\label{eq:ufr.4}
\widetilde E(E_{a_1},E_{a_2}) = \widetilde F (\phi(E_{a_1}),\phi(E_{a_2})) .
\end{equation}
\end{theorem}

\begin{proof}
Let $\varphi $ be a uniform forward bisimulation satisfying (\ref{eq:ufb.1}), and let $\psi\in CR(\varphi )$.~By
definition of $\widetilde E$, $\widetilde F$ and $\widetilde\varphi $, and Corollary
\ref{le:tpsi}, we obtain that
\[
\widetilde E(E_{a_1},E_{a_2}) = E(a_1,a_2)= F(\psi(a_1),\psi(a_2))=
\widetilde F(F_{\psi(a_1)},F_{\psi(a_2)}) = \widetilde F
(\widetilde\varphi (E_{a_1}),\widetilde\varphi (E_{a_2})) ,
\]
and according to Theorem \ref{th:ufbr}, $\widetilde\varphi $ is an isomorphism of
${\cal A}/E$ onto ${\cal B}/F$ satisfying (\ref{eq:ufr.4}).

Conversely, let $\phi :{\cal A}/E \to {\cal B}/F$ be an isomorphism satisfying
(\ref{eq:ufr.4}).~Define functions $\phi_E :A\to A/E $, $\phi_F :B/F\to B$ and $\psi:A\to B$ in the following way:
For any $x\in A$ let $\phi_E(x)=E_x$, for any $\eta\in B/F$ let $\phi_F (\eta )$ be a fixed element
from $\widehat \eta$, and let $\psi=\phi_E\circ\phi\circ\phi_F $.~Define also a fuzzy
relation $\varphi \in {\cal F}(A\times B)$~by
\begin{equation}\label{eq:psi-phi}
\varphi (a,b) = F(\psi(a),b), \qquad \text{for all $a\in A$ and $b\in B$}.
\end{equation}
By the proof of Theorem 3.4 \cite{CIB.09}, $\varphi $ is a uniform fuzzy relation satisfying
(\ref{eq:ufb.1}) and~$\psi\in CR(\varphi )$.~For~an~arbi\-trary $a\in A$~we have that $\phi (E_a)=F_b$, for some $b\in B$, and
\[
\psi (a)=\phi_F(\phi (\phi_E(a)))= \phi_F(\phi (E(a))=  \phi_F(F_b)\in \widehat F_b,
\]
whence it follows that
\[
\widetilde\varphi (E_a)=F_{\psi(a)}=F_b = \phi(E_a),
\]
for every $a\in A$. Therefore, $\widetilde\varphi =\phi $, and according to Theorem
\ref{th:ufbr}, $\varphi $ is a uniform forward bisimulation.
\end{proof}

\section{UFB-equivalent fuzzy automata}\label{sectionUFBeq}

Bisimulations have been studied in various contexts, but they have always been used as a means~to~estab\-lish structural equivalence between states of one or two different automata or related systems.~As we already said in Section~\ref{sectionBisimulations},~two automata can be considered structurally equivalent only if every state of one automaton is equivalent to some state of another automaton, and vice~versa.~There is no real equivalence between two automata if some state is not equivalent to any state of another automaton.~In other words,~equivalence should be modeled by a complete and surjective relation, or in the fuzzy context, by a surjective $\cal L$-function. But, if there is a forward bisimulation between two fuzzy automata $\cal A$ and $\cal B$ which is a surjective~$\cal L$-func\-tion, then the greatest forward bisimulation between $\cal A$ and $\cal B$ also has this property, and according to Theorem \ref{th:GFB}, it is a uniform forward bisimulation.~Therefore, it is absolutely the same to say that there is
a forward bisimulation between $\cal A$ and $\cal B$ which is a surjective~$\cal L$-func\-tion, and to say that there is a uniform forward bisimulation between $\cal A$ and $\cal B$.~For this reason, we introduce the following type of equivalence between fuzzy automata.

Fuzzy automata ${\cal A}=(A,\delta^A,\sigma^A,\tau^A)$ and ${\cal B}=(B,\delta^B,\sigma^B,\tau^B)$ will be called
{\it uniformly forward bisimulation equi\-valent\/}, or briefly {\it UFB-equivalent\/}, in notation ${\cal A}\sim_{UFB}{\cal B}$, if there exists a uniform forward bisimula\-tion between $\cal A$ and $\cal B$.~Analogously we can define {\it uniformly backward bisimulation equivalent\/} fuzzy automata,~but this kind of equivalence will not be considered separately, because of properties that are analogous to the properties of UFB-equiv\-alence.

The next theorem establishes a correspondence between the uniform forward bisimulations and forward bisimulation equivalence relations, analogous to the correspondence between homomorphisms and congruences in algebra.

\begin{theorem}\label{th:nat.uff}
Let ${\cal A}=(A,\delta^A,\sigma^A,\tau^A)$ be a fuzzy automaton, let $E$ be a fuzzy equivalence relation
on $\cal A$, and let ${\cal A}/E=(A/E,\delta^{A/E},\sigma^{A/E},\tau^{A/E})$ be the fuzzy factor automaton
of $\cal A$ with respect to $E$.
\begin{itemize}\parskip=0pt
\item[{\rm (A)}] A fuzzy relation $\varphi\in {\cal F}(A\times A/E)$ defined by
\begin{equation}\label{eq:nat.uff}
\varphi (a_1,E_{a_2})=E(a_1,a_2),\ \ \ \text{for all $a_1,a_2\in A$,}
\end{equation}
is a uniform fuzzy relation such that $E_A^\varphi =E$ and $E_{A/E}^\varphi
$ is a fuzzy equality, and $\varphi $ is both a forward  and back\-ward simulation.
\item[{\rm (B)}] The following conditions are equivalent:
\begin{itemize}\parskip=0pt
\item[{\rm (i)}] $E$ is a forward  bisimulation on $\cal A$;
\item[{\rm (ii)}] $\varphi $ is a forward  bisimulation;
\item[{\rm (iii)}] $\varphi $ is a backward-forward  bisimulation.
\end{itemize}
\end{itemize}
\end{theorem}

\begin{proof} (A) According to Theorem 7.1 \cite{CIB.09}, $\varphi $ is a uniform fuzzy
relation, $E_A^\varphi =E$ and $E_{A/E}^\varphi $ is a fuzzy equality. Moreover,
for arbitrary $x\in X$ and $a,a_1,a_2\in A$ we have that
\begin{equation}
\label{eq:Enati1}
\begin{aligned}
\sigma^A(a)&\leqslant (\sigma^A\circ E \circ E) (a)=\bigvee_{a_3\in A}(\sigma^A\circ E)(a_3)\otimes E(a_3,a)=\bigvee_{a_3\in A}\sigma^{A/E}(E_{a_3})\otimes \varphi^{-1}(E_{a_3},a)\\&=(\sigma^{A/E}\circ \varphi^{-1})(a),
\end{aligned}
\end{equation}
\begin{equation}
\label{eq:Enat1}
\begin{aligned}
(\varphi^{-1}\circ \delta_x^A) (E_{a_1},a_2) &=
\bigvee_{a_3\in A}\varphi^{-1}(E_{a_1},a_3)\otimes \delta_x^A(a_3,a_2) = \bigvee_{a_3\in A}E(a_1,a_3)\otimes \delta_x^A(a_3,a_2) \\
&= (E\circ \delta_x^A )(a_1,a_2) \leqslant  (E\circ \delta_x^A\circ E)(a_1,a_2) = (E\circ \delta_x^A\circ E\circ E)(a_1,a_2) \\
&= \bigvee_{a_3\in A}(E\circ \delta_x^A\circ E)(a_1,a_3)\otimes E(a_3,a_2) = \bigvee_{a_3\in A}\delta_x^{A/E}(E_{a_1},E_{a_3})\otimes \varphi^{-1}(E_{a_3},a_2)\\
&= (\delta_x^{A/E}\circ \varphi^{-1} )(E_{a_1},a_2),
\end{aligned}
\end{equation}
\begin{equation}
\label{Enatt1}
\begin{aligned}
(\varphi^{-1}\circ\tau^A)(E_a)&=\bigvee_{a_3\in A}\varphi^{-1}(E_a,a_3)\otimes \tau^A(a_3)=E(a,a_3)\otimes\tau^A(a_3)=(E\circ \tau^A)(a)=\tau^{A/E}(E_a),
\end{aligned}
\end{equation}

and also,
\begin{equation}\label{eq:Enati2}
\begin{aligned}
(\sigma^A\circ \varphi)(E_a)=\bigvee_{a_1\in A}\sigma^A(a_3)\otimes \varphi(a_1,E_a)=\bigvee_{a_3\in A}\sigma^A(a_1)\otimes E(a_1,a)=(\sigma^A\circ E)(a)=\sigma^{A/E}(E_a),
\end{aligned}
\end{equation}
\begin{equation}\label{eq:Enat2}
\begin{aligned}
(\delta_x^A\circ \varphi) &(a_1,E_{a_2}) =
\bigvee_{a_3\in A}\delta_x^A(a_1,a_3)\otimes  \varphi(a_3,E_{a_2}) =
\bigvee_{a_3\in A}\delta_x^A(a_1,a_3)\otimes  E(a_3,a_2)  \\
&= (\delta_x^A\circ E)(a_1,a_2) \leqslant  (E\circ \delta_x^A\circ E)(a_1,a_2) = (E\circ E\circ \delta_x^A\circ  E)(a_1,a_2) \\
&= \bigvee_{a_3\in A} E(a_1,a_3)\otimes E\circ \delta_x^A\circ E(a_3,a_2) = \bigvee_{a_3\in A} \varphi(a_1,E_{a_3})\otimes\delta_x^{A/E}(E_{a_3},E_{a_2})\\
&=  (\varphi\circ\delta_x^{A/E} )(a_1,E_{a_2}),
\end{aligned}
\end{equation}
\begin{equation}\label{eq:Enatt2}
\begin{aligned}
\tau^A(a)\leqslant (E\circ E\circ \tau^A)(a)=\bigvee_{a_3\in A}E(a,a_3)\otimes E\circ\tau^A(a_3)=\bigvee_{a_3\in A}\varphi(a,E_{a_3})\otimes\tau^{A/E}(E_{a_3})=(\varphi\circ\tau^{A/E})(a).
\end{aligned}
\end{equation}
Therefore, $\varphi $ is both a forward and a backward simulation.

(B) It is evident that the opposite inequalities in (\ref{eq:Enat2}) and (\ref{eq:Enatt2}) hold (i.e., $\varphi^{-1}$
is a forward simulation) if and~only if $E$ is a forward bisimulation on $\cal A$.~Therefore,
we conclude that conditions (i), (ii) and~(iii)  are equivalent.
\end{proof}

The next example shows that a fuzzy relation $\varphi $ defined as in (\ref{eq:nat.uff}) need not be a forward bisimulation (if $E$ is not a forward bisimulation on $\cal A$).

\begin{example}\label{ex:ffr}\rm
Let $\cal L$ be the Boolean structure, and let ${\cal A}=(A,\delta^A,\sigma^A,\tau^A)$ be a fuzzy
automaton over $\cal L$ (i.e., a nondeterministic automaton) with $|A|=4$, $X=\{x\}$ and
\[\small
\delta_x^A=\begin{bmatrix}
1 & 0 & 0 & 0 \\
0 & 0 & 0 & 1 \\
0 & 0 & 0 & 0 \\
0 & 0 & 0 & 0
\end{bmatrix},\ \ \ \
\sigma^A=\begin{bmatrix}
0 & 1 & 0 & 0
\end{bmatrix},\ \ \ \
\tau^A=\begin{bmatrix}
0 \\
0 \\
1 \\
1
\end{bmatrix},
\]
and let $E$ and $F$ be fuzzy equivalence relations on $A$ given by
\[\small
E=\begin{bmatrix}
1 & 0 & 0 & 0 \\
0 & 1 & 0 & 0 \\
0 & 0 & 1 & 1 \\
0 & 0 & 1 & 1
\end{bmatrix},\ \ \ \
F=\begin{bmatrix}
1 & 1 & 0 & 0 \\
1 & 1 & 0 & 0 \\
0 & 0 & 1 & 1 \\
0 & 0 & 1 & 1
\end{bmatrix}.
\]
The factor fuzzy automata ${\cal A}/E=(A/E,\delta^{A/E},\sigma^{A/E},\tau^{A/E})$ and
${\cal A}/F=(A/F,\delta^{A/F},\sigma^{A/F},\tau^{A/F})$ are given as~follows: $|A/E|=3$, $|A/F|=2$, and
\[\small
\delta_x^{A/E}=\begin{bmatrix}
1 & 0 & 0 \\
0 & 0 & 1 \\
0 & 0 & 0
\end{bmatrix},\ \
\sigma^{A/E}=\begin{bmatrix}
0 & 1 & 0
\end{bmatrix},\ \
\tau^{A/E}=\begin{bmatrix}
0 \\
0 \\
1
\end{bmatrix},\ \ \ \
\delta_x^{A/F}=\begin{bmatrix}
1 & 1 \\
0 & 0
\end{bmatrix},\ \
\sigma^{A/F}=\begin{bmatrix}
1 & 0
\end{bmatrix},\ \
\tau^{A/F}=\begin{bmatrix}
0 \\
1
\end{bmatrix}.
\]
It can be easily verified that ${\cal A}/E$ is language equivalent to ${\cal A}$, but ${\cal A}/F$ is not language equivalent to ${\cal A}$ (cf. Example 3.1 \cite{CSIP.10}).~Let $\varphi_E \in {\cal F}(A\times A/E)$ be a fuzzy relation given by
\[\small
\varphi_E=\begin{bmatrix}
1 & 0 & 0 \\
0 & 1 & 0 \\
0 & 0 & 1 \\
0 & 0 & 1
\end{bmatrix}.
\]
It is also easy to check that $\varphi_E$ is a forward bisimulation between $\cal A$ and ${\cal A}/E$, as well as a backward-forward bisimulation between $\cal A$ and ${\cal A}/E$, and hence, fuzzy automata $\cal A$ and ${\cal A}/E$ are UFB-equivalent.

On the other hand, fuzzy automata $\cal A$ and ${\cal A}/F$ are not UFB-equivalent, because they are not language equivalent (cf.~Theorem \ref{th:Lang}), and the fuzzy relation $\varphi_F\in {\cal F}(A\times A/F)$ defined by $\varphi_F(a_1,F_{a_2})=F(a_1,a_2)$, for all $a_1,a_2\in A$ (i.e., defined as in (\ref{eq:nat.uff})), is not a forward bisimulation.~For suspicious,
\[\small
\varphi_F=
\begin{bmatrix}
1 & 0 \\
1 & 0 \\
0 & 1 \\
0 & 1
\end{bmatrix} \quad \text{and}\quad \varphi_F\circ \delta_x^{A/F} =
\begin{bmatrix}
1 & 1 \\
1 & 1 \\
0 & 0 \\
0 & 0
\end{bmatrix} \nleqslant
\begin{bmatrix}
1 & 0 \\
0 & 1 \\
0 & 0 \\
0 & 0
\end{bmatrix} = \delta_x^A\circ\varphi_F .
\]
\end{example}

The following theorem (the part (A)) can be conceived as a version, for fuzzy automata, of the~well-known Second Isomorphism Theorem
from universal algebra.

\begin{theorem}\label{th:G:E}
Let ${\cal A}=(A,\delta^A,\sigma^A,\tau^A)$ be a fuzzy automaton, let $E$ and $G$ be fuzzy equivalence relations
on $\cal A$ such that $E\leqslant G$, and let ${\cal A}/E=(A/E,\delta^{A/E},\sigma^{A/E},\tau^{A/E})$ be the fuzzy factor automaton of $\cal A$ with respect to $E$.~Then
\begin{itemize}\parskip=0pt
\item[{\rm (A)}] A fuzzy relation $G/E$ on ${\cal A}/E$ defined by
\begin{equation}\label{eq:G:E}
G/E (E_{a_1},E_{a_2})=G(a_1,a_2),\ \ \ \text{for all $a_1,a_2\in A$,}
\end{equation}
is a fuzzy equivalence relation on ${\cal A}/E$, and the factor fuzzy automata $({\cal A}/E)/(G/E)$
and ${\cal A}/G$ are isomorphic.
\item[{\rm (B)}] A fuzzy relation $\varphi\in {\cal F}(A\times A/E)$ defined by
\begin{equation}\label{eq:F:E.nat}
\varphi (a_1,E_{a_2})=G(a_1,a_2),\ \ \ \text{for all $a_1,a_2\in A$,}
\end{equation}
is a uniform fuzzy relation satisfying $E_A^\varphi =G$ and $E_{A/E}^\varphi = G/E$.
\end{itemize}
In addition, if $E$ is a forward bisimulation on $\cal A$, then the following is true:
\begin{itemize}\parskip=0pt
\item[{\rm (C)}] $G$ is a forward bisimulation on $\cal A$ if and only if $G/E$ is a forward bisimulation on ${\cal A}/E$.
\item[{\rm (D)}] $G$ is the greatest forward bisimulation on $\cal A$ if and only if $G/E$ is the greatest forward bisimulation on ${\cal A}/E$.
\item[{\rm (E)}] $G$ is a forward bisimulation on $\cal A$ if and only if $\varphi $ is a forward bisimulation between $\cal A$ and ${\cal A}/E$.
\end{itemize}
\end{theorem}

\begin{proof}
The assertion (A) was proved in Theorem 3.1 \cite{CSIP.10}, but without considering fuzzy sets of initial and terminal states, and in Theorem 3.3 \cite{SCI.10}, in a more general context (for fuzzy quasi-orders).

(B) For arbitrary $a_1,a_2,a_3,a_4\in A$ we have that
\[
\varphi (a_1,E_{a_2})\otimes \varphi (a_3,E_{a_2})\otimes \varphi (a_3,E_{a_4})=G(a_1,a_2)\otimes G(a_2,a_3)\otimes G(a_3,a_4)\leqslant G(a_1,a_4) =\varphi (a_1,E_{a_4}),
\]
so $\varphi $ is a partial fuzzy function. It is clear that $\varphi $ is
also a surjective $\cal L$-function, and hence, $\varphi $ is a~uniform fuzzy relation.
Moreover, for arbitrary $a_1,a_2\in A$ we have that
\[
\begin{aligned}
(\varphi\circ\varphi^{-1})(a_1,a_2) &= \bigvee_{a_3\in A}\varphi(a_1,E_{a_3})\otimes
\varphi(a_2,E_{a_3}) = \bigvee_{a_3\in A}G(a_1,a_3)\otimes G(a_3,a_2)=G(a_1,a_2),\\
(\varphi^{-1}\circ\varphi)(E_{a_1},E_{a_2}) &= \bigvee_{a_3\in A}\varphi(a_3,E_{a_1})\otimes
\varphi(a_3,E_{a_2}) =  \bigvee_{a_3\in A}G(a_1,a_3)\otimes G(a_3,a_2) \\
&= G(a_1,a_2) = (G/E)(E_{a_1},E_{a_2}),
\end{aligned}
\]
and we conclude that $E_A^\varphi = \varphi\circ\varphi^{-1}=G$ and $E_{A/E}^\varphi =
\varphi^{-1}\circ \varphi = G/E$.

Further, let $E$ be a forward bisimulation on $\cal A$.~The assertions (C) and (D)
were proved in~Theo\-rem~5.1 \cite{CSIP.10}, but also disregarding fuzzy sets of initial and terminal states,
and in Theorem 7.1 \cite{SCI.10}, in a more general context (for fuzzy quasi-orders).~Here we are going to prove (E).

Since $E\circ \delta_x^A\circ E=\delta_x^A\circ E$, and by $E\leqslant G$ it follows $E\circ G=G\circ E=G$, for any $a_1,a_2\in A$~we have~that
\[
\begin{aligned}
(\varphi^{-1}\circ \delta_x^A )(E_{a_1},a_2)&=\bigvee_{a_3\in A}\varphi (a_3,E_{a_1})\otimes
\delta_x^A(a_3,a_2) =  \bigvee_{a_3\in A}G(a_1,a_3)\otimes \delta_x^A(a_3,a_2) =
(G\circ \delta_x^A)(a_1,a_2), \\
 (\delta_x^{A/E}\circ \varphi^{-1})(E_{a_1},a_2)&=  \bigvee_{a_3\in A} \delta_x^{A/E}(E_{a_1},E_{a_3})\otimes \varphi^{-1}(E_{a_3},a_2)  =\bigvee_{a_3\in A} (E\circ \delta_x^A\circ E)(a_1,a_3)\otimes G(a_3,a_2)   \\
&=(E\circ \delta_x^A\circ
E\circ G)(a_1,a_2)=  (\delta_x^A\circ
E\circ G)(a_1,a_2)=   (\delta_x^A\circ G)(a_1,a_2), \\
(\varphi\circ \delta_x^{A/E})(a_1,E_{a_2}) &= \bigvee_{a_3\in A}\varphi(a_1,E_{a_3})\otimes
\delta_x^{A/E}(E_{a_3},E_{a_2}) =   \bigvee_{a_3\in A}G(a_1,a_3)\otimes
(E\circ \delta_x^A\circ E)(a_3,a_2) \\
&= (G\circ E\circ \delta_x^A\circ E)(a_1,a_2) =(G\circ  \delta_x^A\circ E)(a_1,a_2)  \\
(\delta_x^A\circ \varphi) (a_1,E_{a_2})&=\bigvee_{a_3\in A} \delta_x^A(a_1,a_3)\otimes \varphi(a_3,E_{a_2}) =\bigvee_{a_3\in A} \delta_x^A(a_1,a_3)\otimes G(a_3,a_2) = (\delta_x^A\circ G) (a_1,a_2). \\
\end{aligned}
\]
Now, if $G$ is a forward bisimulation on $\cal A$, then $G\circ \delta_x^A\leqslant \delta_x^A\circ
G$ and $G\circ  \delta_x^A\circ E\leqslant G\circ \delta_x^A\circ G= \delta_x^A\circ G$,
and hence, $\varphi^{-1}\circ \delta_x^A\leqslant \delta_x^{A/E}\circ \varphi^{-1}$ and
$\varphi\circ \delta_x^{A/E}\leqslant \delta_x^A\circ \varphi$, so $\varphi $ is a forward simulation.~Conversely, if~$\varphi $ is a forward bisimulation, then we obtain that
$G\circ \delta_x^A\leqslant \delta_x^A\circ G$, i.e., $G$ is a forward bisimulation on $\cal A$.
\end{proof}

Now we are ready to state and prove the main theorem of this section.

\begin{theorem}\label{th:UFBeq.great}
Let fuzzy automata ${\cal A}=(A,\sigma^A,X,\delta^A,\tau^A)$ and ${\cal B}=(B,\sigma^B,X,\delta^B,\tau^B)$ be UFB-equivalent.

Then there~exists a uniform forward bisimulation $\varphi $ between $\cal A$ and $\cal B$ whose kernel $E_A^\varphi $ is the greatest forward bisimulation on $\cal A$ and the co-kernel $E_B^\varphi $ is the greatest forward bisimulation on $\cal B$.

Moreover, $\varphi $ is the greatest forward bisimulation between $\cal A$ and $\cal B$.
\end{theorem}

\begin{proof}
Let $\chi \in {\cal F}(A\times B)$ be a uniform forward bisimulation, and let $E_A^\chi
=E$ and $E_B^\chi =F$.~According to Theorem \ref{th:ufbr}, $E$ is a forward bisimulation on $\cal A$, $F$
is a forward bisimulation on $\cal B$ and $\widetilde\chi$ is an isomorphism of
${\cal A}/E$ onto ${\cal B}/F$, and by the proof of Theorem \ref{th:ufb.ex}, for
all $a_1,a_2\in A$ we have that
\begin{equation}\label{eq:UFBeq1}
\widetilde E(E_{a_1},E_{a_2})=\widetilde F(\widetilde\chi (E_{a_1}),\widetilde\chi (E_{a_2})).
\end{equation}
Further, let $G$ be the greatest forward bisimulation on $\cal A$, and $H$ the greatest forward bisimulation on $\cal B$, and let
$P$ be the greatest forward bisimulation on ${\cal A}/E$, and $Q$ the greatest forward bisimulation on~${\cal B}/F$. By Theorem \ref{th:G:E} it follows that $P=G/E$
and $Q=H/F$, and by (\ref{eq:UFBeq1}) and the fact that $\widetilde\chi $ is an isomorphism
of ${\cal A}/E$ onto ${\cal B}/F$ we obtain that
\begin{equation}\label{eq:UFBeq2}
\widetilde P(P_{E_{a_1}},P_{E_{a_2}})= P(E_{a_1},E_{a_2})=
Q(\widetilde\chi (E_{a_1}),\widetilde\chi (E_{a_2})) = \widetilde Q(Q_{\widetilde\chi (E_{a_1})},Q_{\widetilde\chi (E_{a_2})}).
\end{equation}
Moreover, $\widetilde\chi $ determines an isomorphism $\xi :({\cal A}/E)/P\to ({\cal B}/F)/Q$ given
by $\xi(P_{E_a})= Q_{\widetilde\chi (E_a)}$, for every~$a\in A$. According to Theorem \ref{th:G:E}, we also
have that there exist uniform forward bisimulations $\varphi_1\in {\cal F}(A\times A/E)$ and
$\varphi_2\in {\cal F}(B\times B/F)$~such that $E_A^{\varphi_1}=G$, $E_{A/E}^{\varphi_1}=G/E$,
$E_B^{\varphi_2}=H$, $E_{B/F}^{\varphi_1}=H/F$, and by Theorems~\ref{th:ufbr} and \ref{th:ufb.ex} we
obtain that $\widetilde\varphi_1$ is an isomorphism of ${\cal A}/G$ onto $({\cal A}/E)/P$, and
$\widetilde\varphi_2$ is an isomorphism of ${\cal B}/H$ onto $({\cal B}/F)/Q$, satisfying
\begin{equation}\label{eq:UFBeq3}
\widetilde G(G_{a_1},G_{a_2})= \widetilde P(\widetilde\varphi_1(G_{a_1}),\widetilde\varphi_1(G_{a_2})),
\ \ \ \widetilde H(H_{b_1},H_{b_2})= \widetilde Q(\widetilde\varphi_2(H_{b_1}),\widetilde\varphi_2(H_{b_2})),
\end{equation}
for all $a_1,a_2\in A$ and $b_1,b_2\in B$.~Next, let $\psi_1:A\to A/E$
and $\psi_2:B\to B/F$ be functions given by $\psi_1(a)=E_a$ and $\psi_2(b)=F_b$,
for all $a\in A$ and $b\in B$.~Then by (\ref{eq:F:E.nat}) it follows that $\psi_1\in
CR(\varphi_1)$~and $\psi_2\in CR(\varphi_2)$, and by (\ref{eq:tphi})
we obtain that
\begin{equation}\label{eq:UFBeq4}
\widetilde\varphi_1(G_a)=P_{\psi_1(a)}=P_{E_a}, \ \ \ \widetilde\varphi_2(H_b)=Q_{\psi_2(b)}=Q_{F_b},
\end{equation}
for all $a\in A$ and $b\in B$.

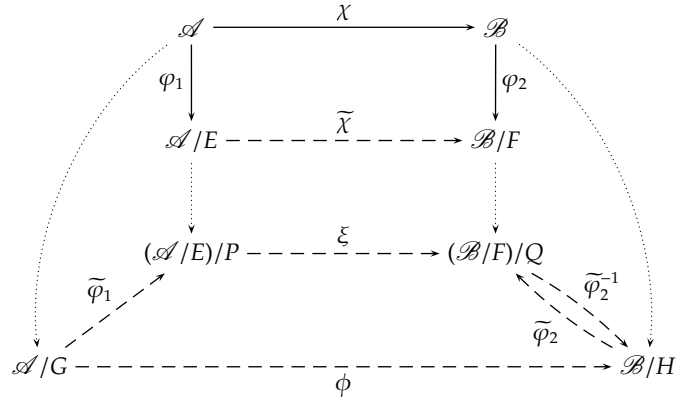
\begin{center}\small
\psset{unit=0.5cm,linewidth=0.5pt}
\newpsobject{showgrid}{psgrid}{subgriddiv=1,griddots=10,gridlabels=6pt}
\begin{pspicture}(-8,0)(8,9)
\rput[cc](-4,9){\rnode{A}{${\cal A}$}}
\rput[cc](4,9){\rnode{B}{${\cal B}$}}
\ncline[nodesep=3pt]{->}{A}{B}\aput[2pt](.50){$\chi $}
\rput[cc](-4,6){\rnode{AE}{${\cal A}/E$}}
\rput[cc](4,6){\rnode{BF}{${\cal B}/F$}}
\ncline[nodesep=3pt,linestyle=dashed]{->}{AE}{BF}\aput[2pt](.50){$\widetilde\chi $}
\rput[cc](-4,3){\rnode{AEP}{$({\cal A}/E)/P$}}
\rput[cc](4,3){\rnode{BFQ}{$({\cal B}/F)/Q$}}
\ncline[nodesep=3pt,linestyle=dashed]{->}{AEP}{BFQ}\aput[2pt](.50){$\xi $}
\rput[cc](-8,0){\rnode{AG}{${\cal A}/G$}}
\rput[cc](8,0){\rnode{BH}{${\cal B}/H$}}
\ncline[nodesep=3pt,linestyle=dashed]{->}{AG}{BH}\bput[2pt](.50){$\phi $}
\ncline[nodesep=3pt]{->}{A}{AE}\bput[2pt](.50){$\varphi_1 $}
\ncline[nodesep=3pt]{->}{B}{BF}\aput[2pt](.50){$\varphi_2 $}
\ncline[nodesep=3pt,linestyle=dotted,dotsep=1.2pt]{->}{AE}{AEP}
\ncline[nodesep=3pt,linestyle=dotted,dotsep=1.2pt]{->}{BF}{BFQ}
\ncline[nodesep=3pt,linestyle=dashed]{->}{AG}{AEP}\aput[2pt](.50){$\widetilde\varphi_1 $}
\ncarc[nodesep=3pt,linestyle=dashed]{->}{BH}{BFQ}\aput[2pt](.50){$\widetilde\varphi_2 $}
\ncarc[nodesep=3pt,linestyle=dashed]{->}{BFQ}{BH}\aput[2pt](.50){$\widetilde\varphi_2^{-1} $}
\ncarc[nodesep=3pt,linestyle=dotted,dotsep=1.2pt,arcangle=-30]{->}{A}{AG}
\ncarc[nodesep=3pt,linestyle=dotted,dotsep=1.2pt,arcangle=30]{->}{B}{BH}
\end{pspicture}
\end{center}
\medskip

Now, let a function $\phi : A/G\to B/H$ be defined by $\phi = \widetilde\varphi_1\circ \xi \circ\widetilde\varphi_2^{-1}$, i.e., let $\phi (G_a)=\widetilde\varphi_2^{-1}(Q_{\widetilde\chi
(E_a)})$, for each $a\in A$. Since $\widetilde\varphi_1$, $\xi$, and $\widetilde\varphi_2^{-1}$
are isomorphisms, we have that $\phi $ is also an isomorphism of ${\cal A}/G$ onto
${\cal B}/H$, and for arbitrary $a_1,a_2\in A$, by (\ref{eq:UFBeq2}), (\ref{eq:UFBeq3})
and (\ref{eq:UFBeq4})  it follows that
\begin{equation}\label{eq:UFBeq5}
\begin{aligned}
\widetilde G(G_{a_1},G_{a_2}) &=\widetilde P(\widetilde\varphi_1(G_{a_1}),\widetilde\varphi_1(G_{a_2}))
= \widetilde P(P_{E_{a_1}},P_{E_{a_2}}) =\widetilde Q(Q_{\widetilde\chi (E_{a_1})},Q_{\widetilde\chi (E_{a_2})}) \\
&= \widetilde H(\widetilde\varphi_2^{-1}(Q_{\widetilde\chi (E_{a_1})}),
\widetilde\varphi_2^{-1}(Q_{\widetilde\chi (E_{a_2})})) = \widetilde H(\phi(G_{a_1}),\phi(G_{a_2})). \end{aligned}
\end{equation}
Thus, by (\ref{eq:UFBeq5}) and Theorem \ref{th:ufb.ex} we obtain that
there exists a uniform forward bisimulation $\varphi\in {\cal F}(A\times B)$  such
that $E_A^\varphi =G$ and $E_B^\varphi =H$.

Next, we are going to prove that $\varphi $ is the greatest forward bisimulation between $\cal A$ and $\cal B$.~According to Theorem \ref{th:GFB}, there exists the greatest forward bisimulation $\theta $ between $\cal A$ and $\cal B$, which
is a partial fuzzy function, and clearly, $\varphi \leqslant \theta $.~Since $\varphi $ is a surjective $\cal L$-function, then $\theta $ is also a surjective $\cal L$-func\-tion, and therefore, $\theta $ is a uniform fuzzy relation.~Now, by Lemma \ref{le:UFR.order} we obtain that $CR(\varphi )\subseteq CR(\theta )$, $E_A^\varphi \leqslant E_A^\theta $,~and $E_B^\varphi \leqslant E_B^\theta $, and by Theorem \ref{th:ufbr} we obtain that $E_A^\theta $ and $E_B^\theta $ are~forward bisimulations~on~$\cal A$~and $\cal B$, respectively. Since $E_A^\varphi =G$ and $E_B^\varphi=H$ are the greatest forward bisimulations on $\cal A$ and $\cal B$, we conclude that $E_A^\varphi = E_A^\theta $ and $E_B^\varphi = E_B^\theta $.~Finally, by (vi) of Theorem \ref{th:ufr}, for arbitrary $a\in A$, $b\in B$, and $\psi\in CR(\varphi )\subseteq CR(\theta )$ we have that
\[
\theta (a,b)=E_B^\theta (\psi(a),b)=E_B^\varphi (\psi(a),b)=\varphi (a,b).
\]
Hence, $\theta=\varphi $, that is, $\varphi $ is the greatest forward bisimulation between $\cal A$ and $\cal B$.
\end{proof}

By Theorems \ref{th:UFBeq.great} and \ref{th:ufb.ex} we obtain the following corollary.

\begin{corollary}\label{cor:equiv.isom}
Let ${\cal A}=(A,\delta^A,\sigma^A,\tau^A)$ and ${\cal B}=(B,\delta^B,\sigma^B,\tau^B)$ be fuzzy automata, and let
$E$ and $F$ be the greatest forward bisimulations on $\cal A$ and $\cal B$, respectively.

Then $\cal A$ and $\cal B$ are UFB-equivalent if and only if there exists an isomorphism $\phi :{\cal
A}/E \to {\cal B}/F$ such that
\begin{equation}\label{eq:equiv.isom}
\widetilde E(E_{a_1},E_{a_2}) = \widetilde F (\phi(E_{a_1}),\phi(E_{a_2})) ,
\end{equation}
for all $a_1,a_2\in A$.
\end{corollary}

Moreover, we have the following.

\begin{corollary}\label{cor:UFBeq}
For arbitrary fuzzy automata $\cal A$, $\cal B$ and $\cal C$ the following is true:
\begin{itemize}\parskip0pt
\item[{\rm (1)}] ${\cal A}\!\sim_{UFB}\!{\cal A}$;
\item[{\rm (2)}] ${\cal A}\!\sim_{UFB}\!{\cal B}$ implies ${\cal B}\!\sim_{UFB}\!{\cal A}$;
\item[{\rm (3)}] ${\cal A}\!\sim_{UFB}\!{\cal B}$ and ${\cal B}\!\sim_{UFB}\!{\cal C}$ imply
${\cal A}\!\sim_{UFB}\!{\cal C}$.
\end{itemize}
\end{corollary}

\begin{proof}
It is clear that (1) and (2) hold, since the identity function is a uniform forward bisimulation between $\cal A$ and itself, and the inverse relation of any uniform forward bisimulation between $\cal A$ and $\cal B$ is a uniform forward bisimulation between $\cal B$ and $\cal A$.

Further, let ${\cal A}\!\sim_{UFB}\!{\cal B}$ and ${\cal B}\!\sim_{UFB}\!{\cal C}$, and let $E$, $F$, and $G$ be respectively the greatest forward bisimulations on $\cal A$, $\cal B$, and $\cal C$.~According to Corollary \ref{cor:equiv.isom}, there are isomorphisms $\phi_1:{\cal A}/E \to {\cal B}/F$ and $\phi_2:{\cal B}/F \to {\cal C}/G$ such that
\[
\widetilde E(E_{a_1},E_{a_2}) = \widetilde F (\phi_1(E_{a_1}),\phi_1(E_{a_2})) \quad\text{and}\quad
\widetilde F(F_{b_1},F_{b_2}) = \widetilde G (\phi_2(F_{b_1}),\phi_2(F_{b_2})),
\]
for all $a_1,a_2\in A$ and $b_1,b_2\in B$, and then, the composition $\phi=\phi_1\circ \phi_2:{\cal A}/E \to {\cal C}/G$ is an isomorphism of ${\cal A}/E$ onto ${\cal C}/G$ satisfying
\[
\widetilde E(E_{a_1},E_{a_2}) = \widetilde G (\phi(E_{a_1}),\phi(E_{a_2}))
\]
for all $a_1,a_2\in A$.~Therefore, ${\cal A}\!\sim_{UFB}\!{\cal C}$.
\end{proof}

According to Corollary \ref{cor:UFBeq}, $\sim_{UFB}$ is an equivalence relation
on the class of all fuzzy automata, which justifies the use of name UFB-equivalent fuzzy automata introduced at the beginning of this section.

Although the composition of two forward bisimulations is also a forward bisimulation (cf.~Lemma~\ref{Composition}), that fact was not helpful in the proof of transitivity of the UFB-equivalence, since the composition of two uniform fuzzy relations need not be a uniform fuzzy relation (cf.~Example 6.1 \cite{CIB.09}).~However, transitivity of the UFB-equivalence~follows from the fact that the composition $\varphi_1\circ\varphi_2$ of the greatest forward bisimulation $\varphi_1$ between fuzzy automata $\cal A$ and $\cal B$, and the greatest forward bisimulation $\varphi_2$ between fuzzy automata $\cal B$ and $\cal C$, is a uniform~forward bisimulation between $\cal A$ and $\cal C$.~Namely, the co-kernel of $\varphi_1$ is equal to the kernel of $\varphi_2$, this is the greatest forward bisimulation on $\cal B$.~Furthermore, $\varphi = \varphi_1\circ\varphi_2$ is the greatest forward bisimulation between $\cal A$ and $\cal C$.~Indeed, if $\theta $ is the greatest forward bisimulation between $\cal A$ and $\cal C$, then $E_C^\theta =E_C^{\varphi_2}$, the greatest forward bisimulation on $\cal C$, and by $\varphi\leqslant \theta $ and Lemma \ref{le:UFR.order} it follows that $CR(\varphi )\subseteq CR(\theta )$.~On the other hand, we have
\[
\begin{aligned}
E_C^\varphi &= \varphi^{-1}\circ\varphi=\varphi_2^{-1}\circ\varphi_1^{-1}\circ \varphi_1 \circ \varphi_2 =
\varphi_2^{-1}\circ \varphi_2\circ \varphi_2^{-1}\circ \varphi_1^{-1}\circ \varphi_1 \circ \varphi_2 =
\varphi_2^{-1}\circ E_B^{\varphi_2}\circ E_B^{\varphi_1} \circ \varphi_2 \\
&= \varphi_2^{-1}\circ E_B^{\varphi_2}\circ \varphi_2 = \varphi_2^{-1}\circ \varphi_2\circ \varphi_2^{-1}\circ \varphi_2 =
\varphi_2^{-1}\circ \varphi_2 = E_C^{\varphi_2}= E_C^\theta ,
\end{aligned}
\]
and as in the proof of Theorem \ref{th:UFBeq.great} we obtain that $\varphi=\theta $.

According to Theorem \ref{th:Lang}, UFB-equivalence implies language-equivalence.~Next we show that language-equivalence does not necessarily imply UFB-equivalence.

\begin{example}\label{ex:lang.UFB}\rm
Let $\cal L$ be the G\"odel structure, and let ${\cal A}=(A,\delta^A,\sigma^A,\tau^A)$ and ${\cal B}=(B,\delta^B,\sigma^B,\tau^B)$ be fuzzy~auto\-mata over $\cal L$, where $|A|=|B|=2$ and $X=\{x,y\}$, given by
\[\small
\delta_x^A=\begin{bmatrix}
1 & 0.5 \\ 0.5 & 1
\end{bmatrix}, \ \
\delta_y^A=\begin{bmatrix}
1 & 0.5 \\ 1 & 0.5
\end{bmatrix}, \ \
\sigma^A=\begin{bmatrix}
1 & 0
\end{bmatrix}, \ \
\tau^A=\begin{bmatrix}
0 \\ 1
\end{bmatrix}, \quad
\delta_x^B=\delta_y^B=\begin{bmatrix}
0 & 1 \\ 0 & 1
\end{bmatrix}, \ \
\sigma^B=\begin{bmatrix}
1 & 0
\end{bmatrix}, \ \
\tau^B=\begin{bmatrix}
0 \\ 0.5
\end{bmatrix}.
\]
Fuzzy automata $\cal A$ and $\cal B$ are language equivalent (cf.~Example 3.1 \cite{ICBP.10}), but they are not UFB-equivalent. Indeed, it can be easily verified that the greatest forward bisimulations on $\cal A$ and $\cal B$ are the equality relations on $\cal A$ and $\cal B$, and hence, the related factor fuzzy automata are isomorphic to $\cal A$ and $\cal B$, respectively.~Clearly, $\cal A$ and $\cal B$ are not isomorphic, and according to Corollary \ref{cor:equiv.isom}, $\cal A$ and $\cal B$ are not UFB-equivalent.

We also have that both $\cal A$ and $\cal B$ are minimal fuzzy automata in the class of all fuzzy automata which are language equivalent to them (again, cf.~Example 3.1 \cite{ICBP.10}).
\end{example}

Let us give few comments on the meaning of Corollary \ref{cor:equiv.isom}.~This result provides a way to test whether two given fuzzy automata $\cal A$ and $\cal B$ are UFB-equivalent.~First, we have to compute the greatest forward bisimulation fuzzy equivalence relations $E$ on $\cal A$ and $F$ on $\cal B$.~In numerous cases this can be done effectively using the algorithm provided in \cite{CSIP.10}.~After that, we construct factor fuzzy automata ${\cal A}/E$ and ${\cal B}/F$, and check if there is an isomorphism between them that satisfies condition (\ref{eq:equiv.isom}).~But, even when we are able to effectively compute the greatest forward bisimulations $E$ and $F$ and construct the factor fuzzy automata ${\cal A}/E$ and ${\cal B}/F$, it may be a very hard problem to determine whether there is an isomorphism between ${\cal A}/E$ and ${\cal B}/F$ that satisfies (\ref{eq:equiv.isom}).

It is important to note that the isomorphism problem for fuzzy automata is closely related to the well-known {\it graph isomorphism problem\/},~the computational problem of determining whether two finite graphs~are isomorphic.~In particular, the isomor\-phism problem for fuzzy automata over the Boolean structure is~exactly the graph isomorphism problem.~Besides its practi\-cal importance, the graph isomorphism problem is a curiosity in computational complexity theory, as it is one of a very small number of problems belonging~to NP that is neither known to be computable in polynomial time nor NP-complete.~Along with integer fac\-torization, it is one of the few important algorith\-mic problems whose rough computational com\-plexity~is still not known, and it is generally accepted that graph isomorphism is a problem that lies between P and NP-complete if P$\ne $NP (cf.~\cite{Skiena.08}).~However, although no worst-case polynomial-time algorithm is known, testing graph isomorphism is usually not very hard in practice.~The basic algorithm examines all $n!$~possible~bijec\-tions between the nodes of two graphs (with $n$ nodes), and tests whether they preserve adjacency of the nodes.~Clearly, the major problem is the rapid growth in the number of bijections when the number of nodes is growing,~which is also the crucial problem in testing isomorphism between fuzzy automata, but the algorithm can be~made more efficient by suitable partitioning of the sets of nodes as described in \cite{Skiena.08}.

Nevertheless, in our case the isomorphism test is applied not to the fuzzy automata $\cal A$ and $\cal B$, but to the factor fuzzy automata of ${\cal A}$ and ${\cal B}$ with respect to the greatest forward bisimulation equivalence
relations. The number of states of these factor fuzzy automata can be much smaller than the number of states of $\cal A$ and~$\cal B$, which can significantly affect the duration of testing.

For a fuzzy automaton ${\cal A}=(A,\delta^A,\sigma^A,\tau^A)$, the class of all fuzzy automata which are UFB-equivalent to $\cal A$ will be denoted by $\mathbb{UFB}(\cal{A})$.~We have the following.

\begin{proposition}\label{cor:minUFB}
Let ${\cal A}=(A,\delta^A,\sigma^A,\tau^A)$ be a fuzzy automaton and let $E$ be the greatest forward bisimulation on~$\cal A$.

Then ${\cal A}/E$ is a minimal fuzzy automaton in $\mathbb{UFB}(\cal{A})$.
\end{proposition}

\begin{proof}
According to Corollary \ref{cor:equiv.isom}, we have that ${\cal A}/E\in \mathbb{UFB}(\cal{A})$, and for an arbitrary ${\cal B}\in \mathbb{UFB}(\cal{A})$ we have that ${\cal A}/E$ is isomorphic to ${\cal B}/F$, where $F$ denotes the greatest forward bisimulation on $\cal B$.~Therefore, $|{\cal A}/E|=|{\cal B}/F|\leqslant |{\cal B}|$, and we conclude that ${\cal A}/E$ is a minimal fuzzy automaton in $\mathbb{UFB}(\cal{A})$.
\end{proof}

The next example shows that minimal fuzzy automata in $\mathbb{UFB}(\cal{A})$ are not necessarily unique up to an isomorphism.

\begin{example}\label{ex:minUFB}\rm
Let $\cal L$ be the G\"odel structure, and let ${\cal A}=(A,\delta^A,\sigma^A,\tau^A)$ and ${\cal B}=(B,\delta^B,\sigma^B,\tau^B)$ be fuzzy~auto\-mata over $\cal L$, where $A=\{a_1,a_2\}$, $B=\{b_1,b_2\}$, and $X=\{x,y\}$, given by
\[\small
\begin{aligned}
&\delta_x^A=\begin{bmatrix}
0.5 & 0 \\ 0.5 & 0
\end{bmatrix}, \ \
\delta_y^A=\begin{bmatrix}
1 & 1 \\ 0 & 0.5
\end{bmatrix}, \ \
\sigma^A=\begin{bmatrix}
1 & 1
\end{bmatrix}, \ \
\tau^A=\begin{bmatrix}
1 \\ 1
\end{bmatrix}, \\
&\delta_x^B=\begin{bmatrix}
0.5 & 0.5 \\ 0.5 & 0.5
\end{bmatrix}, \ \
\delta_y^B=\begin{bmatrix}
1 & 1 \\ 0.5 & 0.5
\end{bmatrix}, \ \
\sigma^B=\begin{bmatrix}
1 & 1
\end{bmatrix}, \ \
\tau^B=\begin{bmatrix}
1 \\ 1
\end{bmatrix}.
\end{aligned}
\]
The greatest forward bisimulation $E$ on $\cal A$ and the greatest forward bisimulation $F$ on $\cal B$ are given by
\[\small
E=\begin{bmatrix}
1 & 0.5 \\ 0.5 & 1
\end{bmatrix}, \ \
F=\begin{bmatrix}
1 & 0.5 \\ 0.5 & 1
\end{bmatrix},
\]
and we have that both ${\cal A}/E$ and ${\cal B}/F$ are isomorphic to $\cal B$ (cf.~Example 5.1 \cite{CSIP.10}).~We also have that
the function $\phi :{\cal A}/E\to {\cal B}/F$ given by $\phi (E_{a_1})=F_{b_1}$ and $\phi (E_{a_2})=F_{b_2}$ is an isomorphism satisfying condition (\ref{eq:equiv.isom}), and by Corollary \ref{cor:equiv.isom} we obtain that fuzzy automata $\cal A$ and $\cal B$ are UFB-equivalent.

We also have that both $\cal A$ and $\cal B$ are minimal fuzzy automata in $\mathbb{UFB}(\cal{A})=\mathbb{UFB}(\cal{B})$ (according to Corollary \ref{cor:minUFB}), but evidently, $\cal A$ and $\cal B$ are not isomorphic.
\end{example}

It is important to note that the fuzzy automata $\cal A$ and $\cal B$ are minimal not only in the class of all~fuzzy~auto\-mata which are UFB-equivalent to them, but are also minimal in the class of all fuzzy automata which are language equivalent to them.

Note that the reason why minimal automata in $\mathbb{UFB}(\cal{A})$ are not isomorphic are fuzzy equalities.~Namely, in contrast to the crisp case, where the factor automaton with respect to the equality is isomorphic to the original automaton, in the fuzzy case this is not true.~The factor fuzzy automaton with respect to a fuzzy equality has the same number of states as the original automaton, but they are not necessarily isomorphic. For instance, in the previous example both $E$ and $F$ are fuzzy equalities, ${\cal B}/F$ is isomorphic to $\cal B$, but ${\cal A}/E$ is not isomorphic to $\cal A$.

\section{Backward-forward bisimulations}\label{sectionBFB}

In the previous sections we discussed the homotypic bisimulations, specifically the forward backward bisimulations.~In this section we deal with heterotypic bisimulations, and our main goal is to underline~similarities and fundamental differences between homotypic and heterotypic bisimulations.

According to Lemma \ref{reverse}, a fuzzy relation $\varphi $ is a backward-forward bisimulation between fuzzy automata ${\cal A}$ and ${\cal B}$ if and only if it is a forward-backward bisimulation between the reverse fuzzy automata $\bar{{\cal A}}$ and $\bar{{\cal B}}$.~Consequently, for any universally valid statement on backward-forward bisimulations there is the~corre\-sponding universally valid statement on forward-backward bisimula\-tions, and in the sequel we will discuss only backward-forward bisimulations.

Let us also note that if $\varphi $ is a forward or backward bisimulation between fuzzy automata $\cal A$ and $\cal B$,~then $\varphi^{-1}$ also has this same property.~If $\varphi $ is a backward-forward or forward-backward bisimulation, then $\varphi^{-1}$ must not have the same property.~However, $\varphi $ is a backward-forward bisimulation between $\cal A$ and $\cal B$ if and only if $\varphi^{-1}$ is a forward-backward bisimulation between $\cal B$ and $\cal A$.

We can easily check that analogues of Lemmas \ref{Composition} and \ref{le:fb.union} also hold for backward-forward bisimulations. In other words, the composition of two backward-forward bisimulations and the join of an arbitrary family of backward-forward bisimulations are also backward-forward bisimulations.~Therefore,~if~there is at least one backward-forward bisimulation between
fuzzy automata ${\cal A}=(A,\delta^A,\sigma^A,\tau^A)$ and ${\cal B}=(B,\delta^B,\sigma^B,\tau^B)$, as in Theorem \ref{th:GFB} we can show that there exists the greatest backward-forward bisimulation $\varphi $ between $\cal A$ and $\cal B$.~However, unlike forward and backward bisimulations, in the case of backward-forward bisimulations we can not prove that $\varphi $ is a partial fuzzy function, because we can not prove that $\varphi\circ\varphi^{-1}\circ \varphi $ is a backward-forward bisimulation.

No matter, we can show that backward-forward bisimulations have some important properties that forward and backward bisimulations have.

\begin{theorem}\label{th:ubfb}
Let ${\cal A}=(A,\delta^A,\sigma^A,\tau^A)$ and ${\cal B}=(B,\delta^B,\sigma^B,\tau^B)$ be fuzzy automata and
let $\varphi \in {\cal F}(A\times B)$ be a uniform fuzzy relation.~Then
 $\varphi $ is a backward-forward bisimulation if and only if the following is true:
\begin{itemize}\parskip=0pt
\item[{\rm (i)}] $E_A^\varphi $ is a forward bisimulation
on $\cal A$;
\item[{\rm (ii)}] $E_B^\varphi $ is a backward bisimulation  on $\cal B$;
\item[{\rm (iii)}]  $\widetilde \varphi $ is an isomorphism of\/ factor fuzzy automata ${\cal A}/E_A^\varphi $ and ${\cal B}/E_B^\varphi $.
\end{itemize}
\end{theorem}

\begin{proof}
For the sake of simplicity set $E=E_A^\varphi $ and $E=E_B^\varphi $.~As $\varphi $ is a uniform fuzzy relation, we have that $E=\varphi\circ\varphi^{-1}$ and $F=\varphi^{-1}\circ\varphi$.

Let $\varphi $ be a backward-forward bisimulation.~Then
\[
\begin{aligned}
&E\circ \delta_x^A\circ E = \varphi\circ\varphi^{-1}\circ\delta_x^A\circ \varphi\circ\varphi^{-1} = \varphi\circ\varphi^{-1}\circ\varphi\circ \delta_x^B\circ \varphi^{-1} = \varphi\circ \delta_x^B\circ \varphi^{-1} =
\delta_x^A\circ \varphi\circ\varphi^{-1} = \delta_x^A\circ E, \\
&E\circ \tau^A =  \varphi\circ\varphi^{-1}\circ \tau^A = \varphi\circ\varphi^{-1}\circ \varphi\circ\tau^B =
\varphi\circ\tau^B = \tau^A, \\
&F\circ \delta_x^B\circ F = \varphi^{-1}\circ\varphi\circ \delta_x^B\circ \varphi^{-1}\circ\varphi =
\varphi^{-1}\circ \delta_x^A\circ\varphi\circ \varphi^{-1}\circ\varphi = \varphi^{-1}\circ \delta_x^A\circ\varphi =
\varphi^{-1}\circ\varphi \circ \delta_x^B = F\circ \delta_x^B , \\
&\sigma^B\circ F = \sigma^B\circ \varphi^{-1}\circ\varphi = \sigma^A\circ\varphi \circ \varphi^{-1}\circ\varphi =
\sigma^A\circ\varphi = \sigma^B ,
\end{aligned}
\]
and therefore, $E=E_A^\varphi$ is a forward bisimulation fuzzy equivalence relation on $\cal A$ and $F=E_B^\varphi$ is a backward bisimulation fuzzy equivalence relation on $\cal B$.~In the same way as in the proof of Theorem \ref{th:ufbr} we prove that $\widetilde{\varphi}$ is an isomorphism between fuzzy automata  ${\cal A}/E$ and ${\cal B}/F$.

Conversely, let (i), (ii), and (iii) hold.~For every $\psi \in CR(\varphi )$, $\xi \in CR(\varphi^{-1})$, $a_1,a_2\in A$, $b_1,b_2\in B$ and $x\in X$, as in the proof of Theorem \ref{th:ufbr} we show that
\[
(E\circ \delta_x^A\circ E)(a_1,a_2)= (F\circ \delta_x^B\circ F) (\psi(a_1),\psi(a_2)) ,\ \
(F\circ \delta_x^B\circ F)(b_1,b_2)= (E\circ \delta_x^A\circ E) (\xi(b_1),\xi(b_2)),
\]
and by (i) and (ii) we obtain that
\[
\begin{aligned}
&\delta_x^A\circ\varphi =\delta_x^A\circ\varphi\circ \varphi^{-1}\circ\varphi = \delta_x^A\circ E\circ\varphi =
E\circ \delta_x^A\circ E\circ\varphi = E\circ \delta_x^A\circ\varphi , \\
&\varphi \circ\delta_x^B= \varphi \circ \varphi^{-1}\circ\varphi \circ\delta_x^B = \varphi \circ F \circ\delta_x^B =
\varphi \circ F \circ\delta_x^B\circ F = \varphi  \circ\delta_x^B \circ F .
\end{aligned}
\]
Now, for all $a\in A$ and $b\in B$ we obtain that
\[
\begin{aligned}
(\delta_x^A\circ \varphi )(a,b)&= (E\circ \delta_x^A\circ\varphi )(a,b) = \bigvee_{a_1\in A}(E\circ \delta_x^A)(a,a_1)\otimes \varphi (a_1,b) = \bigvee_{a_1\in A} (E\circ \delta_x^A)(a,a_1)\otimes E(a_1,\xi (b)) \\
&= (E\circ \delta_x^A\circ E)(a,\xi(b)) = (F\circ \delta_x^B\circ F)(\psi (a),\psi(\xi(b))) = (F\circ \delta_x^B\circ F)(\psi (a),b) \\
&= \bigvee_{b_1\in B} F(\psi(a),b_1)\otimes (\delta_x^B\circ F)(b_1,b) = \bigvee_{b_1\in B} \varphi(a,b_1)\otimes (\delta_x^B\circ F)(b_1,b) = (\varphi\circ \delta_x^B\circ F)(a,b) \\
&= (\varphi\circ \delta_x^B)(a,b),
\end{aligned}
\]
and hence, $\delta_x^A\circ\varphi = \varphi\circ \delta_x^B$.~Therefore, $\varphi $ is a forward-backward bisimulation.
\end{proof}

We also have that the following is true.

\begin{theorem}\label{th:ubfb.ex}
Let ${\cal A}=(A,\delta^A,\sigma^A,\tau^A)$ and ${\cal B}=(B,\delta^B,\sigma^B,\tau^B)$ be fuzzy automata, and
let $E$ be a forward~bisimu\-lation on $\cal A$ and $F$ a backward bisimulation on $\cal B$.

Then there exists a uniform backward-forward bisimulation $\varphi\in {\cal F}(A\times
B)$ such that
\begin{equation}\label{eq:ubfb.1}
E_A^\varphi =E\ \ \mbox{and}\ \ E_B^\varphi =F ,
\end{equation}
if and only if there exists an isomorphism $\phi :{\cal
A}/E \to {\cal B}/F$ such that for all $a_1,a_2\in A$ we have
\begin{equation}\label{eq:ubfb.2}
\widetilde E(E_{a_1},E_{a_2}) = \widetilde F (\phi(E_{a_1}),\phi(E_{a_2})) .
\end{equation}
\end{theorem}

The proof of this theorem is almost the same as the proof of the Theorem \ref{th:ufb.ex} and will be omitted.
\smallskip

In contrast to forward bisimulations, which serve to define an equiv\-alence relation between fuzzy~auto\-mata, backward-for\-ward bisimulations can not be used for this purpose because they can not provide~symmetry.~Regardless of this, backward-forward bisimulations have been widely used in the literature, primarily because of its most important property to ensure the language-equivalence, as well as the fact that they are a natural generalization of homomorphisms (see the next section).

\section{Related concepts}\label{sec:related}

It is well-known that deterministic automata have a natural interpretation as
algebras in which each input symbol is realized as a unary operation.~This interpretation,
advocated by J. R. B¨uchi and J. B. Wright already in the late fifties,
linked automata with universal algebra, and it turned out that many basic notions~of universal algebra have natural interpretations in the theory of automata.~In particular,
homomorphisms~are used for defining ways of simulating an automaton
by another, and congruences are used in reduction~of the number of states.~Closely related to
homomorphisms and congruences are relational morphisms, relations between two algebras (possible different)
which are compatible with algebraic operations.~Relational mor\-phisms were introduced by Tilson \cite[Chapters 11 and 12]{Eil.76} in the theory of~semigroups, to solve some prob\-lems related to the wreath product decomposition of finite semigroups, but it turned out that they can be also useful in the study of recognizable languages (cf.~\cite{Eil.76,Pin.97}).~Later, relational morphisms were generalized to arbitrary universal algebras, and recently they were studied in the fuzzy framework (cf.~\cite{ICB.09}).~Despite~its name, relational morphisms should not be regarded as a generalization of homomorphisms, but rather as a natural extension of the concept of a congruence.~However, there is a key difference between congruences and relational morphisms.~While congruences are required to be compatible equivalence relations, relational morphisms are required only to be compatible, and uniform relational morphisms have proved oneself to be~a~more subtle and convenient extension of the notion of a congruence (cf.~\cite{ICB.09}).

Although nondeterministic, fuzzy and weighted automata can not be regarded as algebras,
many~authors studied certain concepts on these automata which are related
to homomorphisms, congruences and relational morphisms. In this section we give an overview of these concepts and
discuss their relationships with the concepts of bisimulations.

\subsection{Deterministic automata}

Let ${\cal A}=(A,\delta^A,a_0,\tau^A)$ and ${\cal B}=(B,\delta^B,b_0,\tau^B)$ be ordinary deterministic automata, that is, let $\delta_x^A:A\to A$ and $\delta_x^B:B\to B$ be functions, $a_0\in A$, $b_0\in B$, $\tau^A\subseteq A$, and $\tau^B\subseteq B$.~The standard definition of a {\it homomorphism\/} of deterministic automata says that it is a function $\varphi :A\to B$ which satisfies
\begin{align}
&\varphi (a_0)=b_0 , \label{eq:hda1i} \\
&\varphi (\delta_x^A(a)) = \delta_x^B(\varphi (a)), \ \ \text{for all $a\in A$, $x\in X$,}\label{eq:hda1} \\
&a\in \tau^A \ \Leftrightarrow\ \varphi(a)\in \tau^B, \ \ \text{for all $a\in A$.}\label{eq:hda1t}
\end{align}
It is clear that conditions (\ref{eq:hda1i})--(\ref{eq:hda1t}) can be respectively written in the same form as conditions (\ref{eq:bfsi0})--(\ref{eq:bfst0}),~which define a backward-forward bisimulation between fuzzy automata.~In other words, homomorphisms of deterministic automata are precisely those functions which are backward-forward bisimulations.

Let us also note that (\ref{eq:hda1}) is equivalent~to
\begin{equation}\label{eq:hda3}
\delta_x^A(a)=a' \Rightarrow \delta_x^B(\varphi(a))=\varphi(a'), \ \ \text{for all $a,a'\in A$, $x\in
X$,}
\end{equation}
and if we treat $\delta_x^A$ and $\delta_x^B$ as relations, then (\ref{eq:hda3}) can be written as
\begin{equation}\label{eq:hda3a}
(a,a')\in \delta_x^A\ \Rightarrow \ (\varphi(a),\varphi(a'))\in \delta_x^B, \ \ \text{for all $a,a'\in A$, $x\in
X$.}
\end{equation}
The opposite implication in (\ref{eq:hda3}) is not necessarily true, but we have that
\begin{equation}\label{eq:hda4}
\delta_x^B(\varphi(a))=\varphi(a') \Rightarrow (\exists a''\in A)\,\delta_x^A(a)=a'' \ \&\
\varphi(a'')=\varphi(a'),\ \ \text{for all $a,a'\in A$ and $x\in X$.}
\end{equation}
Condition (\ref{eq:hda4}) is not equivalent to (\ref{eq:hda1}), but slightly modifying (\ref{eq:hda4}) we obtain the condition
\begin{equation}\label{eq:hda5}
\delta_x^B(\varphi(a))= b \Rightarrow (\exists a'\in A)\,\delta_x^A(a)=a' \ \&\  \varphi(a')= b, \ \ \text{for all $a\in A$, $b\in B$ and $x\in X$,}
\end{equation}
which is equivalent to (\ref{eq:hda1}).

Another fundamental algebraic concept, which plays an important role in the theory of deterministic automata, is the concept of a congruence.~A {\it congruence\/} on a deterministic automaton ${\cal A}=(A,\delta^A,a_0,\tau^A)$ is defined as a relation $\varrho \subseteq A\times A$ satisfying the conditions
\begin{align}\label{eq:det.cong}
&(a,a')\in \varrho \ \Rightarrow\ (\delta_x^A(a),\delta_x^A(a'))\in \varrho ,\ \ \text{for all $a,a'\in A$ and $x\in X$.} \\
\label{eq:det.cong.t}
&(a,a')\in \varrho \ \Rightarrow\ (\, a\in \tau^A \Leftrightarrow a'\in \tau^A\, ),\ \ \text{for all $a,a'\in A$.}
\end{align}
Condition (\ref{eq:det.cong.t}) means that $\varrho $ is contained in the equivalence relation $\tau^A\times \tau^A\cup (A\setminus \tau^A)\times (A\setminus \tau^A)$ having only two classes $\tau^A$ and $A\setminus \tau^A$, and in some sources it is not included in the definition of a congruence.

The concept of a congruence can be transmitted to the case when we deal with two possible different~auto\-mata ${\cal A}=(A,\delta^A,a_0,\tau^A)$ and ${\cal B}=(B,\delta^B,b_0,\tau^B)$.~For a relation $\varphi \subseteq A\times B$ conditions (\ref{eq:det.cong}) and (\ref{eq:det.cong.t}) become
\begin{align}\label{eq:det.bisim}
&(a,b)\in \varphi \ \Rightarrow\ (\delta_x^A(a),\delta_x^B(b))\in \varphi ,\ \ \text{for all $a\in A$, $b\in B$, and $x\in X$,} \\
\label{eq:det.bisim.t}
&(a,b)\in \varphi \ \Rightarrow\ (\, a\in \tau^A \Leftrightarrow b\in \tau^B\, ),\ \ \text{for all $a\in A$ and $b\in B$,}
\end{align}
and clearly, condition (\ref{eq:det.bisim.t}) means that $\varphi $ is contained in the relation $\tau^A\times \tau^B\cup (A\setminus \tau^A)\times (B\setminus \tau^B)$.~A relation $\varphi $ which satisfies (\ref{eq:det.bisim}) and (\ref{eq:det.bisim.t}) was called by Rutten \cite{Rutten.98} a {\it bisimulation\/} between deterministic auto\-mata~$\cal A$~and~$\cal B$. This concept does not address the initial states, and we will require the relation $\varphi $ to also satisfy the condition
\begin{align}
\label{eq:det.bisim.i}
&(a_0,b_0)\in \varphi .
\end{align}
Following the terminology from \cite{ICB.09}, a relation $\varphi \subseteq A\times B$ which satisfies (\ref{eq:det.bisim}),
(\ref{eq:det.bisim.t}), and (\ref{eq:det.bisim.i}) is called a {\it relational morphism\/} between deterministic auto\-mata $\cal A$ and $\cal B$.~It is a matter of routine to check that condition (\ref{eq:det.bisim})~is~equi\-valent to anyone of the~following two conditions:
\begin{align}\label{eq:bisim.def.d2}
&\varphi^{-1}\circ \delta_x^A \subseteq \delta_x^B\circ \varphi^{-1},\ \ \text{for all $x\in X$,}\\
\label{eq:bisim.def.d3}
&\varphi\circ \delta_x^B \subseteq \delta_x^A\circ \varphi , \ \ \text{for all $x\in X$}.
\end{align}
Evidently, these conditions have exactly the same form as conditions (\ref{eq:fs0}) and (\ref{eq:fs0i}) occurring in the definition of a forward bisimulation between fuzzy automata.~Moreover, we have that condition (\ref{eq:det.bisim.t}) is equivalent~to the conjunction of conditions $\varphi^{-1}\circ \tau^A\subseteq \tau_B$ and $\varphi\circ \tau^B\subseteq \tau^A$, and since $\sigma^A$ and~$\sigma^B$ are singletons, condition~(\ref{eq:det.bisim.i}) is equivalent~to anyone of the conditions $\sigma^B\subseteq \sigma^A\circ \varphi$ and~$\sigma^A\subseteq \sigma^B\circ \varphi^{-1}$.~Therefore, in the terminology~that we use in this~paper, relational morphisms between deterministic automata are precisely forward bisimulations between these automata.~Moreover, considering the case when the automata $\cal A$ and $\cal B$ are the same, we can conclude that congruences on deterministic automata are precisely forward bisimulation equivalence relations on deterministic automata.~On the other hand, it is not hard to find an example of a congruence on~a~deter\-ministic automaton which is not a backward simulation.

The above-discussed concepts concerning deterministic automata can be naturally transmitted to the deterministic fuzzy automata.~Moreover, fuzzy analogues of these concepts can be defined in a similar~way as was done in \cite{ICB.09}.~Ignjatovi\'c et al.~\cite{ICB.09} introduced the concept of a fuzzy relational morphism between algebras, and they applied this concept to deterministic fuzzy automata.~Here~we~give a slightly modified version of the definition stated in \cite[Example 4.1]{ICB.09}.~Let ${\cal A}=(A,\delta^A,a_0,\tau^A)$ and ${\cal B}=(B,\delta^B,b_0,\tau^B)$ be~deterministic fuzzy~automata, which means that $\delta_x^A:A\to A$ and $\delta_x^B:B\to B$ are functions, $a_0\in A$, $b_0\in B$, and $\tau^A\in {\cal F}(A)$ and $\tau^B\in {\cal F}(B)$ are fuzzy~sets.~A fuzzy relation $\varphi \in {\cal F}(A\times B)$ is called a {\it fuzzy relation morphism\/} between $\cal A$ and $\cal B$ if it satisfies the conditions
\begin{align}
\label{eq:frm.cdfa.i}
&\sigma^A\leqslant \sigma^B\circ \varphi^{-1} \ \ \ (\text{equivalently,}\ \sigma^B\leqslant \sigma^A\circ \varphi ),\\
\label{eq:frm.cdfa}
&\varphi(a,b)\leqslant \varphi(\delta_x^A(a),\delta_x^B(b)),\ \ \text{for all $a\in A$, $b\in B$, and $x\in X$},\\
\label{eq:frm.cdfa.t}
&\varphi^{-1}\circ \tau^A\leqslant \tau_B,\ \ \ \varphi\circ \tau^B\leqslant \tau^A .
\end{align}
Since $\sigma^A=\{a_0\}$ and $\sigma^B=\{b_0\}$, condition (\ref{eq:frm.cdfa.i}) is also equivalent to $\varphi(a_0,b_0)=1$.~It is not hard to verify that~in this particular case (\ref{eq:frm.cdfa}) is equivalent both to (\ref{eq:fs0}) and (\ref{eq:fs0i}), and therefore, fuzzy relational morphisms~between deterministic fuzzy automata are precisely forward bisimulations between these automata.

\subsection{Non-deterministic automata}

Let us move from deterministic to nondeterministic automata.~Consider nondeterministic~automata ${\cal A}=(A,\delta^A,\sigma^A,\tau^A)$ and ${\cal B}=(B,\delta^B,\sigma^B,\tau^B)$, where $\delta_x^A\subseteq A\times A$ and $\delta_x^B\subseteq B\times B$ are crisp relations,~and $\sigma^A,\tau^A\subseteq A$ and $\sigma^B,\tau^B\subseteq B$ are crisp sets.~Let us assume that $\varphi :A\to B$ is a function.~In terms of nondeterministic automata condition (\ref{eq:hda3}) can be written as
\begin{equation}\label{eq:mnda1}
(a,a')\in \delta_x^A \Rightarrow (\varphi(a),\varphi(a'))\in \delta_x^B, \ \ \text{for all $a,a'\in
A$ and $x\in X$,}
\end{equation}
and (\ref{eq:hda5}) can be written as
\begin{equation}\label{eq:mnda2}
(\varphi(a),b)\in \delta_x^B \Rightarrow (\exists a'\in A)\, (a,a')\in \delta_x^A\ \&\ \varphi(a')= b, \ \ \text{for all $a\in A$, $b\in B$, and $x\in X$.}
\end{equation}
Conditions (\ref{eq:mnda1}) and (\ref{eq:mnda2}) were first discussed by Bloom and \'Esik \cite[Example 9.4.9]{BE.93}, along with two~conditions on initial and terminal states which can be stated in the same form as conditions (\ref{eq:bfsi0}) and (\ref{eq:bfst0}) (the second one is also equivalent to (\ref{eq:hda1t})).~Conditions (\ref{eq:mnda1}) and (\ref{eq:mnda2}) were also used by Calude et
al.~\cite{CCK.00} in the definition~of~a {\it morphism\/} of nondeterministic automata with outputs, along with a condition relating to the output function.~In fact, they considered condition (\ref{eq:mnda1}) with
equivalence instead of the implication, but the opposite implication is superfluous.~As we have already
noted, this implication~is not necessarily true for deterministic automata.

For~deterministic automata condition (\ref{eq:bfs0}) is equivalent both to (\ref{eq:hda3}) and (\ref{eq:hda5}),
but for nondeterministic~automata (\ref{eq:bfs0}) is not equivalent neither to (\ref{eq:mnda1}) nor to
(\ref{eq:mnda2}), but~it~is~equivalent to the conjunction of (\ref{eq:mnda1}) and (\ref{eq:mnda2}).
Namely, condition (\ref{eq:mnda1}) is equivalent to
\begin{equation}\label{mnda.1a}
\delta_x^A\circ \varphi \subseteq \varphi\circ \delta_x^B, \ \ \text{for all $x\in X$,}
\end{equation}
and (\ref{eq:mnda2}) is equivalent to
\begin{equation}\label{mnda.2a}
\varphi\circ \delta_x^B \subseteq \delta_x^A\circ \varphi, \ \text{for all $x\in X$.}
\end{equation}
Clearly, conditions (\ref{mnda.1a}) and (\ref{mnda.2a}) are precisely conditions (\ref{eq:bs0}) and (\ref{eq:fs0i}) stated in terms of nondeter\-ministic automata.~Conditions (\ref{eq:bfsi0})--(\ref{eq:bfst0}) also served to define simulations between weighted automata (cf.~discussion at the end of this section), and any relation between nondeterministic automata which satisfies (\ref{eq:bfsi0})--(\ref{eq:bfst0}) is also called a {\it simulation\/} between nondeterministic automata \cite{EK.01,EM.10}.~Hence, simulations between nondeterministic automata in the sense of the definitions from \cite{BE.93,EK.01,EM.10} are precisely backward-forward~bisi\-mu\-lations in the terminology that we use in this paper.

It is important to note that a function $\varphi :A\to B$ which satisfies condition (\ref{eq:mnda1}), that is, condition~(\ref{mnda.1a}), along with conditions $\sigma^A\circ \varphi \subseteq \sigma^B$ and $\tau^A\subseteq \varphi\circ \tau^B$, was studied by Lombardy \cite{Lomb.02} and called a~{\it morphism\/} between nondeterministic automata.~Evidently, morphisms between nondeterministic automata~are precisely those functions which are backward simulations in the terminology that we use here.

What are congruences on nondeterministic automata?~In algebra, congruences are required to be equivalence relations  compatible with algebraic operations, and compatibility is needed to provide correct~defini\-tion of algebraic operations on the corresponding factor set and to turn the factor set into an algebra~of~the same type.~Compatibility also ensures transfer of the substantial algebraic properties of the~original~algebra to the factor algebra.~In the case nondeterministic automata, every equivalence relation allows us~to turn the related factor set into a correctly defined nondeterministic automaton, and, for instance, to prove certain analogues of the well-known Homomorphism Theorem and Second Isomorphism Theorem from universal algebra (cf.~Section 3, Theorem \ref{th:nat.uff}, and \cite{CSIP.07,CSIP.10,SCI.10}).~However, not every equivalence relation enables to transmit~all~the~sub\-stantial properties of the original automaton to the factor automaton.~For example, the original automaton and the factor~auto\-maton do not necessarily recognize the same language.~As we~already~mentioned~in~Section 3 for fuzzy automata, the factor automaton recognizes the same language
as the original automaton if and only if the considered equivalence relation is a solution to the general system (the system (\ref{eq:sFRE})).

Therefore, any equivalence relation which is a solution to the general system can be understood as a congruence on a nondeterministic automaton.~However, the general system may consist of infinitely many equations, and finding its nontrivial solutions may be a very difficult task.~In other words, there will be no much~useful to look at congruences as the equivalence relations which are solutions to the general system.~It is more~convenient~to consider some instances of the general system, instead of the general system itself, where an instance~should be understood as a system, built from the same relations,~whose sets of solutions are contained in the set of all solutions to the general system.~These instances have to be as general as possible, but they have to consist of finitely many equations and to be ``easier'' to solve.~Two such instances were studied Ilie, Yu and others \cite{IY.02,IY.03,INY.04,ISY.05},
and their solutions were called {\it right\/} and {\it left invariant equivalence
relations\/}.~In our terminology, these are just forward and backward bisimulation equivalence~relations. Well-behaved equivalence relations studied by Calude~et al.~\cite{CCK.00} also coincide with forward bisimulation equivalence relations.~Both mentioned types of equivalence relations were used in reduction of the number of states of nondeterministic automata.~In state reduction, Ilie, Navaro, and Yu \cite{INY.04} and Champarnaud and Coulon \cite{CC.04} also used natural equivalence relations of right and left invariant quasi-orders (preorders), which are not necessarily right or left invariant equivalence relations, but are solutions to the general system.

It is worth to note that none of the above mentioned two types of equivalence
relations, forward and backward bisimulation equivalence relations, can not be considered better than the other.~There are~cases~where one of~them better reduces the number of states, but there are also other cases where the another one gives a better reduction.~There are also cases where each of them individually causes a polynomial~reduction of the number of states, but alternately using both types of equivalence relations the number of states can be reduced exponentially (cf.~\cite[Section 11]{IY.03}).~Let us also note that forward bisimulations of nondeterministic automata were discussed in a Kozen's book \cite[Suppl.~B]{Kozen.97}.

As we will see later in the discussion of fuzzy automata, there are also other important and ``easily~solvable''~instances~of the general system, whose solutions can also be serious candidates for the title~of~a~congruence on a nondeterministic automaton.~In any case, there is still no a unique definition of a congruence on this type of automata.

\subsection{Fuzzy automata}

Let ${\cal A}=(A,\delta^A,\sigma^A,\tau^A)$ and ${\cal B}=(B,\delta^B,\sigma^A,\tau^B)$ be fuzzy automata,
but let $\varphi :A\to B$ be an~ordinary~function. In terms of fuzzy
automata~conditions (\ref{eq:mnda1}) and (\ref{eq:mnda2}) can be written as
\begin{eqnarray}\label{eq:hfa1}
&\delta_x^A(a,a')  \leqslant \delta_x^B(\varphi(a),\varphi(a')), \ \ \text{for all $a,a'\in A$, $x\in
X$,}& \\
\label{eq:hfa2}
&\delta_x^B(\varphi(a),b) \leqslant \displaystyle\bigvee_{a'\in A,\varphi(a')= b}  \delta_x^A(a,a') ,\ \ \text{for all $a\in A$, $b\in B$, $x\in X$.}&
\end{eqnarray}
We have that (\ref{eq:hfa1}) is equivalent to (\ref{eq:bs0}), and (\ref{eq:hfa2}) is
equivalent to (\ref{eq:fs0i}), so the conjunction of
(\ref{eq:hfa1}) and (\ref{eq:hfa2}) is equivalent to (\ref{eq:bfs0}).~We can also show that the
conjunction of (\ref{eq:hfa1}) and (\ref{eq:hfa2}) is equivalent to
\begin{equation}\label{eq:hfa3}
\delta_x^B(\varphi(a),b) = \bigvee_{a'\in A,\varphi(a')= b}  \delta_x^A(a,a') , \ \ \text{for all $a\in A$, $b\in B$, $x\in X$.}
\end{equation}
Thus, (\ref{eq:bfs0}) is also equivalent to (\ref{eq:hfa3}).~A function $\varphi $ satisfying
condition (\ref{eq:hfa3}) was called by Petkovi\'c~\cite{P.06} a {\it homomorphism\/} of fuzzy
automata.~In fact, she studied fuzzy automata with outputs, and her definition~contains also an additional condition concerning
the fuzzy output function.~However, this definition can be easily turned into a definition of a homomorphism of fuzzy automata with fuzzy sets of initial and terminal states, replacing the condition concerning the fuzzy output function by conditions (\ref{eq:bfsi0})~and~(\ref{eq:bfst0}).~Then homo\-morphisms of fuzzy automata are precisely those functions which are backward-forward bisimulations.

In~a~different context, functions satisfying (\ref{eq:hfa3}) were studied in \cite{MMS.97},
and they were called {\it coverings\/} (see also \cite{CKL.01,KKC.98,KC.02,MM.02}). It is worth noting
that Malik and Mordeson~in~\cite[\P 6.3]{MM.02} used the name homomorphism for a function $\varphi $
satisfying (\ref{eq:hfa1}) (see also \cite{CKL.01,KKC.98,KC.02}), and the name {\it strong homomorphism\/} for a function $\varphi $ satisfying
\begin{equation}\label{eq:shfa}
\delta_x^B(\varphi(a),\varphi(a')) = \bigvee_{a''\in A,\varphi(a'')=\varphi(a')} \delta_x^A(a,a''), \ \ \text{for all $a,a'\in A$, $x\in X$.}
\end{equation}
Condition (\ref{eq:shfa}) is stronger than (\ref{eq:hfa1}), but it is weaker than
(\ref{eq:hfa3}). If $\varphi $ is~surjective, then condition (\ref{eq:hfa2}) is equivalent to
(\ref{eq:shfa}).~It is important to note that there were no conditions concerning fuzzy sets of initial and terminal states, because
the above mentioned papers dealt only with fuzzy automata without initial and terminal states.

The same comments we made in the discussion that dealt with congruences on nondeterministic~auto\-mata, can be also made when working with fuzzy automata.~However, there are some issues that are~spe\-cific to fuzzy automata.~For instance, one of the main questions is whether we should consider congruences as crisp or fuzzy relations.~As in the case of nondeterministic~automata, reduction of the number~of states of fuzzy automata was usually performed on the model of minimization of deterministic automata, by computing and merging~in\-dis\-tin\-guishable states, and it was based on the use of crisp relations (cf.~\cite{BG.02,CM.04,LL.07,MMS.99,MM.02,P.06}). Even the term "minimization"~was~used, which~is~not~adequate because the resulting fuzzy automaton is not necessarily minimal in the class of all fuzzy automata which are equivalent to the original fuzzy automaton.~For a fuzzy automaton ${\cal A}=(A,\delta^A,\sigma^A,\tau^A)$ and a crisp equivalence
relation $\varrho \subseteq A\times A$, we can~show that the condition
\begin{equation}
\label{eq:fbe.rho}
\varrho\circ \delta_x^A \leqslant \delta_x^A\circ \varrho , \ \ \ \text{for every $x\in X$,}
\end{equation}
is equivalent to
\begin{equation}
\label{eq:cong.TP}
\bigvee_{c'\in\varrho_c}\delta_x^A(a,c')=\bigvee_{c'\in\varrho_c}\delta_x^A(b,c'), \ \ \ \text{for all $x\in X$ and $a,b,c\in A$ such that $(a,b)\in \varrho $.}
\end{equation}
Condition (\ref{eq:cong.TP}) was discussed by Petkovi\'c \cite{P.06} in the context of fuzzy automata with outputs, and~equiva\-lence relations satisfying (\ref{eq:cong.TP}), along with a certain condition on the fuzzy output function, were called~{\it congru\-ences\/} on fuzzy automata.~The same concept, formulated in terms of partitions, was called by~Basak~and~Gupta~\cite{BG.02} a {\it partition with substitution property\/}.~If the underlying structure $\cal L$ of membership values is linearly ordered, then conditions (\ref{eq:cong.TP}) and (\ref{eq:fbe.rho}) are equivalent to
\begin{equation}
\label{eq:adm.rel}
(\forall a,b,c\in A)(\forall x\in X)\,\left((a,b)\in \varrho\ \ \&\ \ \delta_x^A(b,c)>0\right)\ \Rightarrow \ \left(
(\exists d\in A)\,\delta_x^A(a,d)\geqslant \delta_x^A(b,c)\ \ \&\ \ (c,d)\in \varrho \right),
\end{equation}
and if the ordering in $\cal L$ is not linear, then (\ref{eq:adm.rel}) implies (\ref{eq:cong.TP}) and (\ref{eq:fbe.rho}), but the opposite implication~does~not necessarily hold.~Condition (\ref{eq:adm.rel}) was considered by Malik and Mordeson~in~\cite[\P 6.4]{MM.02} in the context of fuzzy automata without outputs, and without fixed fuzzy sets of initial and terminal states, and equivalence relations satisfying (\ref{eq:adm.rel}) were called {\it admissible relations\/}.~Therefore, all the mentioned concepts amounts to the concept of a forward bisimulation equivalence relation.

A different approach to state reduction of fuzzy automata was recently proposed in \cite{CSIP.07,CSIP.10,SCI.10}.~The~basic idea is to use fuzzy equivalence
relations instead of the ordinary equivalence relations.~As with nondeterministic automata, the factor fuzzy automaton with respect to a given fuzzy equivalence
relation recognizes the same fuzzy language as the original fuzzy automaton if and only if this fuzzy equivalence relation is a solution to the general system.~Some basic properties of the general system and its numerous instances were studied in \cite{CSIP.07,CSIP.10,SCI.10}.~Among the instances of the general system are the system given by (\ref{Aca1}) and (\ref{Aca1t}), and the system given by (\ref{Aca2})~and~(\ref{Aca2s}), whose solutions in ${\cal E}(A)$ are just forward and backward bisimulation fuzzy~equi\-valence relations.~In \cite{CSIP.07,CSIP.10}, forward and backward bisimulation fuzzy equivalence
relations were called {\it right\/} and {\it left~invar\-iant fuzzy equivalence
relations\/}, and in~\cite{ICB.10}, solutions to similar systems were called {\it right\/} and {\it left regular fuzzy~relations\/}. It was proved that every fuzzy automaton possesses the greatest forward and backward bisimulation~equi\-valence
relations, and iterative algorithms for computing these greatest solutions were provided, which work~when\-ever the underlying structure of membership values is a locally finite complete residuated lattice. Otherwise, some sufficient conditions were determined under which the algorithms work (cf.~\cite{CSIP.07,CSIP.10}).~Moreover, these algorithms were modified so that they compute the greatest crisp forward and backward bisimulation~equi\-va\-lence relations, and these new algorithms work
when the underlying structure of membership values is an~arbi\-trary complete residuated lattice.~However, it~was shown that better reductions can be achieved by using the greatest forward and backward bisimulation fuzzy~equi\-va\-lence
relations than by using their crisp counterparts.

Important instances of the general system are the system
\begin{align}\label{eq:srife}
&E\circ \delta_x^A= \delta_x^A,\qquad\text{for each }x\in X,  &E\circ \tau^A\leqslant \tau^A ,&&&
\end{align}
whose solutions in ${\cal E}(A)$ are called {\it strongly right invariant fuzzy equivalence relations\/}, and the system
\begin{align}\label{eq:slife}
&\delta_x^A\circ E= \delta_x^A,\qquad\text{for each }x\in X, &\sigma^A\circ E\leqslant \sigma^A,&&
\end{align}
whose solutions in ${\cal E}(A)$ are called {\it strongly left invariant fuzzy equivalence relations\/}.~The greatest solutions to systems (\ref{eq:srife}) and (\ref{eq:slife}) can be computed more easily than the greatest solutions to systems (\ref{Aca1})--(\ref{Aca1t}) and (\ref{Aca2})--(\ref{Aca2s}), using effective procedures which are not iterative and work when the underlying structure of membership values is an arbitrary complete residuated lattice.~However, solutions to (\ref{eq:srife}) and (\ref{eq:slife}) form subsets of the sets of solutions to (\ref{Aca1})--(\ref{Aca1t}) and (\ref{Aca2})--(\ref{Aca2s}), which means that the greatest solutions to (\ref{eq:srife}) and (\ref{eq:slife}) can be strictly less than the greatest solutions to (\ref{Aca1})--(\ref{Aca1t}) and (\ref{Aca2})--(\ref{Aca2s}).~Consequently, the greatest strongly right and left invariant fuzzy equivalence relations give worse reductions than the greatest forward and backward bisimulation fuzzy~equi\-va\-lence relations, and even than the greatest forward and backward bisimulation crisp equi\-va\-lence
relations (cf.~\cite{CSIP.07,CSIP.10,SCI.10}).

Instances of the general system that are even more general than the previous ones are the system
\begin{equation}\label{eq:wrife}
E\circ \tau_u^A= \tau_u^A,\qquad\text{for each }u\in X^*,
\end{equation}
where $\tau_u^A=\delta_u^A\circ \tau^A$, whose solutions in ${\cal E}(A)$ are called {\it weakly right invariant fuzzy equivalence relations\/}, and the system
\begin{equation}\label{eq:wlife}
\sigma_u^A\circ E= \sigma_u^A,\qquad\text{for each }u\in X^*,
\end{equation}
where $\sigma_u^A=\sigma^A\circ \delta_u^A$, whose solutions in ${\cal E}(A)$ are called {\it weakly left invariant fuzzy equivalence relations\/} \cite{SCI.10}.~These~two systems have larger sets of solutions than systems (\ref{Aca1})--(\ref{Aca1t}) and (\ref{Aca2})--(\ref{Aca2s}), and also~than systems (\ref{eq:srife})~and (\ref{eq:slife}), and therefore, their greatest solutions give better reductions than the greatest solutions to these other systems.~In some cases, the greatest
weakly right and left invariant fuzzy equivalence relations are even easier to compute than the greatest forward and backward bisimulation
fuzzy equivalence relations.~However, there is a problem concern\-ing the number of equations in (\ref{eq:wrife}) and~(\ref{eq:wlife}).~This number can be infinite, and even if it is finite, it can be exponential in the number of states of the fuzzy automaton $\cal A$.~Namely, the formation of the systems (\ref{eq:wlife}) and~(\ref{eq:wrife}), i.e., the computing of the fuzzy relations $\sigma_u^A$ and $\tau_u^A$, for all $u\in X^*$, amounts to the determinization of the fuzzy automaton $\cal A$ and its reverse fuzzy automaton by means of the Nerode automata (cf.~\cite{CDIV.10,ICB.08,ICBP.10}).

It is worth noting that even better results in the state reduction can be achieved by alternating reductions by means of the greatest forward and backward bisimulation fuzzy equivalence relations, etc.~In addition, in all previous considerations fuzzy equivalence relations can be replaced by fuzzy quasi-orders,~and~appli\-cation of fuzzy quasi-orders can give even better results in the state reduction.

From a different point of view, state reduction and equivalence of fuzzy automata were studied by N.~C. Basak and A. Gupta \cite{BG.02}, W. Cheng and Z. Mo \cite{CM.04}, H. Lei and Y. M. Li \cite{LL.07}, D. S. Malik, J. N. Mordeson and M. K. Sen \cite{MMS.99}, K. Peeva \cite{P.04b}, K. Peeva and Z. Zahariev \cite{PZ.08}, T. Petkovi\'c \cite{P.06}, L. H. Wu and D. W. Qiu \cite{WQ.10}, H. Xing, D.W. Qiu, F. C. Liu and Z. J. Fan \cite{XQLF.07}, as well as in the books by J. N. Mordeson and D. S. Malik \cite{MM.02} and K. Peeva and Y. Kyosev \cite{PK.04}.~The basic idea exploited in these sources was to reduce the number~of states of a fuzzy automaton by computing and merging indistinguish\-able states, based on the model of the minimization algorithm for deterministic automata.~However, in the deterministic case we can effectively detect and merge indistinguishable~states, but in the nondeterministic case we have sets of states and it is seemingly very difficult to decide whether two states are distin\-guish\-able or not.~What we shall
do is to find a superset such that one is certain not to merge state~that should not be merged.~There can always be
states which could be merged but detecting those is too computationally expensive.~In the case of fuzzy automata this
problem is even worse because we work with fuzzy sets of states.~Moreover, detecting and merging indistinguishable states gives a crisp relation between states, but it was shown in \cite{CSIP.07,CSIP.10} that the use of fuzzy relations (fuzzy equivalence relations, fuzzy quasi-orders) in general gives better reductions.~Anyway, relationship between the results concerning the reduction and equivalence of fuzzy automata obtained in the above mentioned sources, especially those from \cite{P.04b,PZ.08,WQ.10,XQLF.07}, and those obtained in \cite{CSIP.07,CSIP.10,SCI.10} and here is not clear enough and should be further studied.

Note that in some of the mentioned sources the term minimization was used instead of the term~reduc\-tion, but this term is not adequate because it does not mean the usual construction of the~minimal automaton in the set of all fuzzy automata recognizing a given fuzzy language, i.e., algorithms that were developed there do not necessar\-ily result in a minimal fuzzy automaton of a fuzzy language.~The actual minimization algorithms were given only in the case of crisp-deterministic fuzzy automata, by J. Ignjatovi\'c, M. \'Ciri\'c, S. Bogdanovi\'c and T. Petkovi\'c \cite{ICBP.10} and Y. M. Li and W. Pedrycz \cite{LP.07}.

\subsection{Weighted automata}

Weighted and fuzzy automata are very similar.~All of them are classical nondeterministic automata in which transitions, initial and final states take values from certain structures.~For weighted~auto\-mata these values are called {\it weights\/}, and are usually taken from semirings, and for fuzzy automata they are called {\it truth values\/} or {\it membership values\/}, and are taken from certain ordered structures, the most often from lattice-ordered structures.~Specifically, a {\it semiring\/} is an algebra ${\cal S}=(S,+,\cdot,0,1)$ such that $(S,+,0)$ is a commutative monoid, $(S,\cdot ,1)$ is a monoid, multiplication $\cdot $ distributes over addition $+$, and $0\cdot x = x\cdot 0=0$, for every $x\in S$.~A {\it weighted automaton over\/} $\cal S$ and an alphabet $X$, or simply a {\it weighted automaton\/}, is a quadruple ${\cal A}=(A,\delta^A,\sigma^A,\tau^A )$, where $A$ is a non-empty finite set of states, $\delta^A:A\times X\times A\to S$ is a {\it weighted transition function\/}, $\sigma^A: A\to S$ is an {\it initial weight vector\/}, and $\tau^A : A\to S$ is a {\it terminal weight vector\/}.~As~weighted automata are defined in the same way as fuzzy automata, all notions and notation that have been previously introduced for fuzzy automata are defined in the same way for weighted automata~(only the operations $\lor $ and $\otimes $ are replaced by the operations $+$ and $\cdot $, respectively).~In fact, every fuzzy automaton over a complete residuated lattice ${\cal L}=(L,\land ,\lor , \otimes ,\to , 0, 1)$ can be treated as a weighted automaton over the semiring ${\cal L}^*=(L,\lor , \otimes ,0, 1)$. The semiring ${\cal L}^*$ is usually called the {\it semiring reduct\/} of $\cal L$.

Let ${\cal A}=(A,\delta^A,\sigma^A,\tau^A)$ and ${\cal B}=(B,\delta^B,\sigma^A,\tau^B)$ be weighted automata.~Bloom and \'Esik \cite{BE.93}, and \'Esik and Kuich \cite{EK.01} defined a {\it simulation\/} between weighted automata $\cal A$ and $\cal B$ as
a weighted relation $\varphi :A\times B\to S$ (a matrix over $\cal S$) which satisfies conditions (\ref{eq:bfsi0})--(\ref{eq:bfst0}).~Simulations were introduced in order to provide a structural characterization of equivalence of weighted automata, and in \cite{EK.01} they were used in the~com\-ple\-teness proof of the iteration semiring theory axioms for regular languages.~Under the same name,~simu\-lations were studied in \cite{EM.10}, and under different names in
\cite{Buch.08,BP.03,BLS.05,BLS.06,LS.05,Sak.09}.~B\'eal and Perrin \cite{BP.03} used the name {\it backward elementary equivalence\/}, B\'eal, Lombardy, and Sakarovitch \cite{BLS.05,BLS.06} the name {\it conjugacy\/}, which originates from applications in symbolic dynamics, and Buchholz \cite{Buch.08} the name ({\it forward\/}) {\it relational simulation\/}.~In our terminology these are precisely the backward-forward bisimulations.

The above-mentioned concept is clearly a generalization of the concept of a relational morphism.~In~order to obtain a concept that is a generalization of a homomorphism, many authors have required that $\varphi $ is also a surjective function from $A$ onto $B$ (a surjective functional matrix).~In \cite{BLS.05,BLS.06}, a surjective function $\varphi :A\to B$ which satisfies (\ref{eq:bfsi0})--(\ref{eq:bfst0}) was called a {\it covering\/} (or, more precisely, a {\it $\cal S$-covering\/}, cf.~also \cite{LS.05,Sak.09}), and in \cite{Buch.08} it was called a ({\it forward\/}) {\it bisimulation\/}.~On the other hand, a weighted relation $\varphi :A\times B\to S$ satisfying (\ref{eq:fbsi0})--(\ref{eq:fbst0}) (a forward-backward bisimulation, in our terminology) was called in \cite{BP.03} a {\it forward elementary~equivalence\/}, and a surjective function $\varphi :A\to B$ which satisfies (\ref{eq:fbsi0})--(\ref{eq:fbst0}) was called in \cite{Buch.08} a {\it backward bisimulation\/}.

It is important to note that B\'eal, Lombardy, and Sakarovitch \cite{BLS.05,BLS.06} found that a semiring $\cal S$ often has~the following property:~two weighted automata over $\cal S$ are equivalent (they define the same power series,~a func\-tion from $X^*$ to $S$) if and only if they are connected by a finite chain of simulations.~Semirings having this property include the Boolean semiring \cite{BE.93}, the semiring
of natural numbers and the ring of integers \cite{BLS.05,BLS.06}, etc.~An example of a semiring which does not have this property is the tropical semiring \cite{EM.10}.

Note that all the concepts that we have discussed above amounts to either backward-forward or forward-backward bisimulations.~Forward and backward bisimulations have never been considered in the context of weighted automata but it is not surprising.~They are defined by inequalities, and semirings are generally~not required to be ordered.~The question is how to define forward and backward bisimulations~for weighted auto\-mata over semirings?~Our ideas are, or to use as a model the equivalent definition of a forward~bi\-simu\-lation given in Theorem~\ref{th:ufbreqcond}~(and~the corresponding definition of a backward bisimulation), or to~search~for an appropriate ordering on a semiring or for a semiring with such ordering, which will enable the development of the theory of forward and~back\-ward bisimulations for weighted automata.~These issues will be the subject of our further research.

\subsection{A few additional comments}

Homomorphisms, congruences, relational morphisms and their generalizations have been studied not only in the context of algebraic structures, deterministic, nondeterministic, fuzzy and weighted automata, but also in the more general context of relational structures.~Here we will not discuss this issue, but we will just point to the articles \cite{BK.86,KK.01,Sch.06,WB.98,Yeh.68,Yeh.73}, where we can find more information about it.

In the introduction we gave some general remarks on bisimulations and their applications, and~in~this section we have made some comments on bisimulations for deterministic and nondeterministic automata. Bisimulations for deterministic and nondeterministic automata are also appearing in many applications, but also it will not be discussed in this paper.~It is worth to point out that bisimulations are also appearing in the context of stochastic, timed and hybrid automata.~For more information about all that we refer to the books and articles \cite{AILS.07,Brihaye.07,GPP.03,HKPV.98,LV.95,Milner.89,Milner.99,RM-C.00,Rutten.98,Sang.09}, as well as to the literature mentioned there.

\section{Concluding remarks}

In this article we have formed a conjunction of bisimulations and uniform fuzzy relations as a very~power\-ful tool in the study of equivalence between fuzzy automata.~In this symbiosis, uniform fuzzy relations serve as fuzzy equivalence relations which relate elements of two possibly different sets, while bisimulations provide compatibility with the transitions, initial and terminal states of fuzzy automata.~We have proved that a uniform fuzzy relation between fuzzy automata $\cal A$ and $\cal B$ is a forward bisimulation if and only if its kernel and co-kernel are forward bisimulation fuzzy equivalence
relations on $\cal A$ and $\cal B$ and there is a special isomorphism between factor fuzzy automata with respect to these fuzzy equivalence relations.~As a consequence~we~get~that~fuzzy auto\-mata
$\cal A$ and $\cal B$ are UFB-equivalent, i.e., there is a uniform forward bisimulation between them, if and only if there is a special isomorphism between the factor fuzzy automata of $\cal A$ and $\cal B$ with respect~to~their greatest forward bisimulation fuzzy equivalence relations.~This result reduces the problem of testing UFB-equivalence to the problem of testing isomorphism of fuzzy automata, which is closely related to the well-known graph isomorphism problem.~We have shown that some similar results are also valid for backward-forward bi\-simulations, but there are many significant differences.~Because of the duality with the studied concepts, backward and forward-backward bisimulations have not been considered separately.

In the penultimate section of the article we have given a comprehensive overview of
various concepts on deterministic, nondeterministic, fuzzy, and weighted automata, which are related to bisimulations, as well as to the algebraic concepts of a homomorphism, congruence, and relational morphism.~We have shown~that all these concepts amount either to forward or to backward-forward bisimulations.~However, this does not mean that forward and backward-forward bisimulations can be considered better than backward and~for\-ward-backward bisimulations.~We have pointed out that in the state reduction of fuzzy automata forward and backward bisimulation fuzzy equivalence relations give equally good results.~Moreover, it was proved in \cite{SCI.10}
that backward bisimulation fuzzy equivalence relations can be successfully applied in the conflict analysis of fuzzy discrete event systems, while forward bisimulation fuzzy equivalence relations can not be used for this purpose.

In further research we will focus on weighted automata and try to answer the question of how to~overcome the lack of suitable ordering in semirings and develop the theory of forward bisimulations.~We~also~intend to further study backward-forward bisimulations for weighted automata.~In the theory of fuzzy auto\-mata we will even more deeply explore relationships between the language-equivalence and various types of structural equivalence between fuzzy automata, and we  will try to apply the methodology developed~here in the study of fuzzy automata with outputs.

\end{document}